\documentclass[conference]{IEEEtran}
\usepackage{xspace}
\usepackage{amsmath}
\usepackage{amsfonts}
\usepackage{amsthm}
\usepackage{paralist}
\usepackage{color}
\usepackage[table]{xcolor}
\usepackage{mathtools}
\usepackage{array}
\usepackage{tikz}
\usetikzlibrary{arrows}
\usetikzlibrary{patterns}
\usepackage{stackrel}
\usepackage{pstricks}
\usepackage[skip=3pt,font=footnotesize]{subcaption}
\usepackage{amssymb}
\usepackage{wrapfig}
\usepackage[utf8]{inputenc}
\usepackage{pgfplots}
\usepackage{booktabs}
\usepackage{expl3}
\usepackage{collcell}
\usepackage[hidelinks]{hyperref}
\usepackage[hyphenbreaks]{breakurl}

\newcommand{\myparagraph}[1]{\smallbreak\noindent\textit{#1}:\xspace}

\newif\iffull  
\fulltrue     

\newcounter{note}[section]

\newcommand{\refs}[2]{\mbox{\ref{#1}--\ref{#2}}\xspace}
\newcommand{\secref}[1]{\mbox{Sec.~\ref{#1}}\xspace}

\newcommand{\lineref}[1]{\mbox{line~\ref{#1}}\xspace}
\newcommand{\linesref}[2]{\mbox{lines~\ref{#1}--\ref{#2}}\xspace}

\newcommand{\secrefstatic}[1]{\mbox{Section~{#1}}}
\newcommand{\figref}[1]{\mbox{Fig.~\ref{#1}}\xspace}
\newcommand{\figrefstatic}[1]{\mbox{Fig.~{#1}}\xspace}
\newcommand{\tblref}[1]{\mbox{Table~\ref{#1}}\xspace}
\newcommand{\eqnref}[1]{\mbox{(\ref{#1})}\xspace}
\newcommand{\eqnsref}[2]{\mbox{(\ref{#1})--(\ref{#2})}\xspace}
\newcommand{\propref}[1]{\mbox{Prop.~\ref{#1}}\xspace}
\newcommand{\proprefs}[2]{\mbox{Props.~\ref{#1}--\ref{#2}}\xspace}
\newcommand{\msgref}[1]{\mbox{message~\ref{#1}}\xspace}

\iffull
\newcommand{\appref}[1]{\mbox{Appendix~\ref{#1}}\xspace}
\fi

\newtheorem{prop}{Proposition}

\newcommand{\mega}{\ensuremath{\mathrm{M}}\xspace}

\newcommand{\gigabits}{\ensuremath{\mathrm{Gb}}\xspace}

\newcommand{\terabytes}{\ensuremath{\mathrm{TB}}\xspace}

\newcommand{\gibibytes}{\ensuremath{\mathrm{GiB}}\xspace}
\newcommand{\gigahertz}{\ensuremath{\mathrm{GHz}}\xspace}
\newcommand{\secs}{\ensuremath{\mathrm{s}}\xspace}

\newcommand{\relstddev}{\ensuremath{\chi}\xspace}

\newcounter{requesterLineNmbr}
\renewcommand{\therequesterLineNmbr}{\ensuremath{\mathsf{r\arabic{requesterLineNmbr}}}}
\newcommand{\requesterLabel}[1]{\refstepcounter{requesterLineNmbr}\label{#1}\therequesterLineNmbr.}

\newcounter{responderLineNmbr}
\newcommand{\responderIndent}{\hspace{4.25em}}
\renewcommand{\theresponderLineNmbr}{\ensuremath{\mathsf{s\arabic{responderLineNmbr}}}}
\newcommand{\responderLabel}[1]{\responderIndent\refstepcounter{responderLineNmbr}\label{#1}\theresponderLineNmbr.}

\newcounter{messageNmbr}
\newcommand{\messageIndent}{\hspace{4.25em}}
\renewcommand{\themessageNmbr}{\ensuremath{\mathsf{m\arabic{messageNmbr}}}}
\newcommand{\messageLabel}[1]{\messageIndent\refstepcounter{messageNmbr}\label{#1}\themessageNmbr.~~}

\newlength{\figureheight}
\newcommand{\figurewidth}{\columnwidth}

\ExplSyntaxOn
\cs_new:Npn \intensity #1
   {\fp_eval:n {\MinIntensity + (1.0-\MinIntensity) * ((({#1}-\value{MinNumber}+1)/(\value{MaxNumber}-\value{MinNumber}+1))^(1/5))}
   }
\ExplSyntaxOff
\newcommand{\MinIntensity}{0.4}   
\newcounter{MinNumber}
\newcounter{MaxNumber}
\newcommand{\ApplyGradientX}[1]{\cellcolor[gray]{\intensity{#1}}{\parbox{2em}{\centering \raggedleft{#1}}}}
\newcommand{\ApplyGradientY}[1]{\cellcolor[gray]{\intensity{#1}}{\parbox{1.5em}{\centering \raggedleft{#1}}}}
\newcolumntype{X}{>{\collectcell\ApplyGradientX}c<{\endcollectcell}}
\newcolumntype{Y}{>{\collectcell\ApplyGradientY}c<{\endcollectcell}}

\newcommand{\genericRV}{\ensuremath{Y}\xspace}
\newcommand{\genericRVAlt}{\ensuremath{Y'}\xspace}
\newcommand{\genericSet}{\ensuremath{Z}\xspace}
\newcommand{\genericVar}{\ensuremath{z}\xspace}
\newcommand{\genericNat}{\ensuremath{z}\xspace}
\newcommand{\nats}{\ensuremath{\mathbb{N}}\xspace}
\newcommand{\reals}{\ensuremath{\mathbb{R}}\xspace}
\newcommand{\setSize}[1]{\ensuremath{|{#1}|}\xspace}

\newcommand{\residues}[1]{\ensuremath{[{#1}]}\xspace}
\newcommand{\getsr}{\;\stackrel{\$}{\leftarrow}\;}
\newcommand{\boolTrue}{\ensuremath{\mathit{true}}\xspace}
\newcommand{\boolFalse}{\ensuremath{\mathit{false}}\xspace}

\newcommand{\distEqual}{\ensuremath{\mathbin{\;\overset{d}{=}\;}}\xspace}
\newcommand{\testEqual}{\ensuremath{\mathbin{\;\overset{?}{=}\;}}\xspace}
\newcommand{\testIn}{\ensuremath{\mathbin{\;\overset{?}{\in}\;}}\xspace}
\newcommand{\prob}[1]{\ensuremath{\mathbb{P}\left({#1}\right)}\xspace}

\newcommand{\cprob}[3]{\ensuremath{\mathbb{P}#1(}{#2}\ensuremath{\;#1|} \ifmmode{\;}\fi {#3}\ensuremath{#1)}\xspace}
\newcommand{\cexpv}[3]{\ensuremath{\mathbb{E}#1(}{#2}\ensuremath{\;#1|} \ifmmode{\;}\fi {#3}\ensuremath{#1)}\xspace}
\newcommand{\negligible}[1]{\ensuremath{\mathit{negl}({#1})}\xspace}
\newcommand{\negligibleFn}{\ensuremath{f}\xspace}
\newcommand{\polynomial}{\ensuremath{\phi}\xspace}

\newcommand{\groupSet}{\ensuremath{\mathbb{G}}\xspace}
\newcommand{\groupOrder}{\ensuremath{r}\xspace}
\newcommand{\groupMult}{\ensuremath{\times_{\groupSet}}\xspace}
\newcommand{\groupElmt}{\ensuremath{m}\xspace}
\newcommand{\groupElmtAlt}{\ensuremath{m'}\xspace}
\newcommand{\groupIdentity}{\ensuremath{1_{\groupSet}}\xspace}
\newcommand{\groupGenerator}{\ensuremath{g}\xspace}
\newcommand{\groupExponent}{\ensuremath{y}\xspace}
\newcommand{\groupRand}{\ensuremath{\$(\groupSet)}\xspace}

\newcommand{\ints}[1]{\ensuremath{\mathbb{Z}_{#1}}\xspace}
\newcommand{\relPrimeInts}[1]{\ensuremath{\ints{#1}^{\ast}}\xspace}
\newcommand{\ringEquiv}[1]{\ensuremath{\equiv_{#1}}\xspace}
\newcommand{\testRingEquiv}[1]{\ensuremath{\overset{?}{\equiv}_{#1}}\xspace}

\newcommand{\encScheme}{\ensuremath{\mathcal{E}}\xspace}
\newcommand{\keygen}{\ensuremath{\mathsf{Gen}}\xspace}
\newcommand{\encrypt}[1]{\ensuremath{\mathsf{Enc}_{#1}}\xspace}
\newcommand{\decrypt}[1]{\ensuremath{\mathsf{Dec}_{#1}}\xspace}
\newcommand{\plaintext}[1]{\ensuremath{m_{#1}}\xspace}
\newcommand{\ciphertext}[1]{\ensuremath{c_{#1}}\xspace}
\newcommand{\ciphertextSpace}[1]{\ensuremath{C_{#1}}\xspace}
\newcommand{\secParam}{\ensuremath{\kappa}\xspace}
\newcommand{\secParamMin}{\ensuremath{\secParam_{0}}\xspace}
\newcommand{\privKey}{\ensuremath{\mathit{sk}}\xspace}
\newcommand{\pubKey}{\ensuremath{\mathit{pk}}\xspace}

\newcommand{\encMult}[1]{\ensuremath{\times_{#1}}\xspace}
\newcommand{\encProd}[3]{\ensuremath{\displaystyle\operatorname*{\textstyle\prod_{\mathrlap{#1}}}_{#2}^{#3}\hphantom{_{#1}}}\xspace}
\newcommand{\encExp}[1]{\ensuremath{\exp_{#1}}\xspace}
\newcommand{\encProdIdx}{\ensuremath{i}\xspace}
\newcommand{\encExponent}{\ensuremath{\genericNat}\xspace}
\newcommand{\encExponentAlt}{\ensuremath{\genericNat'}\xspace}
\newcommand{\cryptoHash}{\ensuremath{H}\xspace}

\newcommand{\requesterTerm}{requester\xspace}
\newcommand{\requestersTerm}{requesters\xspace}
\newcommand{\RequesterTerm}{Requester\xspace}
\newcommand{\responderTerm}{responder\xspace}

\newcommand{\respondersTerm}{responders\xspace}
\newcommand{\directoryTerm}{directory\xspace}
\newcommand{\accountLocationPrivacy}{\textbf{account location privacy}\xspace}
\newcommand{\accountSecurity}{\textbf{account security}\xspace}
\newcommand{\trustedForAccountLocationPrivacy}{TALP\xspace}
\newcommand{\untrustedForAccountLocationPrivacy}{UALP\xspace}

\newcommand{\codeExpt}{\ensuremath{\mathtt{Experiment}~}}

\newcommand{\codeIf}{\ensuremath{\mathtt{if}~}}

\newcommand{\codeReturn}{\ensuremath{\mathtt{return}~}}
\newcommand{\codeAbort}{\ensuremath{\mathtt{abort}~}}

\newcommand{\bloomFilterSize}{\ensuremath{\ell}\xspace}
\newcommand{\bloomFilterHashFns}{\ensuremath{k}\xspace}
\newcommand{\bloomFilterHashFnsOpt}{\ensuremath{\bloomFilterHashFns_{\mathrm{opt}}}\xspace}
\newcommand{\bloomFilterHashFn}[1]{\ensuremath{h_{#1}}\xspace}
\newcommand{\bloomFilterHashFnIdx}{\ensuremath{i}\xspace}
\newcommand{\bloomFilterBitIdx}{\ensuremath{j}\xspace}
\newcommand{\bloomFilterBitIdxAlt}{\ensuremath{j'}\xspace}
\newcommand{\bloomFilterBitCtext}[1]{\ensuremath{c_{#1}}\xspace}
\newcommand{\bloomFilterIndicesToSet}[1]{\ensuremath{J_{#1}}\xspace}

\newcommand{\requester}{\ensuremath{R}\xspace}
\newcommand{\responder}[1]{\ensuremath{S_{#1}}\xspace}
\newcommand{\responderIdx}{\ensuremath{i}\xspace}
\newcommand{\responderIdxAlt}{\ensuremath{i'}\xspace}
\newcommand{\nmbrResponders}[1]{\ensuremath{M_{#1}}\xspace}
\newcommand{\nmbrRespondersQueried}{\ensuremath{m}\xspace}
\newcommand{\accountId}{\ensuremath{a}\xspace}
\newcommand{\accountIdAlt}{\ensuremath{a'}\xspace}
\newcommand{\accountIdSet}{\ensuremath{A}\xspace}
\newcommand{\password}[1]{\ensuremath{\pi_{#1}}\xspace}
\newcommand{\passwordAlt}{\ensuremath{\pi'}\xspace}
\newcommand{\honeyPassword}[2]{\ensuremath{\hat{\pi}_{#1}^{#2}}\xspace}
\newcommand{\similarPasswords}[2]{\ensuremath{P_{#2}({#1})}\xspace}
\newcommand{\nmbrSimilarPasswords}{\ensuremath{n}\xspace}
\newcommand{\allSimilar}{\ensuremath{\mathit{Sim}}\xspace}
\newcommand{\resultCiphertext}[1]{\ensuremath{\rho_{#1}}\xspace}
\newcommand{\nmbrHoneyPasswords}{\ensuremath{d}\xspace}
\newcommand{\honeyPasswordIdx}{\ensuremath{j}\xspace}
\newcommand{\cluster}[1]{\ensuremath{\Psi({#1})}\xspace}
\newcommand{\clusterSize}{\ensuremath{\psi}\xspace}
\newcommand{\nmbrFaultyDirectoryReplicas}{\ensuremath{f}\xspace}
\newcommand{\loginBudget}{\ensuremath{\zeta}\xspace}

\newcommand{\ecPrime}{\ensuremath{p}\xspace}

\newcommand{\elgPrivKey}{\ensuremath{u}\xspace}
\newcommand{\elgPubKey}{\ensuremath{U}\xspace}
\newcommand{\elgEphemeralPrivKey}[1]{\ensuremath{v_{#1}}\xspace}
\newcommand{\elgEphemeralPubKey}[1]{\ensuremath{V_{#1}}\xspace}
\newcommand{\elgCiphertext}[1]{\ensuremath{W_{#1}}\xspace}

\newcommand{\adversary}[1]{\ensuremath{B_{#1}}\xspace}
\newcommand{\adversaryState}{\ensuremath{\phi}\xspace}
\newcommand{\experiment}[2]{\ensuremath{\mathbf{Expt}^{#1}_{#2}}\xspace}
\newcommand{\elgCiphertextLog}[1]{\ensuremath{w_{#1}}\xspace}
\newcommand{\residueConstant}{\ensuremath{\delta}\xspace}
\newcommand{\residueConstantAlt}{\ensuremath{\delta'}\xspace}
\newcommand{\residueValue}{\ensuremath{x}\xspace}
\newcommand{\residueValueAlt}{\ensuremath{x'}\xspace}
\newcommand{\elgEphemeralPrivKeyCoeff}[1]{\ensuremath{\beta_{#1}}\xspace}
\newcommand{\elgEphemeralPrivKeyCoeffAlt}[1]{\ensuremath{\beta_{#1}'}\xspace}
\newcommand{\elgCiphertextLogCoeff}[1]{\ensuremath{\gamma_{#1}}\xspace}
\newcommand{\elgCiphertextLogCoeffAlt}[1]{\ensuremath{\gamma_{#1}'}\xspace}
\newcommand{\elgPrivKeyCoeff}{\ensuremath{\alpha}\xspace}
\newcommand{\elgPrivKeyCoeffAlt}{\ensuremath{\alpha'}\xspace}
\newcommand{\equalityQueries}{\ensuremath{q}\xspace}
\newcommand{\someEqEvent}{\ensuremath{E}\xspace}

\newcommand{\responseTime}{\ensuremath{t}\xspace}
\newcommand{\responseTimeMax}{\ensuremath{t_{\mathrm{goal}}}\xspace}
\newcommand{\coeff}[1]{\ensuremath{\beta_{#1}}\xspace}
\newcommand{\trueDetectionRate}{\ensuremath{\mathsf{tdr}}\xspace}

\begin{document}

\title{How to End Password Reuse on the Web}

\author{
  {\rm Ke Coby Wang}\\
  Department of Computer Science\\
  University of North Carolina at Chapel Hill\\
  {\tt kwang@cs.unc.edu}
  \and
  {\rm Michael K.\ Reiter}\\
  Department of Computer Science\\
  University of North Carolina at Chapel Hill\\
  {\tt reiter@cs.unc.edu}
}

\maketitle

\thispagestyle{plain}
\pagestyle{plain}

\begin{abstract}
  We present a framework by which websites can coordinate to make it
  difficult for users to set similar passwords at these websites, in
  an effort to break the culture of password reuse on the web today.
  Though the design of such a framework is fraught with risks to
  users' security and privacy, we show that these risks can be
  effectively mitigated through careful scoping of the goals for such
  a framework and through principled design.  At the core of our
  framework is a private set-membership-test protocol that enables one
  website to determine, upon a user setting a password for use at it,
  whether that user has already set a similar password at another
  participating website, but with neither side disclosing to the other the
  password(s) it employs in the protocol.  Our framework then layers
  over this protocol a collection of techniques to mitigate the
  leakage necessitated by such a test.  We verify via probabilistic
  model checking that these techniques are effective in maintaining
  account security, and since these mechanisms are consistent with
  common user experience today, our framework should be unobtrusive to
  users who do not reuse similar passwords across websites (e.g., due
  to having adopted a password manager).  Through a working
  implementation of our framework and optimization of its parameters
  based on insights of how passwords tend to be reused, we show that
  our design can meet the scalability challenges facing such a
  service.
\end{abstract}  

\section{Introduction}
\label{sec:introduction}

\medskip
\begin{center}
  \begin{minipage}[t]{0.8\columnwidth}
  \textit{The reuse of passwords is the No.\ 1 cause of harm on the
    internet.}\\
    \null\hfill Alex Stamos~\cite{collins2016:facebook}\\
    \null\hfill Facebook CSO (Jun 2015--Aug 2018)
  \end{minipage}
\end{center}

Password reuse across websites remains a dire problem despite
widespread advice for users to avoid it.  Numerous studies
over the past fifteen years indicate that a large majority of users set
the same or similar passwords across different websites
(e.g.,~\cite{brown2004:reuse, riley2006:reuse, shay2010:reuse,
  das2014:tangled, ion2015:expert, pearman2017:habitat, wang2018:domino}).
As such, a
breach of a password database or a phish of a user's
password often leads to the compromise of user accounts on other
websites.  Such ``credential-stuffing'' attacks are a primary cause of
account takeovers~\cite{williamson2014:breach,manico2015:stuffing},
allowing the attacker to drain accounts of stored value, credit
card numbers, and other personal
information~\cite{manico2015:stuffing}.  Ironically,
stringent password requirements contribute to password
reuse, as users reuse strong passwords across websites to cope with
the cognitive burden of creating and remembering
them~\cite{wash2016:reuse}.
Moreover, notifications to accounts at risk due to
password reuse seem insufficient to cause their owners to
stop reusing passwords~\cite{golla2018:notifications}.

It is tempting to view password reuse as inflicting costs on
only users who practice it.  However,
preventing, detecting, and cleaning up compromised accounts and the
value thus stolen is a significant cost for service providers, as
well.  A recent Ponemon survey~\cite{ponemon:2017:stuffing} of 569 IT
security practitioners estimated that credential-stuffing attacks incur
costs in terms of application downtime, loss of customers, and
involvement of IT security that average \$1.7 million, \$2.7 million
and \$1.6 million, respectively, per organization per year.  Some
companies go so far as to purchase compromised credentials on the
black market to find their vulnerable accounts proactively
(e.g.,~\cite{collins2016:facebook}).  Companies also must develop new
technologies to identify overtaken accounts based on their
use~\cite{collins2016:facebook}.  Even the sheer volume of
credential-stuffing attacks is increasingly a challenge; e.g., in
November 2017, 43\% (3.6 out of 8.3 billion) of all login attempts
served by Akamai involved credential abuse~\cite{akamai:2017:state}.
Finally, the aforementioned Ponemon survey estimated the fraud
perpetrated using overtaken accounts could incur average losses of up
to \$54 million per organization
surveyed~\cite{ponemon:2017:stuffing}.  As such, interfering with
password reuse would not only better protect users, but would also
reduce the considerable costs of credential abuse incurred by
websites.

Here we thus explore a technical mechanism to interfere with password
reuse across websites.  Forcing a user to authenticate to each website
using a site-generated password (e.g.,~\cite{menkus1988passwords})
would accomplish this goal.  However, we seek to retain the same
degree of user autonomy regarding her selection of passwords as she
has today---subject to the constraint that she not reuse them---to
accommodate her preferences regarding the importance of the account,
the ease of entering its password on various devices, etc.  At a high
level, the framework we develop enables a website at which a user is
setting a password, here called a \textit{\requesterTerm}, to ask of
other websites, here called \textit{\respondersTerm}, whether the user
has set a similar password at any of them.  A positive answer can then
be used by the \requesterTerm to ask the user to select a different
password.  As we will argue in \secref{sec:goals}, enlisting a
surprisingly small number of major websites in our framework could
substantially weaken the culture of password reuse.

We are under no illusions that our design, if deployed, will elicit
anything but contempt (at least temporarily) from users who reuse
passwords across websites.  Its usability implications are thus not
unlike increasingly stringent password requirements, to which users
have nevertheless resigned.  However, options for password managers
are plentiful and growing, with a variety of trustworthiness,
usability, and cost properties
(e.g.,~\cite{rubenking2017:managers,ferrill2018:managers}).  Indeed,
experts often list the use of a password manager that supports a
different password per website to be one of the best things a user can
do to reduce her online risk~\cite{ion2015:expert}.  While there might
be users who, despite having a rich online presence, cannot use a
password manager for some reason, we expect them to be few.  Of
course, nearly anyone capable of using a computer should be able to
write down her passwords, as a last resort.  Though historically
maligned, the practice is now more widely accepted, exactly because it
makes it easier to not reuse passwords
(e.g.,~\cite{kotadia2005:jot,hayes2014:passwords}).

There are many technical issues that need to be addressed to make a
framework like the one we propose palatable.  First, such a framework
should not reduce the security of user accounts.  Second, the
framework should also not decay user privacy substantially, in the
sense of divulging the websites at which a user has an account.
Third, it is important that the protocol run between a \requesterTerm
and \respondersTerm should scale well enough to ensure that it does
not impose too much delay for setting a password at a website.

Our framework addresses these challenges as follows.  To minimize risk
to user accounts, we design a protocol that enables the \requesterTerm
to learn if a password chosen by a user is similar to one she set at a
\responderTerm; neither side learns the password(s) the other input to
the protocol, however, even by misbehaving.  Our framework leverages
this protocol, together with other mechanisms to compensate for
leakage necessitated by the protocol's output, to ensure that account
security and privacy are not diminished.  Among other properties, this
framework ensures that the \respondersTerm remain hidden from the
\requesterTerm and vice-versa.  We verify using probabilistic model
checking that the success rate of account takeover attempts is not
materially changed by our framework for users who employ distinct
passwords across websites.  Scalability is met in our framework by
carefully designing it to involve only a single round of interaction
between the \requesterTerm and \respondersTerm.  And, using
observations about password reuse habits, we optimize our framework to
detect similar password use with near-certainty while maximizing its
scalability.

To summarize, our contributions are as follows:
\begin{compactitem}
\item We initiate debate on the merits of interfering with password
  reuse on the web, through coordination among websites.  Our goal in
  doing so is to question the zeitgeist in the computer security
  community that password reuse cannot be addressed by technical means
  without imposing unduly on user security or privacy.  In particular,
  we show that apparent obstacles to a framework for interfering with
  password reuse can be overcome through careful scoping of its goals
  and through reasonable assumptions (\secref{sec:goals}).
\item We propose a protocol for privately testing set membership that
  underlies our proposed framework (\secref{sec:protocol}).  We prove
  security of our protocol in the case of a malicious \requesterTerm
  and against malicious
  \iffull
  \respondersTerm (\appref{sec:proofs}).
  \else
  \respondersTerm.
  \fi
\item We embed this protocol within a framework to facilitate
  \requesterTerm-\responderTerm interactions while hiding the
  identities of protocol participants and addressing risks that cannot
  be addressed by---and indeed, that are necessitated by---the private
  set-membership-test protocol (\secref{sec:framework}).  We
  demonstrate using probabilistic model checking that our framework
  does not materially weaken account security against password
  guessing attacks.
\item We evaluate implementations of our proposed framework with
  differing degrees of trust placed in it (\secref{sec:eval}).  Using
  password-reuse tendencies, we illustrate how to configure our
  framework to minimize its costs while ensuring detection of reused
  passwords with high likelihood.  Finally, we demonstrate its
  scalability through experiments with a working implementation in
  deployments that capture its performance in realistic scenarios.
\end{compactitem}

\section{Related Work}
\label{sec:related}

We are aware of no prior work to enable websites to interfere with
password reuse by the same user.  Instead, server-side approaches to
mitigate risks due to password reuse have set somewhat different
goals.

\myparagraph{Web single sign-on (SSO)} SSO schemes such as OAuth
(\url{https://oauth.net/}), OpenID (\url{http://openid.net}), OpenID
Connect (\url{http://openid.net/connect/}), and Facebook Login
(\url{https://developers.facebook.com/docs/facebook-login/}), enable
one website (an ``identity provider'') to share a user's account
information with other websites (``relying parties''), typically in
lieu of the user creating distinct accounts at those relying parties.
As such, this approach mitigates password reuse by simply not having
the user set passwords at the relying parties.  While convenient, SSO
exposes users to a range of new attacks, leading some to conclude
``the pervasiveness of SSO has created an exploitable
ecosystem''~\cite{ghasemisharif2018:sso}.  In addition, the identity
provider in these schemes typically learns the relying parties visited
by the user~\cite{dey2010:pseudoid}.

\myparagraph{Detecting use of leaked passwords by legitimate users}
As mentioned in \secref{sec:introduction}, some companies
cross-reference account passwords against known-leaked passwords,
either as a service to others (e.g.,
\url{https://www.passwordping.com}, \url{https://haveibeenpwned.com})
or for their own users (e.g.,~\cite{collins2016:facebook}).  While
recommended~\cite{nist2017:800-63B}, this approach can detect only
passwords that are \textit{known} to have been leaked.  Because
password database compromises often go undiscovered for long periods
(as of 2017, 15 months on average~\cite{shape2018:spill}), this
approach cannot identify vulnerable accounts in the interim.

\myparagraph{Detecting leaked passwords by their use in attacks}
Various techniques exist to detect leaked passwords by their attempted
use, e.g., honey accounts~\cite{deblasio2017:tripwire} and honey
passwords~\cite{bojinov2010:kamouflage, juels2013:honeywords,
  erguler2016:flatness}, the latter of which we will leverage as well
(\secref{sec:framework:design:responder}).  Alone, these methods do
little to detect an attacker's use of a leaked, known-good password
for one website at another website where the victim user is known to
have an account.  Defending against such discriminating attacks would
seem to require the victim's use of different passwords at distinct
websites, which we seek to compel here.

\myparagraph{Detecting popular passwords}
Schechter et al.~\cite{schechter2010:popularity} proposed a service at
which sites can check whether a password chosen by a user is popular
with other users or, more specifically, if its frequency of use
exceeds a specified threshold.  Our goals here are different---we seek
to detect the use of similar passwords by the same user at different
sites, regardless of popularity.
  
\myparagraph{Limiting password-based access}
Takada~\cite{takada2017:shutter} proposed to interfere with the misuse
of accounts with shared passwords by adding an ``availability
control'' to password authentication.  In this design, a user disables
the ability to log into her website account at a third-party service
and then re-enables it when needed.  This approach requires that the
attacker be unable to itself enable login, and so requires an
additional authentication at the third-party service to protect this
enabling.

\section{Goals and Assumptions}
\label{sec:goals}

In this section we seek to clarify the goals for our system and the
assumptions on which our design rests.

\subsection{Deployment Goals}
\label{sec:goals:deployment}

It is important to recognize that in order to break the culture of
password reuse, we do not require universal adoption of the framework
we propose here.  Instead, it may be enough to enlist a (surprisingly
small) number of top websites.  To see this, consider just the 20
websites listed in \tblref{tbl:top-websites}.\footnote{User counts
  were retrieved on December 4, 2018 from
  \url{https://www.statista.com/statistics/272014/global-social-networks-ranked-by-number-of-users/},
  \url{https://www.statista.com/statistics/476196/number-of-active-amazon-customer-accounts-quarter/},
  \url{http://blog.shuttlecloud.com/the-most-popular-email-providers-in-the-u-s-a/},
  and
  \url{https://expandedramblings.com/index.php/}\texttt{\{}\href{https://expandedramblings.com/index.php/yahoo-statistics/}{yahoo-statistics/}\texttt{,} \href{https://expandedramblings.com/index.php/taobao-statistics/}{taobao-statistics/}\texttt{,} \href{https://expandedramblings.com/index.php/quora-statistics/}{quora-statistics/}\texttt{\}}.}
For a back-of-the-envelope estimate, suppose that the users of each
website in \tblref{tbl:top-websites} are sampled uniformly at random
from the 3.58 billion global Internet users.\footnote{Estimate of
  Internet users was retrieved from
  \url{https://www.statista.com/statistics/273018/number-of-internet-users-worldwide/}
  on December 4, 2018.}  Then, in expectation an Internet user would
have accounts at more than four of them.  As such, if just these
websites adopted our framework, it would force a large fraction of
users to manage five or more dissimilar passwords, which is already at
the limit of what users are capable of managing themselves: ``If
multiple passwords cannot be avoided, four or five is the maximum for
unrelated, regularly used passwords that users can be expected to cope
with''~\cite{adams1999:enemy}.  We thus believe that enlisting these
20 websites could already dramatically improve password-manager
adoption, and it is conceivable that with modest additional adoption
(e.g., the top 50 most popular websites), password reuse could largely
be brought to an end.

\begin{table}
\centering
{\small
\begin{tabular}{l@{\hspace{0.5em}}r@{\hspace{2.5em}}l@{\hspace{0.5em}}r}
  \toprule
Website & Users (\mega) & Website & Users (\mega) \\
\midrule
Facebook & 2234 & Sina Weibo & 431 \\
YouTube & 1900 & Outlook & 400 \\
WhatsApp & 1500 & Twitter & 335 \\
Wechat & 1058 & Reddit & 330 \\
Yahoo! & 1000 & Amazon & 310 \\
Instagram & 1000 & LinkedIn & 303 \\
QQ & 803 & Quora & 300 \\
iCloud & 768 & Baidu Tieba & 300 \\
Taobao & 634 & Snapchat & 291 \\
Douyin/TikTok & 500 & Pinterest & 250 \\
\bottomrule
\end{tabular}
}
\caption{Estimates of active users for selected websites}
\label{tbl:top-websites}
\end{table}

A user might continue using similar passwords across sites that do not
participate in our framework.  Each such reused password may also be
similar to one she set at a site that \textit{does} participate in our
framework, but likely at only \textit{one} such site.  If this reused
password is compromised at a non-participating site (e.g., due to a
site breach), then the attacker might still use this password in a
credential-stuffing attack against the user's accounts at
participating sites, as it could today.  Again, however, due to our
framework, this attack should succeed at only one participating site,
not many.  Importantly, our framework restricts the attacker from
posing queries about the user's accounts as a \requesterTerm unless it
gains the user's consent to do so (see
\secref{sec:framework:design:responder}).  Even if it tricked the user
into consenting, it could use such a query to confirm that the
compromised password is similar to one set by the same user at
\textit{some} participating site, but not \textit{which} site (see
\secref{sec:goals:security}).  More generally, in
\secref{sec:framework:analysis}, we will show quantitatively that our
framework offers little advantage to an attacker that can pose a
limited number of queries as a \requesterTerm.

\subsection{User Identifiers}
\label{sec:goals:ids}

An assumption of our framework is that there is an identifier for a
user's accounts that is common across websites.  An email address for
the user would be a natural such identifier, and as we will describe
in \secref{sec:framework:design:responder}, this has other uses in our
context, as well.  Due to this assumption, however, a user could reuse
the same password across different websites, despite our framework, if
she registers a different email address at each.

Several methods exist for a user to amass many distinct email
addresses, but we believe they will interfere little with our goals
here.  First, some email providers support multiple addresses for a
single account.  For example, one Gmail account can have arbitrarily
many addresses, since Gmail addresses are insensitive to
capitalization, insertion of periods (`\texttt{.}'), or insertion of a
plus (`\texttt{+}') followed by any string, anywhere before
`\texttt{@gmail.com}'.  As another example, 33mail
(\url{https://33mail.com}) allows a user to receive mail sent to
\texttt{\textit{<alias>}@\textit{<username}>.33mail.com} for any alias
string.  Though these providers enable a user to provide a distinct
email address to each website (e.g.,~\cite{nield2014:infinite}), our
framework could nevertheless extract a canonical identifier for each
user.  For Gmail, the canonical identifier could be obtained by
normalizing capitalization and by eliminating periods and anything
between `\texttt{+}' and `\texttt{@gmail.com}'.  For 33mail,
\texttt{you@\textit{<username}>.33mail.com} should suffice.
Admittedly this requires customization specific to each such provider
domain, though this customization is simple.

Second, some hosting services permit a customer to register a domain
name and then support many email aliases for it (e.g.,
\texttt{\textit{<alias>}@\textit{<domain>}.com}).  For example, Google
Domains (\url{http://domains.google}) supports 100 email aliases per
domain.  Since these domains are custom, it might not be tractable to
introduce domain-specific customizations as above.  However,
registering one's own domain as a workaround to keep using the same
password across websites presumably saves the user little effort or
money (registering domains is not free) in comparison to just
switching to a password manager.  Going further, a user could manually
register numerous email accounts at free providers such as Gmail.
Again, this is presumably at least as much effort as alternatives that
involve no password reuse.  As such, we do not concern ourselves with
such methods of avoiding password reuse detection.

This discussion highlights an important clarification regarding our
goals: we seek to eliminate \textit{easy} methods of reusing passwords
but not ones that require similar or greater effort from the user than
more secure alternatives, of which we take a password manager as an
exemplar.  That is, we do not seek to make it \textit{impossible} for
a user to reuse passwords, but rather to make reusing passwords
about as difficult as not reusing them.  We expect that even this
modest goal, if achieved, will largely eliminate password reuse, since
passwords are reused today almost entirely for convenience.

\subsection{Security and Privacy Goals}
\label{sec:goals:security}

The goals we take as more absolute have to do with the privacy of
users and the security of their accounts.  Specifically, we seek to
ensure the following:
\begin{compactitem}
\item \accountLocationPrivacy: Websites do not learn the identities of
  other websites at which a user has an account.

\item \accountSecurity: Our framework strengthens security of user
  accounts at a site that participates in our framework, by
  interfering with reuse of similar passwords at other participating
  sites.  Moreover, it does not qualitatively degrade user account
  security in other ways.
\end{compactitem}
As we will see, \accountSecurity is difficult to achieve, since our
framework must expose whether \respondersTerm' passwords are similar
to the one chosen by the user at the \requesterTerm.  However,
\accountLocationPrivacy hides from the \requesterTerm each
\responderTerm from which the \requesterTerm learns this information.
As such, if a user attempts to set the same password at a malicious
\requesterTerm that she has also set at some \responderTerm, or if a
malicious \requesterTerm otherwise obtains this password (e.g.,
obtaining it in a breach of a non-participating site), the malicious
\requesterTerm must still attempt to use that password blindly at
participating websites, just as in a credential-stuffing attack today.
(The attacker might succeed, but it would succeed without our
framework, too.)  Moreover, in \secref{sec:framework}
we will detail additional defenses against this leakage to further
reduce the risk of attacks from malicious \requestersTerm to whom the
user does not volunteer this password, and show using formal verification
that these defenses are effective.

We conclude this section by summarizing some potential goals that we
(mostly) omit from consideration in this paper.  From a privacy
perspective, we try to hide neither \textit{when} a password is being
set at some \requesterTerm for an account identifier nor the
\textit{number} of \respondersTerm at which an account has been
established using that account identifier, simply because we are
unaware of common scenarios in which these leakages would have
significant practical ramifications.  And, while we strive to
guarantee \accountLocationPrivacy and \accountSecurity even against a
\requesterTerm and \respondersTerm that misbehave, we generally do not
seek to otherwise detect that misbehavior.  So, for example, each
\requesterTerm and \responderTerm has complete autonomy in determining
the passwords that it provides to the protocol as the candidate
password submitted by the user and the passwords similar to the one
for the account with the same identifier, respectively.  As we will
see in \secref{sec:dos}, such misbehaviors can give rise to
denial-of-service opportunities, for which we propose remedies there.

\section{Privately Testing Set Membership}
\label{sec:protocol}

A building block of our framework is a protocol by which a
\requesterTerm \requester can inquire with a \responderTerm
\responder{} as to whether a password \password{} chosen at \requester
for an account identifier is similar to one already in use at
\responder{} for the same identifier.  If
for an account identifier \accountId, the \responderTerm \responder{}
has a set \similarPasswords{\accountId}{} of passwords similar to that
already set at \responder{}, then the goal of this protocol is for the
\requesterTerm to learn whether the candidate password \password{} is
in \similarPasswords{\accountId}{}.  However, any additional
information leakage to the \requesterTerm (about any passwords in
\similarPasswords{\accountId}{} or even the number of passwords in
\similarPasswords{\accountId}{}) or to the \responderTerm (about
\password{}) should be minimized.

This general specification can be met with a private
set-membership-test (PMT) protocol.  Though several such protocols exist
(e.g.,~\cite{nojima2009:bloom, meskanen2015:private, tamrakar2017:pmt,
  ramezanian2017:private}), we develop a new one here with an
interaction pattern and threat model that is better suited for our
framework.  In particular, existing protocols require special
hardware~\cite{tamrakar2017:pmt} or more rounds of
interaction~\cite{nojima2009:bloom,meskanen2015:private}, or leak more
information in our threat
model~\cite{nojima2009:bloom,meskanen2015:private,ramezanian2017:private}
than the one we present.

In designing this protocol, we sought guidance from the considerable
literature on private set intersection (PSI), surveyed recently by
Pinkas et al.~\cite{pinkas2018:psi}.  Informally, PSI protocols allow
two parties to jointly compute the intersection of the sets that each
inputs to the protocol, and ideally nothing else.  Furthermore, PSI
protocols secure in the malicious adversary model, where one party
deviates arbitrarily from the protocol, have been proposed
(e.g.,~\cite{de2010:mal-psi, dachman2009:mal-psi,
  freedman2016:mal-psi, freedman2004:mal-psi, kamara2014:mal-psi,
  kolesnikov2017:symmetric, rindal2017:mal-psi, rindal2017:mal-psi2}).
Still, while a PSI protocol would allow \requester to determine
whether $\password{} \in \similarPasswords{\accountId}{}$, without
additional defenses it could reveal too much information; e.g., if
\requester input multiple passwords to the protocol, then it would
learn which of these passwords were in
\similarPasswords{\accountId}{}.  Moreover, as Tamrakar et
al.~\cite{tamrakar2017:pmt} argue, PSI protocols are not ideal for
implementing PMT due to their high communication complexity and poor
scalability.

By comparison, two-party private set-intersection \textit{cardinality}
(PSI-CA) protocols are closer to our needs; these protocols output the
size of the intersection of each party's input set, and ideally
nothing else (e.g.,~\cite{davidson2017:psi-ca,
  decristofaro2012:intersection, debnath2015:psi-ca, egert2015:union,
  kissner2005:psi-ca}).  As with PSI protocols, however, using a
PSI-CA protocol without modification to implement PMT would reveal too
much information if \requester input multiple passwords to the
protocol.  As such, our protocol here is an adaptation of a PSI-CA
protocol due to Egert et
al.~\cite[\secrefstatic{4.4}]{egert2015:union}, in which we (i) reduce
the information it conveys to only the results of a membership test,
versus the cardinality of a set intersection, and (ii) analyze its
privacy properties in the face of malicious behavior by a
\requesterTerm or \responderTerm (versus only an honest-but-curious
participant in their work), accounting for leakage intrinsic in the
application for which we use it here.

\subsection{Partially Homomorphic Encryption}
\label{sec:protocol:encryption}

Our protocol builds upon a
multiplicatively homomorphic encryption scheme $\encScheme = \langle
\keygen, \encrypt{}, \decrypt{}, \encMult{[\cdot]}\rangle$ with the
following algorithms.  Below, $\genericVar \getsr \genericSet$ denotes
random selection from set \genericSet and assignment to \genericVar,
and $\genericRV \distEqual \genericRVAlt$ denotes that random
variables \genericRV and \genericRVAlt are distributed identically.

\begin{compactitem}
\item \keygen is a randomized algorithm that on input $1^\secParam$
  outputs a public-key/private-key pair $\langle\pubKey,
  \privKey\rangle \gets \keygen(1^{\secParam})$.  The value of \pubKey
  uniquely determines a \textit{plaintext space} \groupSet where
  $\langle \groupSet, \groupMult\rangle$ denotes a multiplicative,
  cyclic group of order \groupOrder with identity \groupIdentity, and
  where \groupOrder is a \secParam-bit prime.  The randomized function
  \groupRand returns a new, random $\groupElmt \getsr \groupSet$.
  We let $\ints{\groupOrder} = \{0, \ldots, \groupOrder-1\}$ and
  $\relPrimeInts{\groupOrder} = \{1, \ldots, \groupOrder-1\}$, as
  usual.
\item \encrypt{} is a randomized algorithm that on input public key
  \pubKey and plaintext $\plaintext{} \in \groupSet$ produces a
  ciphertext $\ciphertext{} \gets \encrypt{\pubKey}(\plaintext{})$.
  Let $\ciphertextSpace{\pubKey}(\plaintext{})$ denote the set of all
  ciphertexts that $\encrypt{\pubKey}(\plaintext{})$ produces with
  nonzero probability.  Then, $\ciphertextSpace{\pubKey} =
  \bigcup_{\plaintext{}\in\groupSet} \ciphertextSpace{\pubKey}(\plaintext{})$
  is the ciphertext space of the scheme with public key \pubKey.
\item \decrypt{} is a deterministic algorithm that on input a private
  key \privKey and ciphertext $\ciphertext{}
  \in \ciphertextSpace{\pubKey}(\plaintext{})$, for $\plaintext{} \in
  \groupSet$ and \pubKey the public key corresponding to \privKey,
  produces $\plaintext{} \gets \decrypt{\privKey}(\ciphertext{})$.  If
  $\ciphertext{} \not\in \ciphertextSpace{\pubKey}$, then
  $\decrypt{\privKey}(\ciphertext{})$ returns $\bot$.
\item \encMult{[\cdot]} is a randomized algorithm that on input a
  public key \pubKey and ciphertexts $\ciphertext{1}
  \in \ciphertextSpace{\pubKey}(\plaintext{1})$ and $\ciphertext{2}
  \in \ciphertextSpace{\pubKey}(\plaintext{2})$ produces a ciphertext
  $\ciphertext{} \gets \ciphertext{1} \encMult{\pubKey}
  \ciphertext{2}$ chosen uniformly at random from
  $\ciphertextSpace{\pubKey}(\plaintext{1} \plaintext{2})$.  If
  $\ciphertext{1} \not\in \ciphertextSpace{\pubKey}$ or
  $\ciphertext{2} \not\in \ciphertextSpace{\pubKey}$, then
  $\ciphertext{1} \encMult{\pubKey} \ciphertext{2}$ returns $\bot$.
  We use \encProd{\pubKey}{}{} and \encExp{\pubKey} to denote
  multiplication of a sequence and exponentiation using
  \encMult{\pubKey}, respectively, i.e., 
  \begin{align*}
    \encProd{\pubKey}{\encProdIdx=1}{\genericNat} \ciphertext{\encProdIdx}
    & \distEqual \ciphertext{1} \encMult{\pubKey} \ciphertext{2} \encMult{\pubKey}
    \ldots \encMult{\pubKey} \ciphertext{\genericNat} \\
    \encExp{\pubKey}(\ciphertext{}, \genericNat)
    & \distEqual \encProd{\pubKey}{\encProdIdx = 1}{\genericNat} \ciphertext{}
  \end{align*}
\end{compactitem}

\subsection{Protocol Description}
\label{sec:protocol:description}

Our protocol is shown in \figref{fig:protocol}, with the actions by
the \requesterTerm \requester listed on the left
(\linesref{prot:requester:init}{prot:requester:return}), those by the
\responderTerm \responder{} listed on the right
(\refs{prot:responder:checkCiphertexts}{prot:responder:makeResult}),
and messages between them in the middle
(\refs{prot:msg:bloomFilter}{prot:msg:result}).  In
\figref{fig:protocol} and below, \residues{\genericNat} for integer
$\genericNat > 0$ denotes the set $\{0, \ldots, \genericNat-1\}$.

\newcolumntype{R}{>{$}p{1pt}<{$}}
\begin{figure}[t]
  \centering
  \setcounter{requesterLineNmbr}{0}
  \setcounter{responderLineNmbr}{0}
  \setcounter{messageNmbr}{0}
  \framebox{\small
     $\begin{array}{@{}r@{\hspace{2pt}}R@{}r@{}R@{\hspace{2pt}}r@{\hspace{2pt}}l@{}} 
      & \mathrlap{\requester(\accountId, \password{}, \bloomFilterSize, \langle\bloomFilterHashFn{\bloomFilterHashFnIdx}\rangle_{\bloomFilterHashFnIdx \in \residues{\bloomFilterHashFns}})}
      & & & & \multicolumn{1}{c}{\responder{}(\{\similarPasswords{\accountIdAlt}{}\}_{\accountIdAlt \in \accountIdSet})} \\[10pt]
      \requesterLabel{prot:requester:init}
      & \mathrlap{\langle\pubKey, \privKey\rangle \gets \keygen(1^{\secParam})}
      \\[4pt]
      \requesterLabel{prot:requester:calculateIndices}
      & \mathrlap{\displaystyle \bloomFilterIndicesToSet{\requester} \gets \bigcup_{\bloomFilterHashFnIdx \in \residues{\bloomFilterHashFns}} \{\bloomFilterHashFn{\bloomFilterHashFnIdx}(\password{})\}}
      \\
      \requesterLabel{prot:requester:encryptBF}
      & \mathrlap{\forall \bloomFilterBitIdx \in \residues{\bloomFilterSize}:
      \bloomFilterBitCtext{\bloomFilterBitIdx} \gets
      \left\{\begin{array}{@{}l@{\hspace{3pt}}l@{}}
      \encrypt{\pubKey}(\groupRand) & \mbox{if $\bloomFilterBitIdx \in \bloomFilterIndicesToSet{\requester}$}
      \\
      \encrypt{\pubKey}(\groupIdentity) & \mbox{if $\bloomFilterBitIdx \not\in \bloomFilterIndicesToSet{\requester}$}
      \end{array}\right.}
      \\[12pt]
      & & \messageLabel{prot:msg:bloomFilter}
      & \mathrlap{\xrightarrow{\makebox[9.75em]{$\accountId, \pubKey, \langle\bloomFilterHashFn{\bloomFilterHashFnIdx}\rangle_{\bloomFilterHashFnIdx \in \residues{\bloomFilterHashFns}}, \langle \bloomFilterBitCtext{\bloomFilterBitIdx}\rangle_{\bloomFilterBitIdx\in\residues{\bloomFilterSize}}$}}}
      \\
      & & & & \responderLabel{prot:responder:checkCiphertexts}
      & \codeAbort \codeIf \exists\bloomFilterBitIdx\in\residues{\bloomFilterSize}: \bloomFilterBitCtext{\bloomFilterBitIdx} \not\in \ciphertextSpace{\pubKey}
      \\
      & & & & \responderLabel{prot:responder:calculateIndices}
      & \displaystyle \bloomFilterIndicesToSet{\responder{}} \gets \bigcup_{\passwordAlt \in \similarPasswords{\accountId}{}} \bigcup_{\bloomFilterHashFnIdx \in \residues{\bloomFilterHashFns}} \{\bloomFilterHashFn{\bloomFilterHashFnIdx}(\passwordAlt)\}
      \\
      & & & & \responderLabel{prot:responder:generateBlinding}
      & \encExponent \getsr \relPrimeInts{\groupOrder}
      \\
      & & & & \responderLabel{prot:responder:makeResult}
      & \displaystyle \resultCiphertext{} \gets \encExp{\pubKey}\!\left(\!\left(\encProd{\pubKey}{\bloomFilterBitIdx \in \residues{\bloomFilterSize}\setminus\bloomFilterIndicesToSet{\responder{}}}{} \hspace{-0.75em}\bloomFilterBitCtext{\bloomFilterBitIdx}\right)\!, \encExponent\!\right)\!\!\!\!
      \\
      & & \messageLabel{prot:msg:result}
      & \mathrlap{\xleftarrow{\makebox[9.75em]{\resultCiphertext{}}}}
      \\
      \requesterLabel{prot:requester:checkCiphertext}
      & \mathrlap{\codeAbort \codeIf \resultCiphertext{} \not\in \ciphertextSpace{\pubKey}}
      \\
      \requesterLabel{prot:requester:decrypt}
      & \mathrlap{\plaintext{} \gets \decrypt{\privKey}(\resultCiphertext{})}
      \\
      \requesterLabel{prot:requester:return}
      & \mathrlap{\codeReturn (\plaintext{} \testEqual \groupIdentity)}
      \\
    \end{array}$
  }
  \caption{PMT protocol; see
    \secref{sec:protocol:description}.  \RequesterTerm \requester
    returns \boolTrue if password \password{} is similar to another
    password used at \responderTerm \responder{} for the same account
    identifier \accountId, i.e., if $\password{} \in
    \similarPasswords{\accountId}{}$.}
  \label{fig:protocol}
\end{figure}

At a conceptual level, our PMT protocol works as follows.  The
\requesterTerm \requester takes as input an account identifier
\accountId, the user's chosen password \password{}, a
Bloom-filter~\cite{bloom1970:space} length \bloomFilterSize, and the
hash functions $\langle\bloomFilterHashFn{\bloomFilterHashFnIdx}
\rangle_{\bloomFilterHashFnIdx \in \residues{\bloomFilterHashFns}}$
for the Bloom filter (i.e., each
$\bloomFilterHashFn{\bloomFilterHashFnIdx}: \{0,1\}^\ast \rightarrow
\residues{\bloomFilterSize}$).  \requester computes its Bloom filter
containing \password{}, specifically a set of indices
$\bloomFilterIndicesToSet{\requester} \gets
\bigcup_{\bloomFilterHashFnIdx \in \residues{\bloomFilterHashFns}}
\{\bloomFilterHashFn{\bloomFilterHashFnIdx}(\password{})\}$
(\lineref{prot:requester:calculateIndices}).  The \responderTerm
\responder{} receives as input a set
\similarPasswords{\accountIdAlt}{} of passwords similar to the
password for each local account $\accountIdAlt \in \accountIdSet$
(i.e., \accountIdSet is its set of local account identifiers), and
upon receiving \msgref{prot:msg:bloomFilter} computes its own
\bloomFilterSize-sized Bloom filter containing
\similarPasswords{\accountId}{}, i.e., indices
$\bloomFilterIndicesToSet{\responder{}} \gets \bigcup_{\passwordAlt
  \in \similarPasswords{\accountId}{}} \bigcup_{\bloomFilterHashFnIdx
  \in \residues{\bloomFilterHashFns}}
\{\bloomFilterHashFn{\bloomFilterHashFnIdx}(\passwordAlt)\}$
(\lineref{prot:responder:calculateIndices}).\footnote{This assumes
  that all of \similarPasswords{\accountId}{} will ``fit'' in an
  \bloomFilterSize-sized bloom filter.  If not, \responder{} can use
  any subset of \similarPasswords{\accountId}{} it chooses of
  appropriate size.  This will be discussed further in
  \secref{sec:eval:implementation:bloomFilters}.}  The protocol should
return \boolTrue to \requester if $\password{} \in
\similarPasswords{\accountId}{}$, which for a Bloom filter is
indicated by $\bloomFilterIndicesToSet{\requester} \subseteq
\bloomFilterIndicesToSet{\responder{}}$ (with some risk of false
positives, as will be discussed in
\secref{sec:eval:implementation:bloomFilters}).

Our protocol equivalently returns a value to \requester that indicates
whether
$\residues{\bloomFilterSize}\setminus\bloomFilterIndicesToSet{\responder{}}
\subseteq
\residues{\bloomFilterSize}\setminus\bloomFilterIndicesToSet{\requester}$,
where ``$\setminus$'' denotes set difference,
without exposing
\bloomFilterIndicesToSet{\responder{}} to \requester or
\bloomFilterIndicesToSet{\requester} to \responder{}.  To do so, the
\requesterTerm \requester encodes \bloomFilterIndicesToSet{\requester}
as ciphertexts $\langle
\bloomFilterBitCtext{\bloomFilterBitIdx}\rangle_{\bloomFilterBitIdx\in\residues{\bloomFilterSize}}$
where $\bloomFilterBitCtext{\bloomFilterBitIdx} \in
\ciphertextSpace{\pubKey}(\groupIdentity)$ if $\bloomFilterBitIdx \in
\residues{\bloomFilterSize}\setminus\bloomFilterIndicesToSet{\requester}$
and $\bloomFilterBitCtext{\bloomFilterBitIdx} \in
\ciphertextSpace{\pubKey}(\groupElmt)$ for a randomly chosen
$\groupElmt \getsr \groupSet$ if $\bloomFilterBitIdx \in
\bloomFilterIndicesToSet{\requester}$
(\ref{prot:requester:encryptBF}).  In this way, when \responder{}
computes \resultCiphertext{} in
\lineref{prot:responder:makeResult}---i.e., by homomorphically
multiplying \bloomFilterBitCtext{\bloomFilterBitIdx} for each
$\bloomFilterBitIdx \in
\residues{\bloomFilterSize}\setminus\bloomFilterIndicesToSet{\responder{}}$
and then exponentiating by a random $\encExponent \getsr
\relPrimeInts{\groupOrder}$
(\ref{prot:responder:generateBlinding})---\resultCiphertext{} is
in $\ciphertextSpace{\pubKey}(\groupIdentity)$ if
$\residues{\bloomFilterSize}\setminus\bloomFilterIndicesToSet{\responder{}}
\subseteq
\residues{\bloomFilterSize}\setminus\bloomFilterIndicesToSet{\requester}$
and otherwise is almost certainly not in
$\ciphertextSpace{\pubKey}(\groupIdentity)$.  As such, \requester
returns \boolTrue, indicating that \password{} is similar to the
password set at \responder{} for account \accountId, if and only if
$\decrypt{\privKey}(\resultCiphertext{}) = \groupIdentity$
(\refs{prot:requester:decrypt}{prot:requester:return}).

It is important that both \responder{} and \requester check the
validity of the ciphertexts they receive
(lines~\ref{prot:responder:checkCiphertexts} and
\ref{prot:requester:checkCiphertext}, respectively).  For
\responder{}, implicit in this check is that \pubKey is a valid public
key (i.e., capable of being output by \keygen).  For our
implementation described in \secref{sec:eval:implementation},
these checks are straightforward.

\subsection{Security}
\label{sec:protocol:security}

We now reason about the security of the protocol of
\figref{fig:protocol} against malicious \requestersTerm
(\secref{sec:protocol:security:responder}) and against malicious
\respondersTerm (\secref{sec:protocol:security:requester}).  More
specifically, our focus in this section is properties that underlie
\accountSecurity as informally described in \secref{sec:goals};
\accountLocationPrivacy will be discussed in \secref{sec:framework}.
\iffull
Proofs for all propositions in this section are given in
\appref{sec:proofs}.
\else
Proofs for all propositions in this section can be found in our
technical report~\cite{wang2018:reuse}.
\fi

\subsubsection{Security against malicious \requesterTerm}
\label{sec:protocol:security:responder}

\requester learns nothing more from executing the
protocol in \figref{fig:protocol} besides the result \plaintext{}
\testEqual \groupIdentity in \lineref{prot:requester:return} because
no other information is encoded in \resultCiphertext{} if the
\responderTerm follows the protocol (i.e., unconditional security).
First, if
$\resultCiphertext{}\not\in\ciphertextSpace{\pubKey}(\groupIdentity)$
then \resultCiphertext{} is a ciphertext of any $\groupElmt \in
\groupSet \setminus \{\groupIdentity\}$ with equal probability:

\begin{prop}
  If the \responderTerm follows the protocol, then
  $\cprob{\big}{\resultCiphertext{}\!\in\!\ciphertextSpace{\pubKey}(\groupElmt)}{\resultCiphertext{}\!\not\in\!\ciphertextSpace{\pubKey}(\groupIdentity)}
  = \frac{1}{\groupOrder-1}$ for any $\groupElmt
  \in \groupSet \setminus \{\groupIdentity\}$.
  \label{prop:responderSecurityPtext}
\end{prop}

\noindent
Second, if
$\resultCiphertext{}\!\in\!\ciphertextSpace{\pubKey}(\groupElmt)$, it
is uniformly distributed in $\ciphertextSpace{\pubKey}(\groupElmt)$:

\begin{prop}
  If the \responderTerm follows the protocol, then
  $\cprob{\big}{\resultCiphertext{} = \ciphertext{}}{\resultCiphertext{} \in \ciphertextSpace{\pubKey}(\groupElmt})
  = \frac{1}{\setSize{\ciphertextSpace{\pubKey}(\groupElmt)}}$
  for any $\groupElmt \in \groupSet$ and any $\ciphertext{}
  \in \ciphertextSpace{\pubKey}(\groupElmt)$.
\label{prop:responderSecurityCtext}
\end{prop}

\subsubsection{Security against malicious \responderTerm}
\label{sec:protocol:security:requester}

The system of which the protocol in \figref{fig:protocol} is a
component will typically leak the result of the protocol run to the
\responderTerm.  Specifically, if a run of the protocol is immediately
followed by another run of the protocol, then this suggests that the
protocol returned \boolTrue, i.e., that $\password{} \in
\similarPasswords{\accountId}{}$.  We will discuss in
\secref{sec:framework:design:requester} using extra, ``decoy''
protocol runs to obscure this leakage.  However, for the purposes of
this section, we will assume that the result of the protocol is leaked
to the \responderTerm reliably.

The implications of this leakage to the requirements for the
encryption scheme \encScheme are that the \requesterTerm serves as an
oracle for the \responderTerm to learn whether one ciphertext
\resultCiphertext{} of its choosing satisfies $\resultCiphertext{} \in
\ciphertextSpace{\pubKey}(\groupIdentity)$.  The \responderTerm could
potentially use this oracle to determine which of the ciphertexts
$\langle
\bloomFilterBitCtext{\bloomFilterBitIdx}\rangle_{\bloomFilterBitIdx\in\residues{\bloomFilterSize}}$
that it receives in \lineref{prot:msg:bloomFilter} satisfy
$\bloomFilterBitCtext{\bloomFilterBitIdx}
\in \ciphertextSpace{\pubKey}(\groupIdentity)$ and, in turn, gain
information about the password \password{} that the user is trying to
set.  Indeed, some leakage of this form is unavoidable; e.g., the
\responderTerm could simply set $\resultCiphertext{} =
\bloomFilterBitCtext{0}$ and, in doing so, learn whether
$\bloomFilterBitCtext{0}
\in \ciphertextSpace{\pubKey}(\groupIdentity)$.  Similarly, the
\responderTerm could set $\resultCiphertext{} =
\bloomFilterBitCtext{0} \encMult{\pubKey} \bloomFilterBitCtext{1}$; if
the protocol returns \boolTrue, then the \responderTerm can conclude
that both $\bloomFilterBitCtext{0} \in
\ciphertextSpace{\pubKey}(\groupIdentity)$ and
$\bloomFilterBitCtext{1} \in
\ciphertextSpace{\pubKey}(\groupIdentity)$.

To capture this leakage and the properties of our protocol
more formally, we define a \responderTerm-adversary \adversary{} to be
a pair $\adversary{} = \langle\adversary{1}, \adversary{2}\rangle$ of
probabilistic algorithms.  \adversary{1} takes as input \pubKey and
$\langle
\bloomFilterBitCtext{\bloomFilterBitIdx}\rangle_{\bloomFilterBitIdx\in\residues{\bloomFilterSize}}$
and outputs a ciphertext \resultCiphertext{} and a state
\adversaryState.\footnote{We elide the other values in
  \msgref{prot:msg:bloomFilter} from the input to \adversary{1} only
  because they do not contribute the security of the protocol.}
\adversary{2} is provided the oracle response (i.e., whether
$\resultCiphertext{} \in \ciphertextSpace{\pubKey}(\groupIdentity)$)
and the state \adversaryState and then outputs a set
$\bloomFilterIndicesToSet{\adversary{}} \subseteq
\residues{\bloomFilterSize}$.  \adversary{} is said to
\textit{succeed} if $\bloomFilterIndicesToSet{\adversary{}} =
\bloomFilterIndicesToSet{\requester}$, where
\bloomFilterIndicesToSet{\requester} is the set of indices the
\requesterTerm ``set'' in its Bloom filter by encrypting a random
group element (\lineref{prot:requester:encryptBF}).  More
specifically, we define experiment
$\experiment{\responder{}}{\encScheme}(\langle\adversary{1},
\adversary{2}\rangle)$ as follows:
\[
\begin{array}{ll}
  \mathrlap{\codeExpt \experiment{\responder{}}{\encScheme}(\langle\adversary{1}, \adversary{2}\rangle):} \\ 
  \hspace{3em}
  & \langle\pubKey, \privKey\rangle \gets \keygen(1^{\secParam}) \\
  & \bloomFilterIndicesToSet{\requester} \getsr \{\bloomFilterIndicesToSet{} \subseteq \residues{\bloomFilterSize} \mid \setSize{\bloomFilterIndicesToSet{}} = \bloomFilterHashFns\} \\
  & \forall \bloomFilterBitIdx \in \residues{\bloomFilterSize}:
      \bloomFilterBitCtext{\bloomFilterBitIdx} \gets
      \left\{\begin{array}{@{}l@{\hspace{3pt}}l@{}}
      \encrypt{\pubKey}(\groupRand) & \mbox{if $\bloomFilterBitIdx \in \bloomFilterIndicesToSet{\requester}$}
      \\
      \encrypt{\pubKey}(\groupIdentity) & \mbox{if $\bloomFilterBitIdx \not\in \bloomFilterIndicesToSet{\requester}$}
      \end{array}\right. \\
    & \langle \resultCiphertext{}, \adversaryState\rangle \gets \adversary{1}\left(\pubKey, \langle \bloomFilterBitCtext{\bloomFilterBitIdx}\rangle_{\bloomFilterBitIdx\in\residues{\bloomFilterSize}}\right) \\
    & \bloomFilterIndicesToSet{\adversary{}}\gets \adversary{2}\left(\adversaryState, \left(\resultCiphertext{} \testIn \ciphertextSpace{\pubKey}(\groupIdentity)\right)\right) \\
    & \codeReturn (\bloomFilterIndicesToSet{\adversary{}} \testEqual \bloomFilterIndicesToSet{\requester})
\end{array}
\]
Then, we analyze the security of our protocol against
\responderTerm-adversaries \adversary{} that run in time polynomial in
\secParam by bounding
$\prob{\experiment{\responder{}}{\encScheme}(\adversary{}) =
  \boolTrue}$.

\myparagraph{ElGamal encryption}
To prove security against a malicious \responderTerm, we instantiate
the encryption scheme \encScheme as ElGamal
encryption~\cite{elgamal1985:public-key}, which is implemented as
follows.
\begin{compactitem}
\item $\keygen(1^\kappa)$ returns a private key $\privKey = \langle
  \elgPrivKey\rangle$ and public key $\pubKey = \langle
  \groupGenerator, \elgPubKey \rangle$, where $\elgPrivKey \getsr
  \ints{\groupOrder}$, \groupGenerator is a generator of the (cyclic)
  group $\langle \groupSet, \groupMult \rangle$, and $\elgPubKey \gets
  \groupGenerator^{\elgPrivKey}$.  We leave it implicit that the
  public key \pubKey and private key \privKey must include whatever
  other information is necessary to specify \groupSet, e.g., the
  elliptic curve on which the members of \groupSet lie.

\item $\encrypt{\langle\groupGenerator,\elgPubKey\rangle}(\groupElmt)$
  returns $\langle\elgEphemeralPubKey{}, \elgCiphertext{} \rangle$
  where $\elgEphemeralPubKey{} \gets
  \groupGenerator^{\elgEphemeralPrivKey{}}$, $\elgEphemeralPrivKey{}
  \getsr \ints{\groupOrder}$, and $\elgCiphertext{} \gets \groupElmt
  \elgPubKey^{\elgEphemeralPrivKey{}}$.

\item $\decrypt{\langle \elgPrivKey \rangle}(\langle
  \elgEphemeralPubKey{}, \elgCiphertext{}\rangle)$ returns
  $\elgCiphertext{} \elgEphemeralPubKey{}^{-\elgPrivKey}$ if
  $\{\elgEphemeralPubKey{}, \elgCiphertext{}\} \subseteq
  \groupSet$ and returns $\bot$ otherwise.

\item $\encProd{\langle \groupGenerator,
    \elgPubKey\rangle}{\encProdIdx=1}{\genericNat} \langle
  \elgEphemeralPubKey{\encProdIdx}, \elgCiphertext{\encProdIdx}
  \rangle$ returns $\langle \elgEphemeralPubKey{1} \ldots
  \elgEphemeralPubKey{\genericNat} \groupGenerator^\groupExponent,
  \elgCiphertext{1} \ldots \elgCiphertext{\genericNat}
  \elgPubKey^\groupExponent\rangle$ for $\groupExponent \getsr
  \ints{\groupOrder}$ if each $\{\elgEphemeralPubKey{\encProdIdx},
  \elgCiphertext{\encProdIdx} \} \subseteq \groupSet$ and returns
  $\bot$ otherwise.  $\langle \elgEphemeralPubKey{1},
  \elgCiphertext{1} \rangle \encMult{\langle \groupGenerator,
    \elgPubKey\rangle} \langle \elgEphemeralPubKey{2},
  \elgCiphertext{2} \rangle$ is just the special case $\genericNat
  = 2$.
\end{compactitem}

\myparagraph{Generic group model} We prove the security of our
protocol against a \responderTerm-adversary \adversary{} in the
generic group model as presented by Maurer~\cite{maurer2005:abstract}.
The generic group model allows modeling of attacks in which the
adversary \adversary{} cannot exploit the representation of the group
elements used in the cryptographic algorithm.  For some problems, such
as the discrete logarithm problem on general elliptic curves, generic
attacks are currently the best known (though better algorithms exist
for curves of particular forms, e.g.,~\cite{diem2011:ecdl}).  Somewhat
like the random oracle model~\cite{canetti2004:ro}, the generic group
model is idealized, and so even an algorithm proven secure in the
generic group model can be instantiated with a specific group
representation that renders it insecure.  Still, and also like the
random oracle model, it has been used to provide assurance for the
security of designs in numerous previous works; e.g., see Koblitz and
Menezes~\cite{koblitz2007:generic} for a discussion of this
methodology and how its results should be interpreted.

A function $\negligibleFn: \nats \rightarrow \reals$ is said to be
\textit{negligible} if for any positive polynomial
$\polynomial(\secParam)$, there is some \secParamMin such that
$\negligibleFn(\secParam) < \frac{1}{\polynomial(\secParam)}$ for all
$\secParam > \secParamMin$.  We denote such a function by
\negligible{\secParam}.

\begin{prop}
  If \encScheme is ElGamal encryption, then in the generic group
  model,
  \[
  \prob{\experiment{\responder{}}{\encScheme}(\adversary{}) = \boolTrue}
  \le 2{\bloomFilterSize \choose \bloomFilterHashFns}^{-1} +
  \negligible{\secParam}
  \]
  for any \responderTerm-adversary \adversary{} that runs in time
  polynomial in \secParam.
  \label{prop:genericGroup}
\end{prop}

Our proof of \propref{prop:genericGroup}
\iffull
(see \appref{sec:proofs})
\else
(see~\cite{wang2018:reuse})
\fi
depends on disclosing to \adversary{} the result $\resultCiphertext{}
\testIn \ciphertextSpace{\pubKey}(\groupIdentity)$ for only a single
\resultCiphertext{} or, in other words, on the use of a new public key
\pubKey per run of the protocol in \figref{fig:protocol} (see
\lineref{prot:requester:init}).  Since for ElGamal, generating a new
public key costs about the same as an encryption, reusing a public key
saves at most only $1/(\bloomFilterSize+1)$ of the computational cost
for \requester in the protocol, and so we have not prioritized
evaluating the security of such an optimization.

\propref{prop:genericGroup} is tight, i.e., there is a generic
\responderTerm-adversary that achieves its bound (to within a term
negligible in \secParam).  This adversary $\adversary{} = \langle
\adversary{1}, \adversary{2}\rangle$ performs as follows:
\adversary{1} outputs, say, $\resultCiphertext{} \gets
\bloomFilterBitCtext{0}$ and, upon learning $\resultCiphertext{}
\testIn \ciphertextSpace{\pubKey}(\groupIdentity)$, \adversary{2}
guesses \bloomFilterIndicesToSet{\adversary{}} to be a
\bloomFilterHashFns-element subset of \residues{\bloomFilterSize}
where $0 \in \bloomFilterIndicesToSet{\adversary{}}$ iff
$\resultCiphertext{}
\not\in \ciphertextSpace{\pubKey}(\groupIdentity)$.  Once \groupSet is
instantiated in practice, security rests on the \textit{assumption}
that no \responderTerm-adversary can do better, i.e., that given the
decisional Diffie-Hellman (DDH) instances
$\langle\bloomFilterBitCtext{\bloomFilterBitIdx}\rangle_{\bloomFilterBitIdx\in\residues{\bloomFilterSize}}$
for public key $\langle \groupGenerator, \elgPubKey \rangle$, no
adversary can create a single DDH instance \resultCiphertext{} for
which the answer enables it to solve the instances
$\langle\bloomFilterBitCtext{\bloomFilterBitIdx}\rangle_{\bloomFilterBitIdx\in\residues{\bloomFilterSize}}$
with probability better than that given in
\propref{prop:genericGroup}.  Informally, \propref{prop:genericGroup}
says that any such adversary would need to leverage the representation
of \groupSet to do so.

\section{Interfering with Password Reuse}
\label{sec:framework}

In this section, we propose a password reuse detection
framework based on the PMT protocol proposed in \secref{sec:protocol}.

\subsection{Design}
\label{sec:framework:design}

Our password reuse detection framework enables a \requesterTerm
\requester to inquire with multiple \respondersTerm as to whether the
password \password{} chosen by a user for the account at \requester
with identifier \accountId is similar to another password already set
for \accountId at some \responderTerm.  The \requesterTerm does so
with the help of a \textit{\directoryTerm}, which is a (possibly
replicated) server that provides a front-end to \requestersTerm for
this purpose.  The \directoryTerm stores, per identifier \accountId, a
list of addresses (possibly pseudonymous addresses, as we will discuss
below) of websites at which \accountId has been used to set up an
account.  We stress that the \directoryTerm does \textit{not} handle
or observe passwords in our framework.

The \requestersTerm and \respondersTerm need not trust each other in
our framework, and employ the protocol described in
\secref{sec:protocol} to interact via the \directoryTerm.  More
specifically, a user of the \requesterTerm \requester selects a
password \password{} for her account with identifier \accountId, and
submits \password{} to \requester.  \requester sends the message in
\lineref{prot:msg:bloomFilter} of \figref{fig:protocol} to the
\directoryTerm, which it forwards to some subset of
\nmbrRespondersQueried \respondersTerm, out of the
\nmbrResponders{\accountId} total registered as having accounts
associated with \accountId at the \directoryTerm.  (How it selects
\nmbrRespondersQueried is discussed in
\secref{sec:eval:optimization}.)  The response from \responderTerm
\responder{\responderIdx} is denoted
$\ref{prot:msg:result}_{\responderIdx}$ in \figref{fig:framework}.
Once the \directoryTerm collects these responses, it forwards them
back to \requester, after permuting them
randomly to prevent \requester from knowing which
\responderTerm returned which result (see
\secref{sec:framework:design:responder}).
\requester then processes each as in
\linesref{prot:requester:checkCiphertext}{prot:requester:return}; any
of these results that are \boolTrue indicates that some \responderTerm
that was queried has a password similar to \password{} set for account
\accountId.  If any are true, then the \requesterTerm (presumably) rejects
\password{} and asks the user to select a different password (perhaps
with guidance to help her choose one that is likely to not be used
elsewhere).

\begin{figure}[h]
	\centering
	\setlength\figureheight{2.2in}
%
\hspace*{0.1em}
\psscalebox{0.57 0.57} 
{
\begin{pspicture}(0,-2.7)(23.024794,2.7)
\pscircle[linecolor=black, linewidth=0.04, dimen=outer](0.55146116,1.0666667){0.23333333}
\psline[linecolor=black, linewidth=0.04](0.55146116,0.8333333)(0.55146116,0.36666667)(0.55146116,0.36666667)
\psline[linecolor=black, linewidth=0.04](0.55146116,0.36666667)(0.3181278,-0.1)(0.3181278,-0.1)
\psline[linecolor=black, linewidth=0.04](0.55146116,0.36666667)(0.78479445,-0.1)(0.78479445,-0.1)
\psline[linecolor=black, linewidth=0.04](0.55146116,0.6)(1.0181278,0.8333333)(1.0181278,0.8333333)
\psline[linecolor=black, linewidth=0.04](0.55146116,0.6)(0.08479446,0.8333333)(0.08479446,0.8333333)
\psdots[linecolor=black, dotsize=0.1](13.484795,0.8)
\psline[linecolor=black, linewidth=0.04, arrowsize=0.05291667cm 3.0,arrowlength=1.4,arrowinset=0.0]{<-}(7.6847944,1.54)(5.6847944,1.54)
\psline[linecolor=black, linewidth=0.04, arrowsize=0.05291667cm 3.0,arrowlength=1.4,arrowinset=0.0]{<-}(12.483691,2.0928416)(9.67179,1.3218133)
\psline[linecolor=black, linewidth=0.04, linestyle=dashed, dash=0.17638889cm 0.10583334cm, arrowsize=0.05291667cm 3.0,arrowlength=1.4,arrowinset=0.0]{<-}(5.6847944,-0.80102116)(7.684794,-0.8)
\psdots[linecolor=black, dotsize=0.1](13.484795,0.5003448)
\psdots[linecolor=black, dotsize=0.1](13.470415,0.20068963)
\rput[bl](11.084794,2.1){\Large{\ref{prot:msg:bloomFilter}}}
\rput[bl](11.084794,-0.1){\Large{\ref{prot:msg:bloomFilter}}}
\rput[bl](11.084794,0.7){\Large{\ref{prot:msg:result}$_1$}}
\rput[bl](5.884794,-0.6){\Large{\{\ref{prot:msg:result}$_\responderIdx$\}$^\nmbrRespondersQueried_{\responderIdx=1}$}}
\psline[linecolor=black, linewidth=0.04, arrowsize=0.05291667cm 3.0,arrowlength=1.4,arrowinset=0.0]{<-}(3.6847718,1.96)(1.2848171,1.9465474)
\psline[linecolor=black, linewidth=0.04, linestyle=dashed, dash=0.17638889cm 0.10583334cm, arrowsize=0.05291667cm 3.0,arrowlength=1.4,arrowinset=0.0]{<-}(1.2847944,-1.1)(3.6847944,-1.1)
\rput[bl](1.5847945,2.1){\Large{password}}
\rput[bl](4.5247946,0.3){\Large{\requester}}
\rput[bl](7.6847944,-2.7){\Large{\textbf{Directory}}}
\rput[bl](13.284795,1.7){\Large{\responder{1}}}
\rput{-242.57664}(11.974575,-8.877314){\psarc[linecolor=black, linewidth=0.04, dimen=outer, arrowsize=0.05291667cm 3.0,arrowlength=1.4,arrowinset=0.0]{<-}(8.684794,-0.8){0.6}{0.0}{300.0}}
\rput[bl](8.284795,-0.9){\textbf{Perm}}
\rput[bl](1.9847945,-0.2){\Large{accept}}
\rput[bl](2.3847945,-0.5){\Large{or}}
\rput[bl](1.9847945,-1.0){\Large{reject}}
\rput[bl](0.12479446,-2.6){\Large{\textbf{User}}}
\rput[bl](3.6847944,-2.7){\Large{\textbf{Requester}}}
\rput[bl](12.384794,-2.7){\Large{\textbf{Responders}}}
\rput[bl](13.284795,-1.1){\Large{\responder{\nmbrRespondersQueried}}}
\psframe[linecolor=black, linewidth=0.04, dimen=outer](9.484795,2.7)(7.884794,-1.7)
\psframe[linecolor=black, linewidth=0.04, dimen=outer](14.284795,2.7)(12.684794,1.1)
\psline[linecolor=black, linewidth=0.04, linestyle=dashed, dash=0.17638889cm 0.10583334cm, arrowsize=0.05291667cm 3.0,arrowlength=1.4,arrowinset=0.0]{<-}(9.685407,0.9049246)(12.496959,1.6770135)
\psline[linecolor=black, linewidth=0.04, arrowsize=0.05291667cm 3.0,arrowlength=1.4,arrowinset=0.0]{<-}(12.55744,-0.6653243)(9.693124,0.11899191)
\psline[linecolor=black, linewidth=0.04, linestyle=dashed, dash=0.17638889cm 0.10583334cm, arrowsize=0.05291667cm 3.0,arrowlength=1.4,arrowinset=0.0]{<-}(9.611862,-0.28063497)(12.476631,-1.0641012)
\psframe[linecolor=black, linewidth=0.04, dimen=outer](14.284795,-0.1)(12.684794,-1.7)
\rput[bl](11.084794,-1.4){\Large{\ref{prot:msg:result}$_{\nmbrRespondersQueried}$}}
\rput[bl](6.2847943,1.8){\Large{\ref{prot:msg:bloomFilter}}}
\psframe[linecolor=black, linewidth=0.04, dimen=outer](5.4847946,2.7)(3.8847945,-1.7)
\end{pspicture}
}
	\caption{Password reuse detection framework based on the PMT
          protocol introduced in \secref{sec:protocol}.}
	\label{fig:framework}
\end{figure}

There are some additional operations needed to support the framework,
as well.
\begin{compactitem}
\item \textit{Directory entry addition}: After a new account is set
  up, the \requesterTerm sends its address (discussed below) to the
  \directoryTerm to be stored with the account identifier \accountId.
\item \textit{Directory entry deletion}: When an account \accountId on
  a web server (\responderTerm) is no longer used, the \responderTerm
  can optionally update the \directoryTerm to remove the \responderTerm's
  address associated with \accountId.
\item \textit{Password change}: When a user tries to change the
  password of an account, the web server should launch the protocol
  (as a \requesterTerm) before the new password is accepted to replace
  the old.
\end{compactitem}

The \requesterTerm can communicate with the \directoryTerm normally
(e.g., using TLS over TCP/IP), trusting the \directoryTerm to mask its
identity from each \responderTerm to implement \accountLocationPrivacy
(in which case, the \directoryTerm behaves as an anonymizing proxy,
cf.,~\cite{boyan1997:anonymizer,gabber1999:lpwa}).  Or, if the
\requesterTerm does not trust the \directoryTerm to hide its identity,
then it can communicate with the \directoryTerm using an anonymizing
protocol such as Tor
(\url{https://www.torproject.org/},~\cite{dingledine2004:tor}).
Similarly, each \responderTerm address registered at the
\directoryTerm can be a regular TCP/IP endpoint, if it trusts the
\directoryTerm to hide its identity from others, or an anonymous
server address such as a Tor hidden-service
address~\cite{dingledine2004:tor} if it does not.  In the latter case,
the \responderTerm should register a distinct anonymous server address
at the \directoryTerm per account identifier \accountId, to prevent
identifying the \responderTerm by the number of accounts it hosts.

While each website could choose individually whether to trust the
\directoryTerm to hide its identity from others, we will evaluate the
performance of our system only when either all websites trust the
\directoryTerm in this sense or none do.  We refer to these models in
the rest of the paper as the
\textit{\trustedForAccountLocationPrivacy-\directoryTerm} model (short
for ``trusted for \accountLocationPrivacy'') and the
\textit{\untrustedForAccountLocationPrivacy-\directoryTerm} model
(``untrusted for \accountLocationPrivacy''), respectively.  We believe
that the \trustedForAccountLocationPrivacy-\directoryTerm model would
be especially well-suited for deployment by a large cloud operator to
serve its tenants, since these tenants already must trust the cloud
operator.

Our framework is agnostic to the method by which each \responderTerm
generates the set \similarPasswords{\accountId}{} of similar passwords
for an account \accountId.  We envision it doing so by leveraging
existing password guessers (e.g.,~\cite{das2014:tangled,
  wang2018:domino, wang2016:guesser, zhang2010:expiration}), seeded
with the actual password for the account.  In addition, if, say,
Google observes a user \accountId set the password \texttt{google123},
it could add \texttt{twitter123} and \texttt{facebook123} to
\similarPasswords{\accountId}{}.  So as to eliminate the need to store
trivial variations of passwords in \similarPasswords{\accountId}{} and
so reduce its size, the \responderTerm could reduce all such variants
to a single canonical form, e.g., by translating all capital letters
to lowercase, provided that \requestersTerm know to do the same.

\subsubsection{Security for each \responderTerm}
\label{sec:framework:design:responder}

The \responderTerm need not retain the elements of
\similarPasswords{\accountId}{} explicitly, but instead should store in
\similarPasswords{\accountId}{} only the hash of each similar
password, using a time-consuming cryptographic hash function
\cryptoHash, making \similarPasswords{\accountId}{} more costly to
exploit if the site is breached~\cite{spafford1992:opus}.  In
particular, this hash function need not be the same as that used to
hash the real password during normal login, and so can be considerably
more time-consuming.
In addition, \cryptoHash can be salted with a salt computed
deterministically from user identifier \accountId, so that the salts
for \accountId used at different sites are identical.
Going further, the \responderTerm could proactively generate the set
\bloomFilterIndicesToSet{\responder{}} when the password for
\accountId is set at \responder{}, and dispense of
\similarPasswords{\accountId}{} altogether.  However, this
precomputation would require the Bloom filter size \bloomFilterSize
and hash functions $\langle\bloomFilterHashFn{\bloomFilterHashFnIdx}
\rangle_{\bloomFilterHashFnIdx \in \residues{\bloomFilterHashFns}}$ to
be fixed and known to the \responderTerm in advance.

\myparagraph{Protecting \bloomFilterIndicesToSet{\responder{}} from
  disclosure} As shown in \secref{sec:protocol:security:responder},
the only information leaked to the \requesterTerm is the result of the
protocol in \figref{fig:protocol}, i.e., $\resultCiphertext{}
\testIn \ciphertextSpace{\pubKey}(\groupIdentity)$, regardless of the
behavior of the \requesterTerm
(\proprefs{prop:responderSecurityPtext}{prop:responderSecurityCtext}).
Still, however, this information can erode the security of
\bloomFilterIndicesToSet{\responder{}} over multiple queries.  For
example, if a malicious \requesterTerm sets
$\bloomFilterBitCtext{\bloomFilterBitIdx} \gets
\encrypt{\pubKey}(\groupElmt)$ where $\groupElmt \neq \groupIdentity$
for one Bloom-filter index \bloomFilterBitIdx and
$\bloomFilterBitCtext{\bloomFilterBitIdxAlt} \gets
\encrypt{\pubKey}(\groupIdentity)$ for $\bloomFilterBitIdxAlt \neq
\bloomFilterBitIdx$, then the result of $\resultCiphertext{}
\testIn \ciphertextSpace{\pubKey}(\groupIdentity)$ reveals whether
$\bloomFilterBitIdx \in \bloomFilterIndicesToSet{\responder{}}$.
After \bloomFilterSize such queries, the \requesterTerm can learn the
entirety of \bloomFilterIndicesToSet{\responder{}} and then search for
the items stored in the Bloom filter offline.

Our framework mitigates this leakage using three mechanisms.  First,
each \responderTerm serves only PMT queries forwarded through the
\directoryTerm, i.e., by authenticating requests as coming 
from the \directoryTerm.  This step is important for the following
two mitigations to work.

Second, the \directoryTerm randomly permutes the
$\ref{prot:msg:result}_{\responderIdx}$ messages received from
\respondersTerm before returning them to the \requesterTerm, thereby
eliminating any indication (by timing or order) of the \responderTerm
\responder{\responderIdx} from which each
$\ref{prot:msg:result}_{\responderIdx}$ was returned.  This largely
eliminates the information that a malicious \requesterTerm can glean
from multiple PMT queries.  In particular, the method above reveals
nothing to the \requesterTerm except the number of queried
\respondersTerm \responder{\responderIdx} for which
$\bloomFilterBitIdx \in
\bloomFilterIndicesToSet{\responder{\responderIdx}}$, but not the
\respondersTerm for which this is true.

Third, we involve the user to restrict the number of PMT queries that
any \requesterTerm can make.  Assuming \accountId is an email address
or can be associated with one at the \directoryTerm, the
\directoryTerm emails the user upon being contacted by a
\requesterTerm, to confirm that she is trying to (re)set her password
at that website.\footnote{The user can check that the confirmation
  email pertains to the site at which she is (re)setting her password
  if the site generates a nonce that it both displays to the user and
  passes to the \directoryTerm to include in the confirmation email.
  The email should instruct the user to confirm this password (re)set
  only if the nonce displayed by the website matches that received in
  the email.  } This email could be presented to the user much like an
account setup confirmation email today, containing a URL at the
\directoryTerm that user clicks to confirm her attempt to (re)set her
password.  The \directoryTerm simply holds
\msgref{prot:msg:bloomFilter} until receiving this confirmation,
discarding the message if it times out.  (Presumably the
\requesterTerm website would alert the user to check her inbox for
this confirmation email.)  To avoid requiring the user to confirm
multiple attempts to set a password at the \requesterTerm and so
multiple runs of the protocol in \figref{fig:protocol} (which should
occur only if the user is still not using a password manager), the
\directoryTerm could allow one such confirmation to permit queries
from this \requesterTerm for a short window of time, at the risk of
allowing a few extra queries if the \requesterTerm is malicious.
However, except during this window, \requesterTerm queries will be
dropped by the \directoryTerm.

Leveraging the \directoryTerm to permute PMT responses and limit PMT
queries requires that we place trust in the \directoryTerm to do so.
If desired, this trust can be mitigated by replicating the
\directoryTerm using Byzantine fault-tolerance methods to overcome
misbehavior by individual replicas.  Ensuring that only user-approved
PMTs are allowed can be implemented using classic BFT state-machine
replication, for which increasingly practical frameworks exist
(e.g.,~\cite{bessani2014:bft-smart,clement2009:upright}).  Permuting
PMT responses in a way that hides which \responder{\responderIdx}
returned each $\ref{prot:msg:result}_{\responderIdx}$ even from
\nmbrFaultyDirectoryReplicas corrupt \directoryTerm replicas can be
achieved by simply having $\nmbrFaultyDirectoryReplicas + 1$ replicas
permute and re-randomize the $\ref{prot:msg:result}_{\responderIdx}$
messages in sequence before returning them to the \requesterTerm.

\myparagraph{Limiting utility of a
  \bloomFilterIndicesToSet{\responder{}} disclosure} The risk that the
adversary finds the password for user \accountId at \responderTerm
\responder{}, even with \bloomFilterIndicesToSet{\responder{}}, is
small if the user leveraged a state-of-the-art password manager to
generate a password that resists even an offline dictionary attack.
Even if the user is not already using a password manager, obtaining
the account password using this attack should again be expensive if
the cryptographic hash function \cryptoHash is costly to compute.
Moreover, the attacker can utilize a guessed account password only if
it can determine the \responderTerm \responder{} at which it is set
for account \accountId, with which \accountLocationPrivacy interferes.

Still, to counter any remaining risk in case the attacker finds
\bloomFilterIndicesToSet{\responder{}}, we advocate that \responder{}
form its set \similarPasswords{\accountId}{} to include \textit{honey
  passwords}~\cite{bojinov2010:kamouflage, juels2013:honeywords,
  erguler2016:flatness}.  That is, when the password is set (or reset)
at a website for account \accountId, the website chooses a collection
of \nmbrHoneyPasswords honey passwords $\honeyPassword{}{1}, \ldots,
\honeyPassword{}{\nmbrHoneyPasswords}$, as well, via a
state-of-the-art method of doing so.  It then generates a
\textit{cluster} of similar passwords for each of the
$\nmbrHoneyPasswords+1$ passwords---we denote the cluster for the real
password \password{} by \cluster{\password{}} and the honey-password
clusters by $\cluster{\honeyPassword{}{1}}, \ldots,
\cluster{\honeyPassword{}{\nmbrHoneyPasswords}}$---with each cluster
being the same size \clusterSize.  Then, it sets the similar passwords
for account \accountId to be the union of these clusters, i.e.,
$\similarPasswords{\accountId}{} = \cluster{\password{}} \cup
\left(\bigcup_{\honeyPasswordIdx=1}^{\nmbrHoneyPasswords}
\cluster{\honeyPassword{}{\honeyPasswordIdx}}\right)$.

In this way, even if the attacker learns the entire contents of
\bloomFilterIndicesToSet{\responder{}} for a \responderTerm \responder{},
the set \bloomFilterIndicesToSet{\responder{}} will contain at least
$\nmbrHoneyPasswords+1$ passwords that appear to be roughly equally
likely.  If any password in a honey-password cluster is then used in
an attempt to log into the account, the website can lock the account
and force the user to reset her password after authenticating via a
fallback method.  The main cost of using honey passwords is a
linear-in-\nmbrHoneyPasswords growth in the size of
\similarPasswords{\accountId}{}, which reduces the cluster size
\clusterSize that can be accommodated by the Bloom-filter size
\bloomFilterSize (which is determined by the \requesterTerm).  We will
show in \secref{sec:eval:optimization}, however, that this cost has
little impact on interfering with password reuse.

\subsubsection{Security for the \requesterTerm}
\label{sec:framework:design:requester}

Security for the \requesterTerm is more
straightforward, given \propref{prop:genericGroup} that proves the
privacy of \bloomFilterIndicesToSet{\requester} against a malicious
\responderTerm (and from the \directoryTerm) in the generic group
model.  Moreover, the \requesterTerm's identity is hidden from
\respondersTerm either by the \directoryTerm (in the
\trustedForAccountLocationPrivacy-\directoryTerm model) or because the
\requesterTerm contacts the \directoryTerm anonymously (in the
\untrustedForAccountLocationPrivacy-\directoryTerm model).

As discussed in \secref{sec:protocol:security:requester} and accounted
for in \propref{prop:genericGroup}, \respondersTerm (and the
\directoryTerm) learn the outcome of the protocol, since they see if
the \requesterTerm runs the protocol again.  That is, a \boolTrue
result will presumably cause the \requesterTerm to reject the password
and ask the user for another, with which it repeats the protocol.
However, because the password is different in each run (which the
\requesterTerm should enforce), the information leaked to
\respondersTerm does not accumulate over these multiple runs.  And,
the \respondersTerm learn only that \textit{at least one} response
resulted in \boolTrue, not how many or which \respondersTerm' did so.

Still, if the information leaked by the \boolFalse result for the
password \password{} finally accepted at the \requesterTerm is of
concern, it is possible to obfuscate even this information to an
extent, at extra expense.  To do so, the \requesterTerm follows the
acceptance of \password{} with a number of ``decoy'' protocol runs
(e.g., each using a randomly chosen
\bloomFilterIndicesToSet{\requester} set of size \bloomFilterHashFns),
as if the run on \password{} had returned \boolTrue.  The user need
not be delayed while each decoy run is conducted.  That said, because
decoy runs add overhead and because the \responderTerm is limited to
learn information about \password{} in only a single protocol run (and
to learn a limited amount, per \propref{prop:genericGroup}), we do not
consider decoys further here.

\subsection{Analysis via Probabilistic Model Checking}
\label{sec:framework:analysis}

Probabilistic model checking is a formal method to analyze
probabilistic behaviors in a system. In this section, we evaluate the
security of our framework against a malicious \requesterTerm using
Storm, a probabilistic model checker~\cite{dehnert2017:storm}.

Storm supports analysis of a Markov decision process (MDP), by which
we model the attacker targeting a specific account \accountId.  That
is, we specify the adversary as a set of \textit{states} and possible
\textit{actions}.  When in a state, the attacker can choose from among
these actions nondeterministically; the chosen action determines a
probability distribution on the state to which the attacker then
transitions.  These state transitions satisfy the \textit{Markov
  property}: informally, the probability of next transitioning to a
specific state depends only on the current state and the attacker's
chosen action.  Storm exhaustively searches all decisions an attacker
can make to maximize the probability of the attacker succeeding in its
goal.  Here, we define this goal to be gaining access to account
\accountId on any \responderTerm, and so Storm calculates the
probability of the attacker doing so under an optimal strategy.

As is common in formal treatments of password guessing
(e.g.,~\cite{katz2009:efficient}), we parameterize the attacker with a
password \textit{dictionary} of a specified size, from which
\accountId's password \password{\responderIdx} at each \responderTerm
\responder{\responderIdx} is chosen independently and uniformly at
random.  The base-2 logarithm of this size represents the
entropy of the password.
We then vary the size of this dictionary to model the attacker's
knowledge about \accountId's password choices at \respondersTerm.  We
further presume that the similar-password set
\similarPasswords{\accountId}{\responderIdx} at each \responderTerm
\responder{\responderIdx} is contained in this dictionary (or
equivalently we reduce \similarPasswords{\accountId}{\responderIdx} to
the subset that falls into the dictionary).
For simplicity, we assume
that the clusters $\cluster{\password{\responderIdx}},
\cluster{\honeyPassword{\responderIdx}{1}}, \ldots,
\cluster{\honeyPassword{\responderIdx}{\nmbrHoneyPasswords}}$ that
comprise \similarPasswords{\accountId}{\responderIdx} are mutually
disjoint and disjoint across \respondersTerm; so,
$\setSize{\bigcup_{\responderIdx=1}^{\nmbrRespondersQueried}
  \similarPasswords{\accountId}{\responderIdx}} =
\sum_{\responderIdx=1}^{\nmbrRespondersQueried}
\setSize{\similarPasswords{\accountId}{\responderIdx}} =
\nmbrRespondersQueried (\nmbrHoneyPasswords+1) \clusterSize$ where
\clusterSize is the size of each cluster.  Below, we denote by
$\allSimilar = \bigcup_{\responderIdx=1}^{\nmbrRespondersQueried}
\similarPasswords{\accountId}{\responderIdx}$ the union of all
similar-password sets constructed by \respondersTerm.
  
The attacker is limited by two parameters.  First, we presume that
each \responderTerm limits the number of consecutive failed logins per
account before the account locks, as is typical and recommended
(e.g.,~\cite{nist2017:800-63B}); we call this number the \textit{login
  budget} and denote it \loginBudget.  Second, our framework limits
the PMT queries on \accountId's accounts to those approved by user
\accountId when she is (re)setting her password (see
\secref{sec:framework:design:responder}); we model this restriction as
a \textit{PMT budget}.  The login budget is per \responderTerm,
whereas the PMT budget is a global constraint.

We also permit the adversary advantages that he might not have in
practice.  First, he knows the full set of
\respondersTerm, so he can attempt to log into any of them, and the
login budget at each.  Second, if he receives a positive
response to a PMT query with password \passwordAlt{}, then the cluster
containing \passwordAlt becomes completely known to him.
That is, if $\passwordAlt \in \cluster{\password{\responderIdx}}$ for
the actual account-\accountId password \password{\responderIdx} at
\responder{\responderIdx}, then \cluster{\password{\responderIdx}} is
added to the adversary's set of identified clusters, and if
$\passwordAlt \in
\cluster{\honeyPassword{\responderIdx}{\honeyPasswordIdx}}$ for a
honey password \honeyPassword{\responderIdx}{\honeyPasswordIdx} at
\responder{\responderIdx}, then
\cluster{\honeyPassword{\responderIdx}{\honeyPasswordIdx}} is added to
that set.  Critically, however, he learns neither whether
the new cluster is the cluster of a real password or a honey password,
nor the \responderTerm \responder{\responderIdx} at which the cluster
was chosen; both of these remain hidden in our design.  Third, each
failed login attempt at \responder{\responderIdx} provides the
adversary complete information about the attempted password
\passwordAlt, specifically if it is in a honey-password cluster
($\passwordAlt \in \bigcup_{\honeyPasswordIdx=1}^{\nmbrHoneyPasswords}
\cluster{\honeyPassword{\responderIdx}{\honeyPasswordIdx}}$) or simply
incorrect ($\passwordAlt \neq \password{\responderIdx}$).

\subsubsection{Model Description}

A state in our model is defined to include the following items of
information: previous adversary PMT queries and their results; the
number of PMT queries that remain available to the attacker; the
password clusters whose existence in \allSimilar has been confirmed by
the adversary via PMTs, to which we refer as the \textit{confirmed}
clusters; and per website, the previous adversary login attempts,
their results, and the number of login queries remaining at that
website.

\begin{figure}[h]
	\centering
	\setlength\figureheight{2.2in}
%
\psscalebox{0.57 0.59} 
{
\begin{pspicture}(0,-4.2)(48.320538,4.2)
\rput[bl](5.6,3.0){\bf{STATE 2}}
\psellipse[linecolor=black, linewidth=0.06, linestyle=dashed, dash=0.17638889cm 0.10583334cm, dimen=outer](2.6,0.3)(1.2,0.5)
\rput[bl](1.8,0.2){\bf{ACTION 1}}
\psline[linecolor=black, linewidth=0.04, linestyle=dashed, dash=0.17638889cm 0.10583334cm, arrowsize=0.05291667cm 2.0,arrowlength=1.4,arrowinset=0.0]{<-}(3.753216,0.78)(5.2,2.26)
\psline[linecolor=black, linewidth=0.04, linestyle=dashed, dash=0.17638889cm 0.10583334cm, arrowsize=0.05291667cm 2.0,arrowlength=1.4,arrowinset=0.0]{<-}(3.8,-0.2)(5.0,-1.5)
\psline[linecolor=black, linewidth=0.04, arrowsize=0.05291667cm 2.0,arrowlength=1.4,arrowinset=0.0]{->}(3.6757755,1.1078637)(5.2,2.6)
\psline[linecolor=black, linewidth=0.04, arrowsize=0.05291667cm 2.0,arrowlength=1.4,arrowinset=0.0]{<-}(4.969963,-1.8289475)(3.6998484,-0.5)
\psline[linecolor=black, linewidth=0.04, arrowsize=0.05291667cm 2.0,arrowlength=1.4,arrowinset=0.0]{->}(12.8,-0.4)(11.8,-1.2)
\psline[linecolor=black, linewidth=0.04, linestyle=dashed, dash=0.17638889cm 0.10583334cm, arrowsize=0.05291667cm 2.0,arrowlength=1.4,arrowinset=0.0]{<-}(12.0,-0.2)(7.6,-1.8)
\psline[linecolor=black, linewidth=0.04, linestyle=dashed, dash=0.17638889cm 0.10583334cm, arrowsize=0.05291667cm 2.0,arrowlength=1.4,arrowinset=0.0]{<-}(12.0,1.2)(7.6,3.0)
\psline[linecolor=black, linewidth=0.04, linestyle=dashed, dash=0.17638889cm 0.10583334cm, arrowsize=0.05291667cm 2.0,arrowlength=1.4,arrowinset=0.0]{->}(4.8,0.4)(4.0,0.4)
\psline[linecolor=black, linewidth=0.04, linestyle=dashed, dash=0.17638889cm 0.10583334cm, arrowsize=0.05291667cm 2.0,arrowlength=1.4,arrowinset=0.0]{<-}(11.838107,0.12611231)(7.78553,0.0988186)
\psline[linecolor=black, linewidth=0.04, arrowsize=0.05291667cm 2.0,arrowlength=1.4,arrowinset=0.0]{->}(11.756178,0.3890818)(7.773847,0.4)
\rput[bl](1.6,3.4){Probablistic transition}
\psline[linecolor=black, linewidth=0.04, linestyle=dashed, dash=0.17638889cm 0.10583334cm, arrowsize=0.05291667cm 2.0,arrowlength=1.4,arrowinset=0.0]{->}(0.60032535,3.8951342)(1.4,3.9)
\rput[bl](1.6,3.8){Adversary's decision}
\psline[linecolor=black, linewidth=0.04, arrowsize=0.05291667cm 2.0,arrowlength=1.4,arrowinset=0.0]{->}(0.60032535,3.4351342)(1.4,3.44)
\rput[bl](0.0,-2.6){STATE 1: PMT negative response}
\rput[bl](0.0,-3.0){STATE 2: PMT positive response}
\rput[bl](0.0,-3.4){STATE 3: Failed login}
\rput[bl](7.4,-3.8){ACTION 1: Submit a PMT query (\emph{PMT budget$--$})}
\rput[bl](7.4,-4.2){ACTION 2: Try to log in (\emph{login budget$--$})}
\rput[bl](5.6,0.2){\bf{STATE 3}}
\rput[bl](5.6,-2.2){\bf{STATE 1}}
\psellipse[linecolor=black, linewidth=0.06, dimen=outer](10.9,-2.1)(1.1,0.9)
\rput[bl](10.2,-2.2){\bf{STATE 4}}
\psellipse[linecolor=black, linewidth=0.06, linestyle=dashed, dash=0.17638889cm 0.10583334cm, dimen=outer](13.2,0.3)(1.2,0.5)
\rput[bl](12.4,0.2){\bf{ACTION 2}}
\rput[bl](0.0,-3.8){STATE 4: Successful login}
\rput[bl](10.2,3.0){\bf{STATE 5}}
\psline[linecolor=black, linewidth=0.04, arrowsize=0.05291667cm 2.0,arrowlength=1.4,arrowinset=0.0]{->}(12.8,1.0)(12.0,2.0)
\rput[bl](0.0,-4.2){STATE 5: Detected by honey passwords}
\rput{-0.18267249}(-0.006363756,0.0079807425){\rput[bl](2.5,2.0){$\prob{\passwordAlt \in \allSimilar}$}}
\rput{-0.10154439}(0.0024844941,0.0037195955){\rput[bl](2.1,-1.4){$\prob{\passwordAlt \notin \allSimilar}$}}
\rput[bl](12.54,1.34){$\prob{\passwordAlt \in \bigcup_{\honeyPasswordIdx=1}^{\nmbrHoneyPasswords} \cluster{\honeyPassword{\responderIdx}{\honeyPasswordIdx}}}$}
\rput[bl](12.4,-1.2){$\prob{\passwordAlt = \password{\responderIdx}}$}
\rput{-0.13594972}(-0.001259619,0.018271832){\rput[bl](7.7,0.54){\prob{\begin{array}{r} \passwordAlt \neq \password{\responderIdx}~\wedge \\ \passwordAlt \not\in \bigcup_{\honeyPasswordIdx=1}^{\nmbrHoneyPasswords} \cluster{\honeyPassword{\responderIdx}{\honeyPasswordIdx}}\end{array}}}}
\psellipse[linecolor=black, linewidth=0.06, dimen=outer](6.3,-2.1)(1.1,0.9)
\psellipse[linecolor=black, linewidth=0.06, dimen=outer](6.3,0.3)(1.1,0.9)
\psellipse[linecolor=black, linewidth=0.06, dimen=outer](6.3,3.1)(1.1,0.9)
\psellipse[linecolor=black, linewidth=0.06, dimen=outer](10.9,3.1)(1.1,0.9)
\psellipse[linecolor=black, linewidth=0.06, dimen=outer](10.9,3.1)(1.3,1.1)
\psellipse[linecolor=black, linewidth=0.06, dimen=outer](10.9,-2.1)(1.3,1.1)
\psline[linecolor=black, linewidth=0.04, linestyle=dashed, dash=0.17638889cm 0.10583334cm, arrowsize=0.05291667cm 2.0,arrowlength=1.4,arrowinset=0.0]{->}(0.6,1.0)(1.2134445,0.4869804)
\psline[linecolor=black, linewidth=0.04, linestyle=dashed, dash=0.17638889cm 0.10583334cm, arrowsize=0.05291667cm 2.0,arrowlength=1.4,arrowinset=0.0]{->}(15.4,1.0)(14.6,0.38322502)
\psellipse[linecolor=black, linewidth=0.06, dimen=outer](6.3,0.3)(1.3,1.1)
\end{pspicture}
}
	\caption{Abstract MDP automaton for attacker interaction with
          \responder{\responderIdx}.  \password{\responderIdx} is the
          correct account password; \honeyPassword{\responderIdx}{1},
          \ldots, \honeyPassword{\responderIdx}{\nmbrHoneyPasswords}
          are its honey passwords; \passwordAlt is an attacker's
          password guess.  Probabilities are conditioned on attacker
          knowledge gained so far.}
	\label{fig:automaton}
\end{figure}

\figref{fig:automaton} shows an automaton that represents the attacker
interacting with one website \responder{\responderIdx}.  The entire
model includes multiple such automata, one per website, and the
adversary can switch among these automata at each step.  Actions and
states shown in \figref{fig:automaton} represent sets of actions and
states in the actual automaton.  For example, when the adversary tries
to login by submitting a password to the login interface of the
website, the password could be chosen from a ``confirmed'' cluster
list or not, which is determined by the adversary. Though these are
separate actions in our model, we let ACTION 2 serve as an
abbreviation for all such actions in \figref{fig:automaton}, to
simplify the figure.  Similarly, a state shown in
\figref{fig:automaton} represents all states resulting from the same
query response but that differ based on the state variables described
above.
Final states (for interacting with \responder{\responderIdx}) are
indicated by double circles.

If the adversary enters STATE~5 for a website or uses up its login
budget for a website, he must switch to another website to continue
attacking. The adversary wins if he enters STATE~4 on any one of the
websites, while he loses if he uses up the login budget or triggers
account locked-down on all websites.

\subsubsection{Results}

The model-checking results, and in particular the impact of various
parameters on those results, are summarized in \figref{fig:prism}.
This figure plots the attacker's success probability, under an optimal
strategy, as a function of the password entropy.  The
leftmost data point in each graph pertains to a dictionary size equal
to $\setSize{\allSimilar} = \nmbrRespondersQueried
(\nmbrHoneyPasswords+1) \clusterSize$, which is the minimum dictionary
size consistent with our model.
This minimum dictionary size---representing a large amount of attacker
knowledge about the dictionary from which the user chooses her
password---is the reason why the attacker succeeds with such high
probability.
Each graph shows four curves,
corresponding to PMT budgets of 0, 3, 6, and 9.  The PMT budget of 0
provides a baseline curve that shows the security of each
configuration in the absence of our design (though in the optimistic
case where user \accountId nevertheless chose different passwords at
each website).

\begin{figure}[t]

   \vspace*{1em}
   \begin{subfigure}[t]{.1\columnwidth}
  \setlength\figureheight{2in}
  	\begin{minipage}[t]{1\columnwidth}
  	\centering
  	\vspace*{4.4em}
    \resizebox{!}{8.6em}{\hspace*{-1.5em}\begin{tikzpicture}
\node at (0,0)[
  scale=1,
  anchor=north,
  text=black,
  rotate=90,
  style={align=center},
]{Maximum probability\\of attacker success};
\end{tikzpicture}}
    \end{minipage}
  \end{subfigure}

  \vspace*{-13em}
 \begin{subfigure}[t]{.1\columnwidth}
  \setlength\figureheight{2in}
  	\begin{minipage}[t]{1\columnwidth}
  	\centering
  	\vspace*{-1.4em}
    \resizebox{!}{2.4em}{\newenvironment{customlegend}[1][]{%
    \begingroup
    \csname pgfplots@init@cleared@structures\endcsname
    \pgfplotsset{#1}%
}{%
    \csname pgfplots@createlegend\endcsname
    \endgroup
}%

\def\addlegendimage{\csname pgfplots@addlegendimage\endcsname}

\hspace*{5.8em}\begin{tikzpicture}

\definecolor{color0}{rgb}{0.129411764705882,0.380392156862745,0.549019607843137}

\begin{customlegend}[
	legend style={{font={\small}},{draw=none}},
	legend columns=2,
	legend cell align={left},
	legend entries={{$\ $PMT budget = $0\quad$},{$\ $PMT budget = $3\quad$},{$\ $PMT budget = $6\quad$},{$\ $PMT budget = $9\quad$}}]
\addlegendimage{line width = 1.2pt, color0}
\addlegendimage{line width = 1.2pt, color0, dashed}
\addlegendimage{line width = 1.2pt, color0, dotted}
\addlegendimage{line width = 1.2pt, color0, dash pattern=on 1pt off 3pt on 3pt off 3pt}

\end{customlegend}

\end{tikzpicture}}
    \end{minipage}
  \end{subfigure}

  \begin{subfigure}[b]{.40\columnwidth}
  \setlength\figureheight{2in}
    \begin{minipage}[b]{1\textwidth}
  	\centering
  	\vspace*{0em}\resizebox{!}{7.5em}{

\hspace*{0.9em}\begin{tikzpicture}
\definecolor{color0}{rgb}{0.129411764705882,0.380392156862745,0.549019607843137}

\pgfplotsset{every axis/.append style={
					compat=1.3,
                    label style={font=\fontsize{1}{3.5}\selectfont},
                    tick label style={font=\fontsize{1}{3.5}\selectfont}  
                    }}

\begin{axis}[
xmin=3, xmax=17,
ymin=0, ymax=0.5,
width=\figurewidth,
height=0.75\figurewidth,
tick align=outside,
tick pos=left,
xtick={3,5,7,9,11,13,15,17},
ytick={0,0.1,0.2,0.3,0.4,0.5},
xticklabels={},
xmajorgrids,
x grid style={lightgray!92.02614379084967!black},
ymajorgrids,
y grid style={lightgray!92.02614379084967!black}
]
\addplot [line width=0.5pt, color0]
table {%
3.90689059560851 0.390057883
4 0.375507878
5 0.228583121
6 0.126460103
7 0.066626728
8 0.034215364
9 0.017340383
10 0.008729306
11 0.004379552
12 0.002193516
13 0.001097695
14 0.000549082
15 0.0002746
16 0.000137314
17 6.87e-05
18 3.43e-05
};
\addplot [line width=0.5pt, color0, dashed]
table {%
3.90689059560851 0.390057883
4 0.387061316
5 0.300872728
6 0.190094561
7 0.106773201
8 0.056558351
9 0.029103052
10 0.014761422
11 0.007433679
12 0.003730144
13 0.001868406
14 0.000935037
15 0.000467727
16 0.000233916
17 0.000116971
18 5.85e-05
};
\addplot [line width=0.5pt, color0, dotted]
table {%
3.90689059560851 0.390057883
4 0.390057883
5 0.352039935
6 0.244596701
7 0.144090988
8 0.078122387
9 0.040661779
10 0.020741386
11 0.010474623
12 0.005263458
13 0.002638286
14 0.001320785
15 0.000660803
16 0.000330504
17 0.000165278
18 8.26e-05
};
\addplot [line width=0.5pt, color0, dash pattern=on 1pt off 3pt on 3pt off 3pt]
table {%
3.90689059560851 0.390057883
4 0.390057883
5 0.378605867
6 0.289394495
7 0.178555243
8 0.098910738
9 0.052017513
10 0.026669356
11 0.013502403
12 0.00679346
13 0.003407336
14 0.001706324
15 0.000853827
16 0.00042708
17 0.000213581
18 0.000106801
};
\end{axis}
\end{tikzpicture}

\hspace*{-0.9em}\begin{tikzpicture}

\definecolor{color0}{rgb}{0.129411764705882,0.380392156862745,0.549019607843137}

\pgfplotsset{every axis/.append style={
					compat=1.3,
                    label style={font=\tiny},
                    tick label style={font=\tiny}  
                    }}
                    
\begin{axis}[
xmin=3, xmax=17,
ymin=0, ymax=0.5,
width=\figurewidth,
height=0.75\figurewidth,
tick align=outside,
tick pos=left,
xtick={3,5,7,9,11,13,15,17},
ytick={0,0.1,0.2,0.3,0.4,0.5},
xticklabels={},
yticklabels={},
xmajorgrids,
x grid style={lightgray!92.02614379084967!black},
ymajorgrids,
y grid style={lightgray!92.02614379084967!black}
]
\addplot [line width=0.5pt, color0]
table {%
3.90689059560851 0.487679608
4 0.487060714
5 0.425171456
6 0.293530743
7 0.174830227
8 0.095844678
9 0.050246564
10 0.025734519
11 0.013024055
12 0.006551739
13 0.003285862
14 0.001645437
15 0.000823346
16 0.00041183
17 0.000205954
18 0.000102987
};
\addplot [line width=0.5pt, color0, dashed]
table {%
3.90689059560851 0.487679608
4 0.4872523
5 0.447632479
6 0.334017445
7 0.207670735
8 0.116163751
9 0.0614782
10 0.031630723
11 0.016043809
12 0.008079724
13 0.004054406
14 0.00203085
15 0.001016338
16 0.000508398
17 0.000254256
18 0.000127142
};
\addplot [line width=0.5pt, color0, dotted]
table {%
3.90689059560851 0.487679608
4 0.48744863
5 0.463444872
6 0.367755856
7 0.237949093
8 0.135733257
9 0.07250929
10 0.037475182
11 0.019050428
12 0.009604401
13 0.00482212
14 0.002416055
15 0.001209279
16 0.000604952
17 0.000302554
18 0.000151297
};
\addplot [line width=0.5pt, color0, dash pattern=on 1pt off 3pt on 3pt off 3pt]
table {%
3.90689059560851 0.487679608
4 0.487647199
5 0.474049049
6 0.395502721
7 0.265773514
8 0.154566324
9 0.083341444
10 0.043268097
11 0.022043938
12 0.011125773
13 0.005589004
14 0.002801053
15 0.001402167
16 0.000701494
17 0.000350849
18 0.00017545
};
\end{axis}
\end{tikzpicture}}
    \hspace*{5.1em}\begin{minipage}[t]{8em}
	\vspace*{-2.5em}\caption{$\nmbrRespondersQueried\!=\!3, \loginBudget\!=\!3, \clusterSize\!=\!1$ \label{fig:prism:baseline}}
	\end{minipage}%
	\hspace*{2.4em}\begin{minipage}[t]{8em}
	\vspace*{-2.5em}\caption{$\nmbrRespondersQueried\!=\!3, \loginBudget\!=\!9, \clusterSize\!=\!1$ \label{fig:prism:loginBudget}}
	\end{minipage}%
    \end{minipage}
  \end{subfigure}%
  
  \begin{subfigure}[b]{.40\columnwidth}
  \setlength\figureheight{2in}
    \begin{minipage}[b]{1\textwidth}
  	\centering
  	\vspace*{-0.8em}\resizebox{!}{8.0em}{
\hspace*{0.9em}\begin{tikzpicture}

\definecolor{color0}{rgb}{0.129411764705882,0.380392156862745,0.549019607843137}

\pgfplotsset{every axis/.append style={
					compat=1.3,
                    label style={font=\tiny},
                    tick label style={font=\tiny}  
                    }}

\begin{axis}[
xmin=3, xmax=17,
ymin=0, ymax=0.5,
width=\figurewidth,
height=0.75\figurewidth,
tick align=outside,
tick pos=left,
xtick={3,5,7,9,11,13,15,17},
ytick={0,0.1,0.2,0.3,0.4,0.5},
xmajorgrids,
x grid style={lightgray!92.02614379084967!black},
ymajorgrids,
y grid style={lightgray!92.02614379084967!black}
]
\addplot [line width=0.5pt, color0]
table {%
5.49185309632967 0.438329724
6 0.335728669
7 0.187221809
8 0.099225285
9 0.051131101
10 0.025960858
11 0.013081309
12 0.006566137
13 0.003289472
14 0.001646341
15 0.000823572
16 0.000411887
17 0.000205969
18 0.000102991
};
\addplot [line width=0.5pt, color0, dashed]
table {%
5.49185309632967 0.438329764
6 0.3897839
7 0.259137568
8 0.15006543
9 0.080850616
10 0.041977575
11 0.021389845
12 0.010796861
13 0.005424128
14 0.002718515
15 0.001360874
16 0.000680841
17 0.000340522
18 0.000170286
};
\addplot [line width=0.5pt, color0, dotted]
table {%
5.49185309632967 0.438329764
6 0.425057632
7 0.320719043
8 0.197283014
9 0.109479907
10 0.057693589
11 0.029619308
12 0.015007304
13 0.007553647
14 0.003789396
15 0.001897851
16 0.000949715
17 0.000609567
18 0.000237577
};
\addplot [line width=0.5pt, color0, dash pattern=on 1pt off 3pt on 3pt off 3pt]
table {%
5.49185309632967 0.438329764
6 0.436142625
7 0.367497589
8 0.240337134
9 0.136988752
10 0.073109374
11 0.037770099
12 0.01919754
13 0.009678041
14 0.004858986
15 0.002434504
16 nan
17 nan
18 0.000304862
};
\end{axis}
\end{tikzpicture}

\hspace*{-1.2em}\begin{tikzpicture}

\definecolor{color0}{rgb}{0.129411764705882,0.380392156862745,0.549019607843137}

\pgfplotsset{every axis/.append style={
					compat=1.3,
                    label style={font=\tiny},
                    tick label style={font=\tiny}  
                    }}

\begin{axis}[
xmin=3, xmax=17,
ymin=0, ymax=0.5,
width=\figurewidth,
height=0.75\figurewidth,
tick align=outside,
tick pos=left,
xtick={{3,5,7,9,11,13,15,17}},
ytick={0,0.1,0.2,0.3,0.4,0.5},
yticklabels={},
xmajorgrids,
x grid style={lightgray!92.02614379084967!black},
ymajorgrids,
y grid style={lightgray!92.02614379084967!black}
]
\addplot [line width=0.5pt, color0]
table {%
5.906890596 0.145395888
6 0.136410428
7 0.068955421
8 0.034780374
9 0.01747963
10 0.008763875
11 0.00439
12 0.0022
13 0.0011
};
\addplot [line width=0.5pt, color0, dashed]
table {%
5.906890596 0.145395888
6 0.145395888
7 0.126976564
8 0.084473377
9 0.048470631
10 0.025919465
11 0.013396578
12 0.006809482
13 0.003432787
};
\addplot [line width=0.5pt, color0, dotted]
table {%
5.906890596 0.145395888
6 0.145395888
7 0.142175471
8 0.11347881
9 0.072713413
10 0.041170715
11 0.0219005
12 0.011293547
13 0.005734446
};
\addplot [line width=0.5pt, color0, dash pattern=on 1pt off 3pt on 3pt off 3pt]
table {%
5.906890596 0.145395888
6 0.145395888
7 0.145021508
8 0.129545723
9 0.091472102
10 0.054695118
11 0.029923356
12 0.015650868
13 0.00800359
};
\end{axis}
\end{tikzpicture}}
    \hspace*{5.1em}\begin{minipage}[t]{8em}
	\vspace*{-1.8em}\caption{$\nmbrRespondersQueried\!=\!9, \loginBudget\!=\!3, \clusterSize\!=\!1$ \label{fig:prism:nmbrRespondersQueried}}
	\end{minipage}%
	\hspace*{2.4em}\begin{minipage}[t]{8em}
	  \vspace*{-1.8em}\caption{$\nmbrRespondersQueried\!=\!3, \loginBudget\!=\!3, \clusterSize\!=\!4$ \label{fig:prism:clusterSize}}
	\end{minipage}%
    \end{minipage}
  \end{subfigure}%
  
  \begin{subfigure}[b]{.43\columnwidth}
  \setlength\figureheight{2in}
    \begin{minipage}[b]{1\textwidth}
  	\centering
  	\vspace*{-0.0em}
    \resizebox{!}{1.40em}{\hspace*{10.6em}\begin{tikzpicture}
\node at (0,0)[
  scale=1,
  anchor=north,
  text=black,
  rotate=0
]{Password entropy (bits)};
\end{tikzpicture}}\vspace*{-0.4em}
    \end{minipage}
  \end{subfigure}
  \vspace*{0.6em}
  \caption{Maximum probability with which attacker logs into account
    at some \responderTerm, as a function of password
    entropy.  Subfigures show different settings for the number
    \nmbrRespondersQueried of \respondersTerm queried, the login
    budget \loginBudget, and the cluster size \clusterSize.  The
    number of honey-password clusters is $\nmbrHoneyPasswords = 4$.
    All subfigures have the same axes.}
  \label{fig:prism}
\end{figure}

\begin{figure}[t]
	\centering
	\vspace*{-0.75em}
	\setlength\figureheight{2.0in}
\hspace*{-0.4em}\begin{tikzpicture}

\definecolor{color0}{rgb}{0.129411764705882,0.380392156862745,0.549019607843137}

\pgfplotsset{every axis/.append style={
					style={align=center},
					xlabel={Password entropy (bits)},
					ylabel={Maximum probability\\of attacker success},
					compat=1.3,
                    label style={font=\small},
                    tick label style={font=\small},
                    yticklabel style={ 
							/pgf/number format/fixed, 
							/pgf/number format/precision=3 
									} 
                    }}
\begin{axis}[
xmin=7, xmax=20,
ymin=0, ymax=0.5,
xtick={7,8,9,10,11,12,13,14,15,16,17,18,19,20},
ytick={0,0.1,0.2,0.3,0.4,0.5},
width=0.95\figurewidth,
height=0.95\figureheight,
tick align=outside,
tick pos=left,
scaled ticks=false,
xmajorgrids,
x grid style={lightgray!92.02614379084967!black},
ymajorgrids,
y grid style={lightgray!92.02614379084967!black},
legend style={at={(0.99,0.99)}, anchor=north east, draw=white!80.0!black},
legend cell align={left},
legend style={{font={\fontsize{8pt}{12}\selectfont}},{draw=none}},
legend entries={{PMT budget = 9},{PMT budget = 6},{PMT budget = 3},{PMT budget = 0},}
]
\addlegendimage{line width = 1.25pt, color0, dash pattern=on 1pt off 3pt on 3pt off 3pt}
\addlegendimage{line width = 1.25pt, color0, dotted}
\addlegendimage{line width = 1.25pt, color0, dashed}
\addlegendimage{line width = 1.25pt, color0}
]
\addplot [line width=1.25pt, color0]
table {%
7.906890596 0.359577313
8 0.341405004
9 0.188728336
10 0.099611777
11 0.051228826
12 0.025985416
13 0.013087463
14 0.006567677
15 0.003289858
16 0.001646438
17 0.000823597
18 0.000411893
19 0.00020597
20 0.000102991
};
\addplot [line width=1.25pt, color0, dashed]
table {%
7.906890596 0.359577313
8 0.355690148
9 0.265715776
10 0.161839727
11 0.089212857
12 0.046825487
13 0.023986844
14 0.012139439
15 0.00610654
16 0.003062515
17 0.001533574
18 0.000767367
19 0.000383828
20 0.00019195
};
\addplot [line width=1.25pt, color0, dotted]
table {%
7.906890596 0.359577313
8 0.359319485
9 0.315589791
10 0.214793616
11 0.12477899
12 0.067070906
13 0.034740521
14 0.017675265
15 0.008914308
16 0.004476373
17 0.002242997
18 0.001122702
19 0.000561652
20 0.000280901
};
\addplot [line width=1.25pt, color0, dash pattern=on 1pt off 3pt on 3pt off 3pt]
table {%
7.906890596 0.359577313
8 0.359573046
9 0.341259993
10 0.256577613
11 0.157312707
12 0.086608971
13 0.045331677
14 0.023172865
15 0.011712862
16 0.005887973
17 nan
18 0.001477899
19 0.000739441
20 0.000369844
};
\end{axis}
\end{tikzpicture}
	\caption{Maximum probability with which attacker logs
          into account at some \responderTerm, as a function of
          password entropy, where $\nmbrRespondersQueried =
          12$, $\loginBudget = 9$, $\clusterSize = 4$, and
          $\nmbrHoneyPasswords = 4$.}
	\label{fig:prism-large}
        \vspace*{-0.5ex}
\end{figure}

\figref{fig:prism:baseline} shows a baseline with small parameters;
the other subgraphs show the effect of increasing one parameter at a
time.  \figref{fig:prism:loginBudget} and
\figref{fig:prism:nmbrRespondersQueried} show the impacts of
increasing the per-website login budget \loginBudget and the number
\nmbrRespondersQueried of \respondersTerm queried in each PMT,
respectively, which both increase the attacker's probability of
success somewhat.  \figref{fig:prism:clusterSize} shows that
increasing the size \clusterSize of each password (including
honey-password) cluster suppresses the success probability.

These graphs show that while growing the PMT budget increases the
attacker's probability of success, the amount by which it does so is
modest and diminishes as the password entropy grows.
\figref{fig:prism-large} shows somewhat more realistic parameters
(though we were limited in growing these calculations to truly
realistic sizes by the computational expense of doing so).  As shown
there, any attacker advantage gained by up to 9 PMT queries all but
disappears with a dictionary of size only $2^{14}$.

\section{Evaluation and Parameter Optimization}
\label{sec:eval}

\subsection{Implementation}
\label{sec:eval:implementation}

We built a prototype of our framework to evaluate its performance and
scalability, and to inform its parameterization (see
\secref{sec:eval:optimization}).  We realized the cryptographic parts
of our protocol in C and other parts using Go.

\subsubsection{Cryptography}
\label{sec:eval:implementation:crypto}

We used the ElGamal cryptosystem in an elliptic-curve
group (EC-ElGamal) as the multiplicatively homomorphic
scheme \encScheme in \figref{fig:protocol}.  We realized all
cryptographic operations using MIRACL
(\url{https://github.com/miracl/MIRACL}).  Our implementation includes
four standardized elliptic curves: secp160r1, secp192r1 (NIST P-192),
secp224r1 (NIST P-224) and secp256r1 (NIST P-256)
\cite{certicom2000:sec2v1,nist2013:fips186-4}.  Elliptic-curve
cryptosystems based on these curves can provide security roughly
equivalent to RSA with key lengths of 1024, 1536, 2048 and 3072 bits,
respectively.  The generator \groupGenerator used with each curve has
a cofactor of $1$~\cite{certicom2000:sec2v1}, so that the group
includes all curve points. This allows the \requesterTerm and
\respondersTerm to check the validity of ciphertexts (i.e.,
lines~\ref{prot:responder:checkCiphertexts} and
\ref{prot:requester:checkCiphertext} in \figref{fig:protocol}) by
checking if each ciphertext component is a valid point on the elliptic
curve (or the point at infinity).

To make messages shorter and save bandwidth, we enable \textit{point
  compression} in our implementation. Point compression
(e.g.,~\cite[\secrefstatic{A.9.6}]{ieee2000:1363}) is a technique that
compresses each elliptic-curve point to half its original size by
using only $y \bmod 2$ in place of its $y$ coordinate value.
Correspondingly, \textit{point decompression} reconstructs the point
by recovering the $y$ coordinate based on the $x$ coordinate and $y
\bmod 2$.

\subsubsection{Bloom filters}
\label{sec:eval:implementation:bloomFilters}
A Bloom filter has a false positive rate of $\approx (1 -
e^{-\bloomFilterHashFns\cdot\frac{\nmbrSimilarPasswords}{\bloomFilterSize}})^{\bloomFilterHashFns}$
where $\nmbrSimilarPasswords =
\setSize{\similarPasswords{\accountId}{}}$ denotes the number elements
to be inserted into the Bloom filter by the \responderTerm,
\bloomFilterSize denotes the length of the Bloom filter and
\bloomFilterHashFns denotes the number of hash functions (e.g.,
see~\cite[pp.\ 109--110]{mitzenmacher2005:probability}).  As such, the
number of hash functions that minimizes false positives is
$\bloomFilterHashFnsOpt =
\frac{\bloomFilterSize}{\nmbrSimilarPasswords}\cdot\ln2$ and in this
case, the minimized false positive rate is
$2^{-\bloomFilterHashFnsOpt} =
2^{-\frac{\bloomFilterSize}{\nmbrSimilarPasswords}\cdot\ln2} \approx
(0.6185)^\frac{\bloomFilterSize}{\nmbrSimilarPasswords}$.
In our framework, \bloomFilterHashFns and \bloomFilterSize are decided
by the \requesterTerm, while \nmbrSimilarPasswords is determined by
each \responderTerm with the knowledge of \bloomFilterHashFns and
\bloomFilterSize received from the \requesterTerm.  In our
implementation, the \requesterTerm chooses $\bloomFilterHashFns = 20$
by default, and so each \responderTerm then generates a set
\similarPasswords{\accountId}{} of size $\nmbrSimilarPasswords \le
\frac{\bloomFilterSize}{\bloomFilterHashFns}\cdot\ln2$ to ensure a
false positive rate of $\approx 2^{-20}$.

\subsubsection{Precomputation}
\label{sec:eval:implementation:precomputation}
We use precomputation to optimize the creation of ciphertexts
\bloomFilterBitCtext{\bloomFilterBitIdx} by the \requesterTerm in our
protocol.  Specifically, the \requesterTerm precomputes private key
\elgPrivKey, public key \elgPubKey, and values
$\{\elgEphemeralPubKey{\bloomFilterBitIdx}\}_{\bloomFilterBitIdx \in
  \residues{\bloomFilterSize}}$ and
$\{\elgCiphertext{\bloomFilterBitIdx}\}_{\bloomFilterBitIdx \in
  \residues{\bloomFilterSize}}$, where each $\langle\elgPubKey,
\elgEphemeralPubKey{\bloomFilterBitIdx},
\elgCiphertext{\bloomFilterBitIdx}\rangle$ is a valid Diffie-Hellman
triple, i.e., $\langle \elgEphemeralPubKey{\bloomFilterBitIdx},
\elgCiphertext{\bloomFilterBitIdx}\rangle
\in \ciphertextSpace{\langle\groupGenerator,
  \elgPubKey\rangle}(\groupIdentity)$.  To create a ciphertext
\bloomFilterBitCtext{\bloomFilterBitIdx} of a different group element
$\groupElmt \neq \groupIdentity$, the \requesterTerm need only
multiply \elgCiphertext{\bloomFilterBitIdx} by \groupElmt; thus,
\lineref{prot:requester:encryptBF} is completed in at most one
multiplication per $\bloomFilterBitIdx \in
\residues{\bloomFilterSize}$.  In practice, this precomputation could
begin once the user enters the account registration web page and
continue during idle periods until a password is successfully set.

\subsection{Response Time}
\label{sec:eval:performance}

In this section, we evaluate the response time of our prototype system
as seen by the \requesterTerm (and in the absence of any user
interaction, such as that described in
\secref{sec:framework:design:responder}), with two goals in mind.
First, we want to systematically measure the effects of various
parameter settings on our prototype implementation, to inform the
selection of these parameters through an optimization process
discussed in \secref{sec:eval:optimization}.  We mainly explore two
different parameters of our framework: The maximum number of similar
passwords $\nmbrSimilarPasswords =
\setSize{\similarPasswords{\accountId}{}}$ per \responderTerm (as
determined by setting the Bloom filter size $\bloomFilterSize =
\lceil\frac{20\nmbrSimilarPasswords}{\ln 2}\rceil$ in the protocol),
and the number \nmbrRespondersQueried of \respondersTerm.
\iffull
In \appref{sec:microbenchmarks}, we also explore the impact of EC-ElGamal
key length on the protocol response time and bandwidth.
\fi
The second
main goal of our experiments here is to compare the performance of our
prototype with and without leveraging Tor for implementing
\accountLocationPrivacy, i.e., the
\untrustedForAccountLocationPrivacy-\directoryTerm and
\trustedForAccountLocationPrivacy-\directoryTerm models, respectively.
In doing so, we hope to shed light on the performance costs of
adopting a more pessimistic trust model in which the \directoryTerm is
not trusted to hide the websites where each account identifier
\accountId has been used to register an account.

\subsubsection{Experimental setup}
\label{sec:eval:performance:setup}

In our evaluations, we set up one \requesterTerm, one \directoryTerm,
and up to 128 \respondersTerm, spread across six machines located in
our department.  The \requesterTerm and the \directoryTerm ran on
separate machines with the same specification: 2.67\gigahertz $\times$
8 physical cores, 72\gibibytes RAM, Ubuntu 14.04 x86\_64.  The (up to)
128 \respondersTerm were split evenly across four other, identical
machines: 2.3\gigahertz $\times$ 32 physical cores with
hyper-threading enabled (and so 64 logical cores), 128\gibibytes RAM,
Ubuntu 16.04 x86\_64.  Each of the \respondersTerm sharing one machine
was limited to two logical cores, and had its own exclusive data
files, processes, and network sockets.
The six machines were on the same $1\gigabits/\secs$ network.  Thus,
while our results might be pessimistic due to resource sharing among
\respondersTerm, they might also be somewhat optimistic in our
\trustedForAccountLocationPrivacy-\directoryTerm experiments due to
leveraging LAN communication.  (The
\untrustedForAccountLocationPrivacy-\directoryTerm case is discussed
further below.)

Parameters were set to the following defaults unless otherwise
specified: $\nmbrRespondersQueried =
64$, and elliptic-curve key length of $192$ bits.  In particular,
$\nmbrRespondersQueried = 64$ is conservative based on recent studies.
For example, a 2017 study with 154 participants found that users have
a mean of $26.3$ password-protected web
accounts~\cite{pearman2017:habitat}, which is quite consistent with
other studies (e.g.,~\cite{florencio2007:reuse, stobert2014:reuse}).

Because the public Tor network is badly under-provisioned for its
level of use and so its performance varies significantly over time, in
our tests for the \untrustedForAccountLocationPrivacy-\directoryTerm
model, we utilized a private Tor network with nodes distributed across
North America and Europe.  Our private Tor network
consisted of three Tor authorities, eight normal
onion routers, and two special onion routers.  The eight normal onion
routers were running on eight different Amazon EC2 (m4.large)
instances, one located in each of the eight Amazon AWS regions in
North America and Europe.  Among these onion routers, three were also
running as Tor authorities, with one in Europe, one in U.S.\ West, and
the other in U.S.\ East.  Two special onion routers were running on
the machine in our department hosting the \directoryTerm; one
(``Exit'' in \figref{fig:tor}) exclusively served as the exit node of
Tor circuits from \requestersTerm, and the other (``RP'' in
\figref{fig:tor}) served exclusively as the ``rendezvous point''
picked by the \directoryTerm to communicate with Tor hidden services,
i.e., the \respondersTerm.  As shown in \figref{fig:tor}, each circuit
included two more onion routers (``OR'' in
\figref{fig:tor}) chosen at random from among the eight normal onion
routers already described.

\begin{figure}[h]
	\centering 
%
\psscalebox{0.5 0.5} 
{
\begin{pspicture}(0,-1.0)(15.24,1.0)
\psframe[linecolor=black, linewidth=0.04, dimen=outer](2.4,1.0)(0.0,-0.6)
\psframe[linecolor=black, linewidth=0.04, dimen=outer](10.0,1.0)(5.2,-1.0)
\psframe[linecolor=black, linewidth=0.04, dimen=outer](15.2,1.0)(12.8,-0.6)
\rput[bl](0.3,0.0){\large{\textbf{Requester}}}
\rput[bl](6.7,0.0){\large{\textbf{Directory}}}
\rput[bl](7.0,-0.4){\large{\textbf{server}}}
\rput[bl](13.06,0.0){\large{\textbf{Responder}}}
\psline[linecolor=black, linewidth=0.04](2.4,0.2)(2.8,0.2)(2.8,0.2)
\psframe[linecolor=black, linewidth=0.04, dimen=outer](3.6,0.6)(2.8,-0.2)
\psframe[linecolor=black, linewidth=0.04, dimen=outer](4.8,0.6)(4.0,-0.2)
\psframe[linecolor=black, linewidth=0.04, dimen=outer](11.2,0.6)(10.4,-0.2)
\psframe[linecolor=black, linewidth=0.04, dimen=outer](12.4,0.6)(11.6,-0.2)
\psframe[linecolor=black, linewidth=0.04, linestyle=dashed, dash=0.17638889cm 0.10583334cm, dimen=outer](6.4,0.6)(5.6,-0.2)
\psframe[linecolor=black, linewidth=0.04, linestyle=dashed, dash=0.17638889cm 0.10583334cm, dimen=outer](9.6,0.6)(8.8,-0.2)
\rput[bl](2.9,-0.7){\large{\textbf{OR}}}
\rput[bl](4.1,-0.7){\large{\textbf{OR}}}
\rput[bl](10.5,-0.7){\large{\textbf{OR}}}
\rput[bl](11.7,-0.7){\large{\textbf{OR}}}
\rput[bl](5.6,-0.7){\large{\textbf{Exit}}}
\rput[bl](8.9,-0.7){\large{\textbf{RP}}}
\psline[linecolor=black, linewidth=0.04](3.6,0.2)(4.0,0.2)(4.0,0.2)
\psline[linecolor=black, linewidth=0.04](4.8,0.2)(5.6,0.2)(5.6,0.2)
\psline[linecolor=black, linewidth=0.04](9.6,0.2)(10.4,0.2)(10.4,0.2)
\psline[linecolor=black, linewidth=0.04](11.2,0.2)(11.6,0.2)(11.6,0.2)
\psline[linecolor=black, linewidth=0.04](12.4,0.2)(12.533334,0.2)(12.8,0.2)
\end{pspicture}
}
	\caption{Topology of our
          \untrustedForAccountLocationPrivacy-\directoryTerm
          experimental setup}
	\label{fig:tor}
\end{figure}

All datapoints reported in the graphs below are averaged from 50
executions.  Relative standard deviations, denoted \relstddev, are
reported in figure captions.

\subsubsection{Results}
\label{sec:eval:performance:responseTime}

A measure of primary concern for our framework is the response time
witnessed by the \requesterTerm, since this delay will be imposed on
the user experience while setting her password.
\figref{fig:responseTime:woTor} shows the response time in the
\trustedForAccountLocationPrivacy-\directoryTerm model, where the
\requesterTerm connects directly to the \directoryTerm and the
\directoryTerm connects directly with each \responderTerm.  In
contrast, \figref{fig:responseTime:Tor} shows the response time in the
\untrustedForAccountLocationPrivacy-\directoryTerm model, and so
connections are performed through Tor.  Precomputation costs (see
\secref{sec:eval:implementation:precomputation}) are not included in
\figref{fig:responseTime}, as these costs are expected to be borne off
the critical path of interacting with the user.  Tor circuit setup
times are amortized over the 50 runs contributing to each datapoint in
\figref{fig:responseTime:Tor}.  In practice, we expect this setup cost
to be similarly amortized over attempts needed by the user to choose
an acceptable (not reused) password, or relegated to a precomputation
stage when the user first accesses the \requesterTerm's account
creation/password reset page.

\begin{figure}[t]

\vspace*{-2.5em}
   \begin{subfigure}[t]{.1\columnwidth}
  \setlength\figureheight{2in}
  	\begin{minipage}[t]{1\columnwidth}
  	\centering
  	\vspace*{4.4em}
    \resizebox{!}{10.5em}{\hspace*{-1em}\begin{tikzpicture}
\node at (0,0)[
  scale=1,
  anchor=north,
  text=black,
  rotate=90,
  style={align=center},
]{Protocol running time (s)};
\end{tikzpicture}}
    \end{minipage}
  \end{subfigure}

  \vspace*{-12em}
 \begin{subfigure}[t]{.1\columnwidth}
  \setlength\figureheight{2in}
  	\begin{minipage}[t]{1\columnwidth}
  	\centering
  	\vspace*{-0.4em}
    \resizebox{!}{2.32em}{\newenvironment{customlegend}[1][]{%
    \begingroup
    \csname pgfplots@init@cleared@structures\endcsname
    \pgfplotsset{#1}%
}{%
    \csname pgfplots@createlegend\endcsname
    \endgroup
}%

\def\addlegendimage{\csname pgfplots@addlegendimage\endcsname}

\hspace*{6em}\begin{tikzpicture}

\definecolor{color0}{rgb}{0.129411764705882,0.380392156862745,0.549019607843137}

\begin{customlegend}[
	legend style={{font={\fontsize{10pt}{12}\selectfont}},{draw=none}},
	legend cell align={left},
	legend columns=3,
	legend entries={{\nmbrRespondersQueried = $1\quad$},{\nmbrRespondersQueried = $32\quad$},{\nmbrRespondersQueried = $64\quad$},{\nmbrRespondersQueried = $96\quad$},{\nmbrRespondersQueried = $128$}}]
\addlegendimage{line width=1pt, color0}
\addlegendimage{line width=1pt, color0, dotted}
\addlegendimage{line width=1pt, color0, dash pattern=on 1pt off 3pt on 3pt off 3pt}
\addlegendimage{line width=1pt, color0, dashed}
\addlegendimage{line width=2pt, color0, dashed}

\end{customlegend}

\end{tikzpicture}}
    \end{minipage}
  \end{subfigure}
  
  \begin{subfigure}[b]{.43\columnwidth}
  \setlength\figureheight{2in}
    \begin{minipage}[b]{1\textwidth}
  	\centering
  	\vspace*{0em}\resizebox{!}{12.5em}{
\hspace*{0.9em}\begin{tikzpicture}

\definecolor{color0}{rgb}{0.129411764705882,0.380392156862745,0.549019607843137}

\pgfplotsset{every axis/.append style={
					compat=1.3,
                    label style={font=\tiny},
                    tick label style={font=\tiny}  
                    }}

\begin{axis}[
xmin=128, xmax=4096,
ymin=0, ymax=60,
xmode=log,
log basis x={2},
xtick={128,256,512,1024,2048,4096},
ytick={0,10,20,30,40,50,60},
width=1.1\figurewidth,
height=1.1\figurewidth,
tick align=outside,
tick pos=left,
xmajorgrids,
minor tick num=1,
x grid style={lightgray!92.026143790849673!black},
ymajorgrids,
y grid style={lightgray!92.026143790849673!black},
]
\addplot [line width=0.5pt, color0]
table {%
128 0.30897422
256 0.60976572
512 1.21648104
1024 2.4220586
2048 4.7702893
4096 9.76527904
};
\addplot [line width=0.75pt, color0, dotted]
table {%
128 0.43634948
256 0.78842218
512 1.52416292
1024 2.97469348
2048 5.82659822
4096 11.88247176
};
\addplot [line width=0.75pt, color0, dash pattern=on 1pt off 3pt on 3pt off 3pt]
table {%
128 0.54709104
256 1.00667884
512 1.96562922
1024 3.82718008
2048 7.352575
4096 14.66240848
};
\addplot [line width=0.75pt, color0, dashed]
table {%
128 0.66607212
256 1.18544074
512 2.27556698
1024 4.43273222
2048 8.66786214
4096 17.45420922
};
\addplot [line width=1.25pt, color0, dashed]
table {%
128 0.82103076
256 1.44377364
512 2.70793158
1024 5.15631922
2048 9.89019532
4096 20.49675554
};
\end{axis}
\end{tikzpicture}

\hspace*{-1.25em}\begin{tikzpicture}

\definecolor{color0}{rgb}{0.129411764705882,0.380392156862745,0.549019607843137}

\pgfplotsset{every axis/.append style={
					compat=1.3,
                    label style={font=\tiny},
                    tick label style={font=\tiny}  
                    }}

\begin{axis}[
xmin=128, xmax=4096,
ymin=0, ymax=60,
xmode=log,
log basis x={2},
xtick={128,256,512,1024,2048,4096},
ytick={0,10,20,30,40,50,60},
yticklabels={},
width=1.1\figurewidth,
height=1.1\figurewidth,
tick align=outside,
tick pos=left,
minor tick num=1,
xmajorgrids,
x grid style={lightgray!92.026143790849673!black},
ymajorgrids,
y grid style={lightgray!92.026143790849673!black},
]
\addplot [line width=0.5pt, color0]
table {%
128 1.80276424
256 2.67040142
512 4.49651676
1024 7.47617528
2048 13.1238372
4096 25.08883928
};
\addplot [line width=0.75pt, color0, dotted]
table {%
128 2.7080117
256 3.70559346
512 5.53991662
1024 9.39000844
2048 17.0989824
4096 33.26568904
};
\addplot [line width=0.75pt, color0, dash pattern=on 1pt off 3pt on 3pt off 3pt]
table {%
128 3.3022177
256 4.66240986
512 6.7232246
1024 11.0877671
2048 19.73507796
4096 38.0223746
};
\addplot [line width=0.75pt, color0, dashed]
table {%
128 3.42254084
256 4.69043296
512 7.05244956
1024 12.43998506
2048 22.71355764
4096 44.51036384
};
\addplot [line width=1.25pt, color0, dashed]
table {%
128 3.58230948
256 4.56667798
512 7.52800982
1024 13.6724325
2048 25.92064192
4096 50.80265902
};
\end{axis}
\end{tikzpicture}}
    \hspace*{4.0em}\begin{minipage}[t]{8em}
	\vspace*{-0.5em}\caption{\trustedForAccountLocationPrivacy \directoryTerm \\ \centering($\relstddev < 6\%$) \label{fig:responseTime:woTor}}
	\end{minipage}%
	\hspace*{4.5em}\begin{minipage}[t]{8em}
	\vspace*{-0.5em}\caption{\untrustedForAccountLocationPrivacy \directoryTerm \\ \centering($\relstddev < 28\%$) \label{fig:responseTime:Tor}}
	\end{minipage}%
    \end{minipage}
  \end{subfigure}%
  
  \vspace*{-3.1em}
  \begin{subfigure}[b]{.43\columnwidth}
  \setlength\figureheight{2in}
    \begin{minipage}[b]{1\textwidth}
  	\centering
    \resizebox{!}{1.19em}{\hspace*{12.2em}\begin{tikzpicture}
\node at (0,0)[
  scale=1,
  anchor=south,
  text=black,
  rotate=0
]{\nmbrSimilarPasswords};
\end{tikzpicture}}\vspace*{-0.4em}
    \end{minipage}
  \end{subfigure}
  \vspace*{2.4em}
  \caption{Response time for various \nmbrSimilarPasswords and
    \nmbrRespondersQueried}
  \label{fig:responseTime}
\end{figure}

One observation from \figref{fig:responseTime} is that the
response-time cost of mistrusting the \directoryTerm and so of relying
on Tor to implement \accountLocationPrivacy, is typically $\ge
2\times$ for the parameters evaluated there.  Recall that in
\figref{fig:responseTime:Tor}, both the \requesterTerm--\directoryTerm
and \directoryTerm--\responderTerm communications were routed through
two onion routers chosen randomly from Amazon datacenter locations in
North America and Europe (see \figref{fig:tor}), in contrast to LAN
communication in \figref{fig:responseTime:woTor}.  The costs of these
long-haul hops and Tor-specific processing increased as
\nmbrSimilarPasswords grew, due to growth in query
message
\iffull
size (see \appref{sec:microbenchmarks}).
\else
size.
\fi

\figref{fig:responseTime} also shows the impact
of more \respondersTerm (larger \nmbrRespondersQueried)
on the response time witnessed by the \requesterTerm.  The main
underlying cause of this effect is the variance in the speeds with
which the \respondersTerm return responses to the \directoryTerm.
This variance is small when communication is direct, but it grows
substantially when Tor is used, due to the differences in
routes taken between the \directoryTerm and each \responderTerm.

These effects are also illustrated in \figref{fig:requesterCDF}, which
shows the response time observed by the \requesterTerm when the
\directoryTerm returned the proportion of $\nmbrRespondersQueried=64$
responses on the vertical axis as soon as that proportion was
available to it.  For example, \figref{fig:requesterCDF:Tor} shows
that when $\nmbrSimilarPasswords=2^{10}$, if the \directoryTerm waited
for $75\%$ of the responses ($48$ responses) before returning
them to the \requesterTerm, the \requesterTerm observed an average
response time of $9.55\secs$ (since $(9.55, 0.75)$ is a point on the
$\nmbrSimilarPasswords = 2^{10}$ curve).

\begin{figure}[t]

\vspace*{-2.5em}
   \begin{subfigure}[t]{.1\columnwidth}
  \setlength\figureheight{2in}
  	\begin{minipage}[t]{1\columnwidth}
  	\centering
  	\vspace*{4.4em}
    \resizebox{!}{10em}{\hspace*{-1em}\begin{tikzpicture}
\node at (0,0)[
  scale=1,
  anchor=north,
  text=black,
  rotate=90,
  style={align=center},
]{Proportion of responses};
\end{tikzpicture}}
    \end{minipage}
  \end{subfigure}

  \vspace*{-12.5em}
 \begin{subfigure}[t]{.1\columnwidth}
  \setlength\figureheight{2in}
  	\begin{minipage}[t]{1\columnwidth}
  	\centering
  	\vspace*{-0.4em}
    \resizebox{!}{2.7em}{\newenvironment{customlegend}[1][]{%
    \begingroup
    \csname pgfplots@init@cleared@structures\endcsname
    \pgfplotsset{#1}%
}{%
    \csname pgfplots@createlegend\endcsname
    \endgroup
}%

\def\addlegendimage{\csname pgfplots@addlegendimage\endcsname}

\hspace*{8em}\begin{tikzpicture}

\definecolor{color0}{rgb}{0.129411764705882,0.380392156862745,0.549019607843137}

\begin{customlegend}[
	legend style={{font={\fontsize{10pt}{12}\selectfont}},{draw=none}},
	legend columns=3,
	legend cell align={left},
	legend entries={{\nmbrSimilarPasswords = $2^7\quad$},{\nmbrSimilarPasswords = $2^8\quad$},{\nmbrSimilarPasswords = $2^9\quad$},{\nmbrSimilarPasswords = $2^{10}\quad$},{\nmbrSimilarPasswords = $2^{11}\quad$},{\nmbrSimilarPasswords = $2^{12}\quad$}}]
\addlegendimage{line width=1pt, densely dotted, color0}
\addlegendimage{line width=1pt, dashed, color0}
\addlegendimage{line width=1pt, dash pattern=on 1pt off 3pt on 3pt off 3pt, color0}
\addlegendimage{line width=2pt, densely dotted, color0}
\addlegendimage{line width=2pt, dashed, color0}
\addlegendimage{line width=2pt, dash pattern=on 1pt off 3pt on 3pt off 3pt, color0}
\addlegendimage{line width=2pt, dash pattern=on 1pt off 3pt on 3pt off 3pt, color0}
\end{customlegend}

\end{tikzpicture}}
    \end{minipage}
  \end{subfigure}

  \begin{subfigure}[b]{.43\columnwidth}
  \setlength\figureheight{2in}
    \begin{minipage}[b]{1\textwidth}
  	\centering
  	\vspace*{0em}\resizebox{!}{12.5em}{
\hspace*{0.8em}\begin{tikzpicture}

\definecolor{color0}{rgb}{0.129411764705882,0.380392156862745,0.549019607843137}

\pgfplotsset{every axis/.append style={
					compat=1.3,
                    label style={font=\tiny},
                    tick label style={font=\tiny}  
                    }}

\begin{axis}[
xmin=-0.341182282, xmax=20,
ymin=0, ymax=1,
width=0.8\figurewidth,
height=\figurewidth,
tick align=outside,
tick pos=left,
minor tick num=1,
xtick={0,10,20},
ytick={0,0.2,0.4,0.6,0.8,1},
yticklabels={0.0,0.2,0.4,0.6,0.8,1.0},
xmajorgrids,
x grid style={lightgray!92.026143790849673!black},
ymajorgrids,
y grid style={lightgray!92.026143790849673!black},
]
\addplot [line width=0.75pt, color0, densely dotted]
table {%
0.36015626 0.015625
0.42198774 0.25
0.44679104 0.5
0.4913151 0.75
0.54709104 1
};
\addplot [line width=0.75pt, color0, dashed]
table {%
0.71444446 0.015625
0.81054832 0.25
0.85180636 0.5
0.9052984 0.75
1.00667884 1
};
\addplot [line width=0.75pt, color0, dash pattern=on 1pt off 3pt on 3pt off 3pt]
table {%
1.44133564 0.015625
1.60472306 0.25
1.67925706 0.5
1.74537042 0.75
1.96562922 1
};
\addplot [line width=1.25pt, color0, densely dotted]
table {%
2.81623364 0.015625
3.15505028 0.25
3.27191674 0.5
3.38851834 0.75
3.82718008 1
};
\addplot [line width=1.25pt, color0, dashed]
table {%
5.6356579 0.015625
6.24064148 0.25
6.42928326 0.5
6.67110868 0.75
7.352575 1
};
\addplot [line width=1.25pt, color0, dash pattern=on 1pt off 3pt on 3pt off 3pt]
table {%
11.5156484 0.015625
12.68392394 0.25
13.07516718 0.5
13.41802364 0.75
14.66240848 1
};
\end{axis}

\end{tikzpicture}

\hspace*{-1.2em}\begin{tikzpicture}

\definecolor{color0}{rgb}{0.129411764705882,0.380392156862745,0.549019607843137}

\pgfplotsset{every axis/.append style={
					compat=1.3,
                    label style={font=\tiny},
                    tick label style={font=\tiny}  
                    }}

\begin{axis}[
xmin=-0.341182282, xmax=40,
ymin=0, ymax=1,
width=1.2\figurewidth,
height=\figurewidth,
tick align=outside,
tick pos=left,
minor tick num=1,
xtick={0,10,20,30,40},
ytick={0,0.2,0.4,0.6,0.8,1},
yticklabels={},
xmajorgrids,
x grid style={lightgray!92.026143790849673!black},
ymajorgrids,
y grid style={lightgray!92.026143790849673!black},
]
\addplot [line width=0.75pt, densely dotted, color0]
table {%
1.48565376 0.015625
1.95511426 0.25
2.201871 0.5
2.82597838 0.75
3.3022177 1
};
\addplot [line width=0.75pt, dashed, color0]
table {%
2.3766118 0.015625
2.862802 0.25
3.2699948 0.5
3.72270524 0.75
4.66240986 1
};
\addplot [line width=0.75pt, dash pattern=on 1pt off 3pt on 3pt off 3pt, color0]
table {%
3.53786544 0.015625
4.20764612 0.25
5.4170094 0.5
5.73985296 0.75
6.7232246 1
};
\addplot [line width=1.25pt, densely dotted, color0]
table {%
6.13149926 0.015625
7.47828284 0.25
9.12166248 0.5
9.54902008 0.75
11.0877671 1
};
\addplot [line width=1.25pt, dashed, color0]
table {%
11.41012414 0.015625
13.72834182 0.25
16.736895 0.5
17.73859294 0.75
19.73507796 1
};
\addplot [line width=1.25pt, dash pattern=on 1pt off 3pt on 3pt off 3pt, color0]
table {%
22.91918708 0.015625
26.64385434 0.25
31.54082418 0.5
34.23714754 0.75
38.0223746 1
};
\end{axis}

\end{tikzpicture}}
    \hspace*{4.0em}\begin{minipage}[t]{8em}
	\vspace*{0em}\caption{\trustedForAccountLocationPrivacy \directoryTerm \\ \centering($\relstddev < 4\%$) \label{fig:requesterCDF:woTor}}
	\end{minipage}%
	\hspace*{3.0em}\begin{minipage}[t]{8em}
	\vspace*{0em}\caption{\untrustedForAccountLocationPrivacy \directoryTerm \\ \centering($\relstddev < 33\%$) \label{fig:requesterCDF:Tor}}
	\end{minipage}%
    \end{minipage}
  \end{subfigure}%
  
  \vspace*{-4.0em}
  \begin{subfigure}[b]{.43\columnwidth}
  \setlength\figureheight{2in}
    \begin{minipage}[b]{1\textwidth}
  	\centering
    \resizebox{!}{1.5em}{\hspace*{10em}\begin{tikzpicture}
\node at (0,0)[
  scale=1,
  anchor=south,
  text=black,
  rotate=0
]{Response time (\secs)};
\end{tikzpicture}}\vspace*{2.5em}
    \end{minipage}
  \end{subfigure}
  \caption{Response time (horizontal axis) when the proportion of
    responses on the vertical axis is returned by \directoryTerm as
    soon as it is available ($\nmbrRespondersQueried = 64$)}
  \label{fig:requesterCDF}
\end{figure}

Recall that the \directoryTerm forwards the \textit{same}
\msgref{prot:msg:bloomFilter} to all \respondersTerm in our framework.
In the \untrustedForAccountLocationPrivacy-\directoryTerm model, using
an anonymous communication system that exploits this one-to-many
multicast pattern to gain efficiencies while still hiding the
multicast recipients (e.g.,~\cite{perng2006:m2}) could presumably
reduce the delays before the \directoryTerm receives responses, and
their variance.  We leave this extension to future work.

\subsection{Parameter Optimization}
\label{sec:eval:optimization}

At first glance, the results of \secref{sec:eval:performance} are
perhaps discouraging, since they suggest that the response time of
testing with a large number \nmbrSimilarPasswords of similar passwords
and at a large number \nmbrRespondersQueried of queried
\respondersTerm is potentially large, especially in the
\untrustedForAccountLocationPrivacy-\directoryTerm model
(\figref{fig:responseTime:Tor}).  In this section we describe an
approach to select optimal parameters for use in our framework,
specifically parameter values \nmbrRespondersQueried and
\nmbrSimilarPasswords that maximize the likelihood of detecting the
use of a similar password, subject to a response-time goal.  As we
will see, the results are not discouraging at all---a high true
detection rate can be achieved within reasonable response-time limits
with a surprisingly small \nmbrSimilarPasswords and while querying a
modest number \nmbrRespondersQueried of \respondersTerm from among the
total number of \respondersTerm \nmbrResponders{\accountId} registered
at the \directoryTerm for account \accountId.

The reason behind this initially surprising result is the typical
manner in which people create new passwords by applying simple,
predictable transforms to existing passwords.  Numerous studies
(e.g.,~\cite{zhang2010:expiration,wang2018:domino}) have found very
low variation in the transforms that users leverage to modify their
passwords (when they modify their passwords at all).  Provided that
\responderTerm \responder{\responderIdx} populates
$\cluster{\password{\responderIdx}} \subseteq
\similarPasswords{\accountId}{\responderIdx}$ (see
\secref{sec:framework:design:responder}) by applying these common
transforms to its account-\accountId password
\password{\responderIdx}, the probability that the user's chosen
password \passwordAlt at a \requesterTerm is contained within
\cluster{\password{\responderIdx}} at a randomly chosen \responderTerm
\responder{\responderIdx} is approximately as shown in
\figref{fig:pwdguesser}
(cf.,~\cite[\figrefstatic{7}]{wang2018:domino}), as a function of
$\clusterSize =
\setSize{\cluster{\password{\responderIdx}}}$.\footnote{\figref{fig:pwdguesser}
  is a log-normal CDF fitted to points selected from
  \cite[\figrefstatic{7}]{wang2018:domino} by manual inspection, as we
  could not obtain the source data for that figure.}  As we can see,
this probability is already substantial for very small \clusterSize.
For example, this probability is $\approx 0.34$ for even $\clusterSize
= 1$; in other words, users on average employ the same password at
$\approx 34\%$ of the websites where they have accounts.  Moreover,
this probability grows quickly as \clusterSize is increased only
slightly.

\begin{figure}
  \centering
  \setlength\figureheight{2in}
\begin{tikzpicture}

\pgfplotsset{every axis/.append style={
					xlabel={\clusterSize},
					ylabel={\prob{\passwordAlt \in \cluster{\password{\responderIdx}}}},
					compat=1.3,
                    label style={font=\small},
                    tick label style={font=\small}  
                    }}

\definecolor{color0}{rgb}{0.129411764705882,0.380392156862745,0.549019607843137}

\begin{axis}[
xmin=0, xmax=5000,
ymin=0.34, ymax=0.48,
ytick={0.34,0.36,0.38,0.40,0.42,0.44,0.46,0.48},
width=0.95\figurewidth,
height=\figureheight,
tick align=outside,
tick pos=left,
xmajorgrids,
x grid style={lightgray!92.026143790849673!black},
ymajorgrids,
y grid style={lightgray!92.026143790849673!black}
]
\addplot [very thick, color0, dashed, forget plot]
table {%
0 0.343
10 0.409053
20 0.415332
30 0.419089
40 0.421787
50 0.423895
60 0.425627
70 0.427097
80 0.428374
90 0.429503
100 0.430514
110 0.431431
120 0.432269
130 0.43304
140 0.433755
150 0.434421
160 0.435044
170 0.43563
180 0.436183
190 0.436705
200 0.437201
210 0.437673
220 0.438123
230 0.438553
240 0.438965
250 0.43936
260 0.43974
270 0.440105
280 0.440457
290 0.440796
300 0.441124
310 0.441442
320 0.441749
330 0.442046
340 0.442335
350 0.442615
360 0.442888
370 0.443152
380 0.44341
390 0.443661
400 0.443906
410 0.444144
420 0.444377
430 0.444604
440 0.444826
450 0.445043
460 0.445255
470 0.445463
480 0.445666
490 0.445865
500 0.446059
510 0.44625
520 0.446438
530 0.446621
540 0.446801
550 0.446978
560 0.447151
570 0.447322
580 0.447489
590 0.447654
600 0.447815
610 0.447974
620 0.448131
630 0.448285
640 0.448436
650 0.448585
660 0.448732
670 0.448876
680 0.449018
690 0.449159
700 0.449297
710 0.449433
720 0.449567
730 0.449699
740 0.44983
750 0.449958
760 0.450085
770 0.450211
780 0.450334
790 0.450456
800 0.450577
810 0.450696
820 0.450813
830 0.450929
840 0.451044
850 0.451157
860 0.451269
870 0.45138
880 0.451489
890 0.451597
900 0.451704
910 0.451809
920 0.451913
930 0.452017
940 0.452119
950 0.45222
960 0.45232
970 0.452419
980 0.452516
990 0.452613
1000 0.452709
1010 0.452804
1020 0.452898
1030 0.452991
1040 0.453083
1050 0.453174
1060 0.453264
1070 0.453354
1080 0.453442
1090 0.45353
1100 0.453617
1110 0.453703
1120 0.453788
1130 0.453873
1140 0.453956
1150 0.454039
1160 0.454122
1170 0.454203
1180 0.454284
1190 0.454364
1200 0.454444
1210 0.454522
1220 0.454601
1230 0.454678
1240 0.454755
1250 0.454831
1260 0.454907
1270 0.454982
1280 0.455056
1290 0.45513
1300 0.455203
1310 0.455275
1320 0.455347
1330 0.455419
1340 0.45549
1350 0.45556
1360 0.45563
1370 0.455699
1380 0.455768
1390 0.455837
1400 0.455904
1410 0.455972
1420 0.456038
1430 0.456105
1440 0.456171
1450 0.456236
1460 0.456301
1470 0.456365
1480 0.456429
1490 0.456493
1500 0.456556
1510 0.456619
1520 0.456681
1530 0.456743
1540 0.456804
1550 0.456865
1560 0.456926
1570 0.456986
1580 0.457046
1590 0.457106
1600 0.457165
1610 0.457223
1620 0.457282
1630 0.45734
1640 0.457397
1650 0.457454
1660 0.457511
1670 0.457568
1680 0.457624
1690 0.45768
1700 0.457735
1710 0.45779
1720 0.457845
1730 0.4579
1740 0.457954
1750 0.458008
1760 0.458061
1770 0.458114
1780 0.458167
1790 0.45822
1800 0.458272
1810 0.458324
1820 0.458376
1830 0.458427
1840 0.458479
1850 0.458529
1860 0.45858
1870 0.45863
1880 0.45868
1890 0.45873
1900 0.458779
1910 0.458829
1920 0.458878
1930 0.458926
1940 0.458975
1950 0.459023
1960 0.459071
1970 0.459118
1980 0.459166
1990 0.459213
2000 0.45926
2010 0.459306
2020 0.459353
2030 0.459399
2040 0.459445
2050 0.459491
2060 0.459536
2070 0.459581
2080 0.459626
2090 0.459671
2100 0.459716
2110 0.45976
2120 0.459804
2130 0.459848
2140 0.459892
2150 0.459936
2160 0.459979
2170 0.460022
2180 0.460065
2190 0.460108
2200 0.46015
2210 0.460192
2220 0.460234
2230 0.460276
2240 0.460318
2250 0.46036
2260 0.460401
2270 0.460442
2280 0.460483
2290 0.460524
2300 0.460564
2310 0.460605
2320 0.460645
2330 0.460685
2340 0.460725
2350 0.460764
2360 0.460804
2370 0.460843
2380 0.460882
2390 0.460921
2400 0.46096
2410 0.460999
2420 0.461037
2430 0.461076
2440 0.461114
2450 0.461152
2460 0.46119
2470 0.461227
2480 0.461265
2490 0.461302
2500 0.46134
2510 0.461377
2520 0.461414
2530 0.46145
2540 0.461487
2550 0.461523
2560 0.46156
2570 0.461596
2580 0.461632
2590 0.461668
2600 0.461703
2610 0.461739
2620 0.461774
2630 0.46181
2640 0.461845
2650 0.46188
2660 0.461915
2670 0.46195
2680 0.461984
2690 0.462019
2700 0.462053
2710 0.462087
2720 0.462121
2730 0.462155
2740 0.462189
2750 0.462223
2760 0.462256
2770 0.46229
2780 0.462323
2790 0.462356
2800 0.46239
2810 0.462423
2820 0.462455
2830 0.462488
2840 0.462521
2850 0.462553
2860 0.462586
2870 0.462618
2880 0.46265
2890 0.462682
2900 0.462714
2910 0.462746
2920 0.462777
2930 0.462809
2940 0.46284
2950 0.462872
2960 0.462903
2970 0.462934
2980 0.462965
2990 0.462996
3000 0.463027
3010 0.463057
3020 0.463088
3030 0.463118
3040 0.463149
3050 0.463179
3060 0.463209
3070 0.463239
3080 0.463269
3090 0.463299
3100 0.463329
3110 0.463358
3120 0.463388
3130 0.463417
3140 0.463447
3150 0.463476
3160 0.463505
3170 0.463534
3180 0.463563
3190 0.463592
3200 0.463621
3210 0.46365
3220 0.463678
3230 0.463707
3240 0.463735
3250 0.463764
3260 0.463792
3270 0.46382
3280 0.463848
3290 0.463876
3300 0.463904
3310 0.463932
3320 0.463959
3330 0.463987
3340 0.464014
3350 0.464042
3360 0.464069
3370 0.464097
3380 0.464124
3390 0.464151
3400 0.464178
3410 0.464205
3420 0.464232
3430 0.464258
3440 0.464285
3450 0.464312
3460 0.464338
3470 0.464365
3480 0.464391
3490 0.464417
3500 0.464444
3510 0.46447
3520 0.464496
3530 0.464522
3540 0.464548
3550 0.464574
3560 0.464599
3570 0.464625
3580 0.464651
3590 0.464676
3600 0.464702
3610 0.464727
3620 0.464752
3630 0.464778
3640 0.464803
3650 0.464828
3660 0.464853
3670 0.464878
3680 0.464903
3690 0.464928
3700 0.464952
3710 0.464977
3720 0.465002
3730 0.465026
3740 0.465051
3750 0.465075
3760 0.465099
3770 0.465124
3780 0.465148
3790 0.465172
3800 0.465196
3810 0.46522
3820 0.465244
3830 0.465268
3840 0.465292
3850 0.465315
3860 0.465339
3870 0.465363
3880 0.465386
3890 0.46541
3900 0.465433
3910 0.465456
3920 0.46548
3930 0.465503
3940 0.465526
3950 0.465549
3960 0.465572
3970 0.465595
3980 0.465618
3990 0.465641
4000 0.465664
4010 0.465687
4020 0.465709
4030 0.465732
4040 0.465755
4050 0.465777
4060 0.4658
4070 0.465822
4080 0.465844
4090 0.465867
4100 0.465889
4110 0.465911
4120 0.465933
4130 0.465955
4140 0.465977
4150 0.465999
4160 0.466021
4170 0.466043
4180 0.466065
4190 0.466086
4200 0.466108
4210 0.46613
4220 0.466151
4230 0.466173
4240 0.466194
4250 0.466216
4260 0.466237
4270 0.466258
4280 0.46628
4290 0.466301
4300 0.466322
4310 0.466343
4320 0.466364
4330 0.466385
4340 0.466406
4350 0.466427
4360 0.466448
4370 0.466469
4380 0.466489
4390 0.46651
4400 0.466531
4410 0.466551
4420 0.466572
4430 0.466592
4440 0.466613
4450 0.466633
4460 0.466653
4470 0.466674
4480 0.466694
4490 0.466714
4500 0.466734
4510 0.466755
4520 0.466775
4530 0.466795
4540 0.466815
4550 0.466835
4560 0.466854
4570 0.466874
4580 0.466894
4590 0.466914
4600 0.466934
4610 0.466953
4620 0.466973
4630 0.466992
4640 0.467012
4650 0.467031
4660 0.467051
4670 0.46707
4680 0.46709
4690 0.467109
4700 0.467128
4710 0.467147
4720 0.467167
4730 0.467186
4740 0.467205
4750 0.467224
4760 0.467243
4770 0.467262
4780 0.467281
4790 0.4673
4800 0.467319
4810 0.467337
4820 0.467356
4830 0.467375
4840 0.467394
4850 0.467412
4860 0.467431
4870 0.467449
4880 0.467468
4890 0.467486
4900 0.467505
4910 0.467523
4920 0.467542
4930 0.46756
4940 0.467578
4950 0.467596
4960 0.467615
4970 0.467633
4980 0.467651
4990 0.467669
5000 0.467687
};
\end{axis}

\end{tikzpicture}
  \vspace{-2ex}
  \caption{Estimate of $\prob{\passwordAlt \in
      \cluster{\password{\responderIdx}}}$ for account-\accountId
      password \password{\responderIdx} at \responder{\responderIdx}
      and candidate password \passwordAlt selected by user \accountId
      at \requester, per cluster size $\clusterSize =
      \setSize{\cluster{\password{\responderIdx}}}$ and taken with
        respect to random selection of the user \accountId and
        \responderTerm \responder{\responderIdx}; based on
        \cite[\figrefstatic{7}]{wang2018:domino}}
  \label{fig:pwdguesser}
\end{figure}

The key insight here is that if a user chooses its candidate password
\passwordAlt as users typically do, then using
a large \clusterSize provides little additional power
(\figref{fig:pwdguesser}) but, since $\nmbrSimilarPasswords =
(\nmbrHoneyPasswords+1) \clusterSize$ where \nmbrHoneyPasswords is the
number of honey-password clusters, imposes much greater cost
(\figref{fig:responseTime}) than using a small \clusterSize.
  Moreover, suppose we model the true detection rate
  when querying \nmbrRespondersQueried randomly chosen
  \respondersTerm\footnote{The \directoryTerm should retain its same
    random choice of \nmbrRespondersQueried \respondersTerm across the
    user's failed attempts to select a password that she has not
    reused, lest she simply retry the same or a closely related
    password until a set of \nmbrRespondersQueried \respondersTerm at
    which it is not used is chosen.  Alternatively, the \requesterTerm
    can be charged with ensuring that the user's attempted passwords
    are sufficiently different from one another.}  as
  $\trueDetectionRate = 1 - (\prob{\passwordAlt \not\in
    \cluster{\password{\responderIdx}}})^{\nmbrRespondersQueried}$,
  i.e., ignoring the probability of false detections due to the use of
  a Bloom filter and assuming that the events $\passwordAlt \not\in
  \cluster{\password{\responderIdx}}$ and $\passwordAlt \not\in
  \cluster{\password{\responderIdxAlt}}$ are independent if
  $\responderIdx \neq \responderIdxAlt$ (which is perhaps reasonable
  since the user is forced to set dissimilar passwords at
  \responder{\responderIdx} and \responder{\responderIdxAlt} by our
  framework).  Then, increasing \nmbrRespondersQueried provides more
  detection power.

To balance these parameters and the
response time of the protocol, we model the response time using
\begin{align*}
  \responseTime(\nmbrRespondersQueried, \nmbrSimilarPasswords) =
  \coeff{0} + \coeff{1} \cdot \nmbrSimilarPasswords +
  \coeff{2} \cdot \nmbrRespondersQueried  +
  \coeff{3} \cdot \nmbrSimilarPasswords \cdot
  \nmbrRespondersQueried
\end{align*}
Regression analysis using the data in \secref{sec:eval:performance}
yields $\coeff{0} = 1.5507$, $\coeff{1} = 5.8834 \times 10^{-3}$,
$\coeff{2} = 2.6209 \times 10^{-3}$ and $\coeff{3} = 4.7135 \times
10^{-5}$ in the \untrustedForAccountLocationPrivacy-\directoryTerm
case (root-mean-square error
$\mathit{RMSE} = 0.4547$) and $\coeff{0} = 6.4595 \times
10^{-3}$, $\coeff{1} = 2.2885 \times 10^{-3}$, $\coeff{2} = 1.0271
\times 10^{-3}$ and $\coeff{3} = 2.0336 \times 10^{-5}$ in the
\trustedForAccountLocationPrivacy-\directoryTerm case ($\mathit{RMSE}
= 0.1276$).  Then, the \requesterTerm chooses \nmbrRespondersQueried
and \nmbrSimilarPasswords using the following optimization:
\begin{align*}
\stackrel[\nmbrRespondersQueried , \nmbrSimilarPasswords]{}{\text{maximize}}~
& \trueDetectionRate = 1 - (\prob{\passwordAlt \not\in
  \cluster{\password{\responderIdx}}})^{\nmbrRespondersQueried} \\
\text{subject to~}
& \responseTime(\nmbrRespondersQueried, \nmbrSimilarPasswords) \leq \responseTimeMax \\
& 1 \leq \clusterSize = \nmbrSimilarPasswords/(\nmbrHoneyPasswords+1) \\
& 1 \leq \nmbrRespondersQueried \le \nmbrResponders{\accountId}
\end{align*}
where \responseTimeMax is the \requesterTerm's desired response time
and \nmbrResponders{\accountId} is the number of
\respondersTerm registered at the \directoryTerm as having an account
for identifier \accountId.  The \directoryTerm can send
\nmbrResponders{\accountId} to the \requesterTerm in an initial
negotiation round before \msgref{prot:msg:bloomFilter}.

This optimization, together with using the curve in
\figref{fig:pwdguesser} to estimate \prob{\passwordAlt \in
  \cluster{\password{\responderIdx}}} and the regression
results above to estimate $\responseTime(\nmbrRespondersQueried,
\nmbrSimilarPasswords)$, yields results like those shown in
\tblref{tbl:optimal}.  In these optimizations, we set
$\nmbrResponders{\accountId} = 26$, because recent work found the mean
number of password-protected online accounts per user is
$26$~\cite{pearman2017:habitat}.  The response-time goals
\responseTimeMax used in \tblref{tbl:optimal} were chosen simply to
show how the optimal \nmbrRespondersQueried and \nmbrSimilarPasswords
vary under stringent response-time constraints.  As shown there, for
many response-time goals \responseTimeMax, a true detection rate
$\trueDetectionRate \approx 1$ can be achieved with very small values
of \nmbrSimilarPasswords.

\begin{table}
  \begin{subfigure}[b]{\columnwidth}
    \centering
    {\scriptsize
    \begin{tabular}{@{}l@{\hspace{0.5em}}r@{\hspace{1.5em}}*{9}{>{\raggedleft\arraybackslash}p{1.8em}@{\hspace{0.9em}}}>{\raggedleft\arraybackslash}p{1.8em}@{}}
    \toprule
    & & \multicolumn{10}{c}{\responseTimeMax (\secs)} \\
    & & .01 & .02 & .03 & .04 & .05 & .06 & .07 & .08 & .09 & .10 \\
    \midrule
    \underline{$\nmbrHoneyPasswords = 0$}
    & \nmbrSimilarPasswords  & 1 & 1 & 2 & 2 & 5 & 9 & 13 & 16 & 20 & 23 \\
    & \nmbrRespondersQueried & 1 & 10 & 17 & 26 & 26 & 26 & 26 & 26 & 26 & 26 \\
    & \trueDetectionRate     & .343 & .985 & $\approx$1 & $\approx$1 & $\approx$1 & $\approx$1 & $\approx$1 & $\approx$1 & $\approx$1 & $\approx$1  \\
    \midrule
    \underline{$\nmbrHoneyPasswords = 4$}
    & \nmbrSimilarPasswords  & - & 5 & 5 & 5 & 5 & 10 & 10 & 15 & 20 & 20 \\
    & \nmbrRespondersQueried & - & 1 & 10 & 19 & 26 & 24 & 26 & 26 & 26 & 26 \\
    & \trueDetectionRate     & - & .343 & .985 & $\approx$1 & $\approx$1 & $\approx$1 & $\approx$1 & $\approx$1 & $\approx$1 & $\approx$1  \\
    \midrule
    \underline{$\nmbrHoneyPasswords = 9$}
    & \nmbrSimilarPasswords  & - & - & - & 10 & 10 & 10 & 10 & 10 & 20 & 20 \\
    & \nmbrRespondersQueried & - & - & - & 8 & 16 & 24 & 26 & 26 & 26 & 26 \\
    & \trueDetectionRate     & - & - & - & .965 & .999 & $\approx$1 & $\approx$1 & $\approx$1 & $\approx$1 & $\approx$1  \\
    \bottomrule
  \end{tabular}
  }
  \caption{\trustedForAccountLocationPrivacy \directoryTerm}
    \label{tbl:optimal:woTor}
  \end{subfigure}
  \\[10pt]
  \begin{subfigure}[b]{\columnwidth}
    \centering
    {\scriptsize
    \begin{tabular}{@{}l@{\hspace{0.5em}}r@{\hspace{1.5em}}*{9}{>{\raggedleft\arraybackslash}p{1.8em}@{\hspace{0.9em}}}>{\raggedleft\arraybackslash}p{1.8em}@{}}
    \toprule
    & & \multicolumn{10}{c}{\responseTimeMax (\secs)} \\
    & & 1.60 & 1.62 & 1.64 & 1.66 & 1.68 & 1.70 & 1.72 & 1.74 & 1.76 & 1.78 \\
    \midrule
    \underline{$\nmbrHoneyPasswords = 0$}
    & \nmbrSimilarPasswords  & 1 & 2 & 2 & 5 & 8 & 11 & 14 & 17 & 19 & 22\\
    & \nmbrRespondersQueried & 16 & 21 & 26 & 26 & 26 & 26 & 26 & 26 & 26 & 26 \\
    & \trueDetectionRate     & .999 & $\approx$1 & $\approx$1 & $\approx$1 & $\approx$1 & $\approx$1 & $\approx$1 & $\approx$1 & $\approx$1 & $\approx$1 \\
    \midrule
    \underline{$\nmbrHoneyPasswords = 4$}
    & \nmbrSimilarPasswords  & 5 & 5 & 5 & 5 & 5 & 10 & 10 & 15 & 15 & 20\\
    & \nmbrRespondersQueried & 6 & 13 & 20 & 26 & 26 & 26 & 26 & 26 & 26 & 26 \\
    & \trueDetectionRate     & .920 & .996 & $\approx$1& $\approx$1 & $\approx$1 & $\approx$1 & $\approx$1 & $\approx$1 & $\approx$1 & $\approx$1 \\
    \midrule
    \underline{$\nmbrHoneyPasswords = 9$}
    & \nmbrSimilarPasswords  & - & 10 & 10 & 10 & 10 & 10 & 10 & 10 & 20 & 20\\
    & \nmbrRespondersQueried & - & 3 & 9 & 16 & 22 & 26 & 26 & 26 & 25 & 26 \\
    & \trueDetectionRate     & - & .716 & .977 & .999 & $\approx$1 & $\approx$1 & $\approx$1 & $\approx$1 & $\approx$1 & $\approx$1 \\
    \bottomrule
  \end{tabular}
  }
  \caption{\untrustedForAccountLocationPrivacy \directoryTerm}
    \label{tbl:optimal:Tor}
  \end{subfigure}
  \caption{Choices for \nmbrRespondersQueried and
    \nmbrSimilarPasswords computed using optimization in
    \secref{sec:eval:optimization} with $\nmbrResponders{\accountId}=26$}
  \label{tbl:optimal}
\end{table}

As such, the full range of parameter settings explored in
\secref{sec:eval:performance} will rarely be needed.  This is
fortunate, since small values of \nmbrSimilarPasswords improve the
throughput of \requesterTerm-\responderTerm interactions, especially
in the \trustedForAccountLocationPrivacy-\directoryTerm model.  To see
this, \tblref{tbl:throughput} shows the throughput of our
implementation, measured as the largest number of \textit{qualifying}
responses achieved as the requests per second were increased, as a
function of \nmbrSimilarPasswords and \nmbrRespondersQueried.  In
\tblref{tbl:throughput:woTor} and \tblref{tbl:throughput:Tor}, a
response was \textit{qualifying} if its response time was $\le 5\secs$
and $\le 8\secs$, respectively.

\begin{table}
  \centering
  \begin{subfigure}[b]{0.5\columnwidth}
  \setcounter{MinNumber}{25}%
  \setcounter{MaxNumber}{4304}%
  \centering
      {\scriptsize
        \setlength{\tabcolsep}{.16667em}
        \begin{tabular}{cr|@{\hspace{.16667em}}|@{\hspace{1.5pt}}*{6}{X}@{\hspace{1.5pt}}}
          & & \multicolumn{6}{c}{\nmbrRespondersQueried} \\
          & \nmbrSimilarPasswords ~~
          & \multicolumn{1}{c}{1}
          & \multicolumn{1}{c}{6}
          & \multicolumn{1}{c}{11}
          & \multicolumn{1}{c}{16}
          & \multicolumn{1}{c}{21}
          & \multicolumn{1}{c}{26}\\
          \cline{2-8} \\[-2.35ex]
          &  1 &  4304 &  1013 &  492 &  325 &  237 & 174  \\
          & 10 &  2415 &  549 &  277 &  188 &  155 & 122  \\
          & 20 &  1478 &  336 &  182 &  129 &  98 & 78  \\
          & 30 &  1076 &  243 &  124 &  86 &  63 & 53  \\
          & 40 &  788 &  187 &  94 &  67 &  49 & 40  \\
          & 50 &  683 &  159 &  76 &  52 &  39 & 33  \\
          & 60 &  611 &  132 &  63 &  43 &  32 & 25
        \end{tabular}
      }
      \caption{\trustedForAccountLocationPrivacy \directoryTerm}
      \label{tbl:throughput:woTor}
  \end{subfigure}
  \hspace{1em}
  \begin{subfigure}[b]{0.425\columnwidth}
  \setcounter{MinNumber}{10}%
  \setcounter{MaxNumber}{95}%
  \centering
      {\scriptsize
        \setlength{\tabcolsep}{.16667em}
        \begin{tabular}{cr|@{\hspace{.16667em}}|@{\hspace{1.5pt}}*{6}{Y}@{\hspace{1.5pt}}}
          & & \multicolumn{6}{c}{\nmbrRespondersQueried} \\
          & \nmbrSimilarPasswords ~~
          & \multicolumn{1}{c}{1}
          & \multicolumn{1}{c}{6}
          & \multicolumn{1}{c}{11}
          & \multicolumn{1}{c}{16}
          & \multicolumn{1}{c}{21}
          & \multicolumn{1}{c}{26}\\
          \cline{2-8} \\[-2.35ex]
          &  1 &  95 &  61  &  42 &  33 &  27 & 22  \\
          & 10 &  87 &  59  &  40 &  31 &  25 & 20  \\
          & 20 &  78 &  54  &  37 &  28 &  23 & 19  \\
          & 30 &  71 &  51  &  35 &  27 &  20 & 16  \\
          & 40 &  62 &  44  &  32 &  24 &  18 & 14  \\
          & 50 &  53 &  39  &  26 &  20 &  15 & 11  \\
          & 60 &  42 &  31  &  20 &  16 &  10 & 10
        \end{tabular}
      }
      \caption{\untrustedForAccountLocationPrivacy \directoryTerm}
      \label{tbl:throughput:Tor}
  \end{subfigure}
  \caption{Maximum qualifying responses per second}
  \label{tbl:throughput}
\end{table}

This $3\secs$ difference between the standards for \textit{qualifying}
in the two tests was needed because we constructed the
\untrustedForAccountLocationPrivacy-\directoryTerm test to capture as
faithfully as possible the Tor costs that a real deployment would
incur.  Notably, even though the \nmbrRespondersQueried
\respondersTerm queried per request were chosen from only 64
\respondersTerm in total (the configuration was the same as in
\secref{sec:eval:performance}), no two requests were allowed to use
the same Tor circuit, since they would be unable to do so in a real
deployment, where different addresses for the same \responderTerm are
stored for different user accounts at the \directoryTerm.  (The
exception is if the requests were for the same user and at the same
\responderTerm.)  So, each request necessitated construction of new
Tor circuits to its \respondersTerm, which increased response times
commensurately.

To put \tblref{tbl:throughput} in context, a throughput of 50
qualifying responses per second is enough to enable each of the 312
million Internet users in the U.S.\footnote{This estimate was
  retrieved from
  \url{https://www.statista.com/topics/2237/internet-usage-in-the-united-states/}
  on December 4, 2018.} to setup or change passwords on about 5
accounts per year.  Moreover, we believe the numbers in
\tblref{tbl:throughput} to be pessimistic, in that in each request,
the \nmbrRespondersQueried \respondersTerm were chosen from only
$\nmbrResponders{\accountId}=64$ \respondersTerm in total, versus from
likely many more in practice.  Still, based on
\tblref{tbl:throughput:Tor}, a deployment using the
\untrustedForAccountLocationPrivacy-\directoryTerm model would
presumably require adaptations of Tor for our use-case
(e.g.,~\cite{perng2006:m2}) and distribution of the \directoryTerm.

We note, however, that even a non-replicated \directoryTerm should
easily handle the \textit{storage} requirements of our design.  With
3.58 billion active Internet users worldwide and an average of 26
password-protected accounts per user, the storage of a Tor hidden
service address for each user account at each website amounts to only
$\approx 1.5 \terabytes$ of state.  In the
\trustedForAccountLocationPrivacy-\directoryTerm model, the storage
requirements would be even less.

\section{Denials of Service}
\label{sec:dos}

Our design introduces denial-of-service opportunities for misbehaving
\requestersTerm, \respondersTerm, or the \directoryTerm.  We discuss
these risks here, as well as methods to remedy them.

Perhaps the most troubling is a \responderTerm who returns
$\resultCiphertext{} \in \ciphertextSpace{\pubKey}(\groupIdentity)$
regardless of the request ciphertexts
$\langle\bloomFilterBitCtext{\bloomFilterBitIdx}\rangle_{\bloomFilterBitIdx\in\residues{\bloomFilterSize}}$
in \msgref{prot:msg:bloomFilter}, thereby giving the \requesterTerm
reason to reject the user's chosen password even when the user's
chosen password is not similar to others she set elsewhere.  This
denial-of-service attack would frustrate users, but
fortunately a \responderTerm that misbehaves in this way can be
caught by simple audit mechanisms.  For example, at any point, the
\directoryTerm could generate a \msgref{prot:msg:bloomFilter} in which
each $\bloomFilterBitCtext{\bloomFilterBitIdx}
\in \ciphertextSpace{\pubKey}\setminus\ciphertextSpace{\pubKey}(\groupIdentity)$
and for which it knows the private key \privKey corresponding to
\pubKey; if a \responderTerm responds with
$\resultCiphertext{} \in
\ciphertextSpace{\pubKey}(\groupIdentity)$, then the \directoryTerm
has proof that the \responderTerm is lying and, e.g., can simply
remove the \responderTerm from future queries.  In principle, a
\requesterTerm could also generate such audit queries, though doing so
would require the \directoryTerm to suspend the user-consent mechanism
in \secref{sec:framework:design:responder}.  In this case, a detection
would enable the \requesterTerm to learn that either one of the
\respondersTerm is misbehaving (but it would need help from the
\directoryTerm to figure out which one) or that the \directoryTerm is
misbehaving (in which case it would need to report it to some managing
authority).

Other misbehaviors can render our framework silently ineffective while
they persist.  For example, a malicious \directoryTerm could simply
not query \respondersTerm at all, instead forging the response
\resultCiphertext{\responderIdx} purportedly from each
\responder{\responderIdx} to indicate \textit{no} password reuse
(i.e., $\resultCiphertext{\responderIdx} \in \ciphertextSpace{\pubKey}
\setminus \ciphertextSpace{\pubKey}(\groupIdentity))$.  Again, a
simple audit (knowingly attempting to reuse a password at a
\requesterTerm) can detect such misbehavior.  Presuming such
misbehaviors will occur rarely and be remedied quickly, we believe our
framework will suffice to discourage password reuse even if it
\textit{usually} works.

As our framework enables the \requesterTerm to perform precomputation
to reduce its costs on the critical path of protocol execution, the
critical-path computation cost of the protocol is greater for the
\responderTerm than it is for the 
\iffull
\requesterTerm (see \appref{sec:microbenchmarks}).
\else
\requesterTerm.
\fi
This is even more true for misbehaving
\requestersTerm that replay the same request, in an effort to 
occupy \directoryTerm and \responderTerm resources.  Of course, this
concern is not unique to our framework, and various techniques to stem
such denials of service exist that would be amenable to adoption in
our framework (e.g.,~\cite{dwork1993:pricing,abadi2005:memory}).  In
addition, steps detailed in \secref{sec:framework:design:responder} to
require user consent (through clicking on a confirmation URL) to
complete the protocol could interfere with such attacks.  In the worst
case, however, \respondersTerm and the \directoryTerm can refuse
requests until the flood subsides, albeit temporarily reducing the
utility of our framework to the status quo today.

\section{Conclusion}
\label{sec:conclusion}

Adams and Sasse famously declared, ``Users are not the
enemy''~\cite{adams1999:enemy}.  While we do not mean to suggest
otherwise, it has also long been understood in a variety of contexts
that users must be compelled to adhere to security policies, as
otherwise they will not do so.  Despite decades of haranguing users to
stop reusing passwords, their adoption of methods to manage passwords
more effectively has been painfully slow.  This, in turn, has given
rise to credential abuses that inflict considerable costs on service
operators (see \secref{sec:introduction}).

We believe it is now time to consider imposing technical measures to
interfere with the use of similar passwords across websites.  In this
paper we have presented one possible method for doing so, by
coordinating password selection across websites so that similar
passwords cannot be used for the same account identifier.  Our
framework combines a set-membership-test protocol
(\secref{sec:protocol}) with a variety of other defenses
(\secref{sec:framework:design}) to implement \accountSecurity and
\accountLocationPrivacy, the former of which we confirm via
probabilistic model checking (\secref{sec:framework:analysis}).
Finally, we leveraged tendencies of how users reuse passwords to
optimize the parameters for our framework, enabling it to be effective
with surprisingly modest costs (\secref{sec:eval:optimization}).

\section*{Acknowledgment}
We are grateful for comments on previous versions of this paper from
Prof.\ Marina Blanton and anonymous reviewers.  This work was
supported in part by NSF grant 1330599.

\bibliographystyle{IEEEtranS}
\bibliography{IEEEabrv,tight,main}

\iffull
\appendices

\section{Proofs}
\label{sec:proofs}

\begin{proof}[Proof of \propref{prop:responderSecurityPtext}]
Note that $\resultCiphertext{} \in \ciphertextSpace{\pubKey}$ because
each $\bloomFilterBitCtext{\bloomFilterBitIdx}
\in \ciphertextSpace{\pubKey}$, by
\lineref{prot:responder:checkCiphertexts}.  If
\[
\left(\encProd{\pubKey}{\bloomFilterBitIdx \in \residues{\bloomFilterSize}\setminus\bloomFilterIndicesToSet{\responder{}}}{} \hspace{-0.75em}\bloomFilterBitCtext{\bloomFilterBitIdx}\right) \in
\ciphertextSpace{\pubKey}(\groupElmtAlt)
\]
in \lineref{prot:responder:makeResult}, then
$\resultCiphertext{}
\in \ciphertextSpace{\pubKey}((\groupElmtAlt)^\encExponent)$ for
\encExponent chosen in
\lineref{prot:responder:generateBlinding}.  So, if
$\resultCiphertext{}\not\in\ciphertextSpace{\pubKey}(\groupIdentity)$
or, in other words, $(\groupElmtAlt)^\encExponent \neq
\groupIdentity$, then \encExponent such that
$(\groupElmtAlt)^\encExponent = \groupElmt$ for a specific $\groupElmt
\in \groupSet\setminus\{\groupIdentity\}$ is chosen in
\lineref{prot:responder:generateBlinding} with probability
$\frac{1}{\groupOrder-1}$.
\end{proof}

\begin{proof}[Proof of \propref{prop:responderSecurityCtext}]
This follows immediately since for $\ciphertext{1}
\in \ciphertextSpace{\pubKey}(\plaintext{1})$ and $\ciphertext{2}
\in \ciphertextSpace{\pubKey}(\plaintext{2})$, the value
$\ciphertext{} \gets \ciphertext{1} \encMult{\pubKey}
\ciphertext{2}$ is chosen uniformly at random from
$\ciphertextSpace{\pubKey}(\plaintext{1} \plaintext{2})$.
\end{proof}

For the proof of \propref{prop:genericGroup}, we leverage the generic
group model as presented by Maurer~\cite{maurer2005:abstract}, which
Jager and Schwenk~\cite{jager2008:equivalence} have shown to be
equivalent to the other common generic model, due to
Shoup~\cite{shoup1997:bounds}.  In the Maurer model, \adversary{}
(i.e., \adversary{1} in experiment
\experiment{\responder{}}{\encScheme}) is provided only the group
order \groupOrder and \textit{black-box} access (i.e., oracle access)
to the group elements \groupGenerator, \elgPubKey,
$\{\elgEphemeralPubKey{\bloomFilterBitIdx}\}_{\bloomFilterBitIdx \in
  \residues{\bloomFilterSize}}$,
$\{\elgCiphertext{\bloomFilterBitIdx}\}_{\bloomFilterBitIdx \in
  \residues{\bloomFilterSize}}$, where each
$\bloomFilterBitCtext{\bloomFilterBitIdx} = \langle
\elgEphemeralPubKey{\bloomFilterBitIdx},
\elgCiphertext{\bloomFilterBitIdx}\rangle$, rather than receiving
these group elements as inputs.  Because the group representation is
never exposed to \adversary{}, each group element is equivalently
represented as its base-\groupGenerator discrete logarithm.  So, the
oracle holds integers $1$, \elgPrivKey,
$\{\elgEphemeralPrivKey{\bloomFilterBitIdx}\}_{\bloomFilterBitIdx \in
  \residues{\bloomFilterSize}}$,
$\{\elgCiphertextLog{\bloomFilterBitIdx}\}_{\bloomFilterBitIdx \in
  \residues{\bloomFilterSize}}$ to represent \groupGenerator,
\elgPubKey,
$\{\elgEphemeralPubKey{\bloomFilterBitIdx}\}_{\bloomFilterBitIdx \in
  \residues{\bloomFilterSize}}$,
$\{\elgCiphertext{\bloomFilterBitIdx}\}_{\bloomFilterBitIdx \in
  \residues{\bloomFilterSize}}$, respectively, where $\elgPubKey =
\groupGenerator^{\elgPrivKey}$, each
$\elgEphemeralPubKey{\bloomFilterBitIdx} =
\groupGenerator^{\elgEphemeralPrivKey{\bloomFilterBitIdx}}$, and each
$\elgCiphertext{\bloomFilterBitIdx} =
\groupGenerator^{\elgCiphertextLog{\bloomFilterBitIdx}}$.  The oracle
stores each of these values in an array at an index known to the
adversary.  Moreover, the oracle supports creation of new values via
\textit{computation queries} reflecting the application of the group
operation \groupMult to two existing group elements represented at
indices specified in the query; the values so created are appended to
the array but not returned.  Specifically, in each computation query,
\adversary{} specifies two indices, and the oracle applies the
\groupMult operator to the group elements
$\groupGenerator^{\residueValue}$,
$\groupGenerator^{\residueValueAlt}$ represented by the values
\residueValue and \residueValueAlt at those indices, resulting in
$\residueValue + \residueValueAlt$ being stored in the array to
represent $\groupGenerator^{\residueValue + \residueValueAlt}$.

In addition to computation queries, \adversary{} can also perform
\textit{equality queries}, where it asks whether the group elements
represented by two indices it specifies are the same.  Finally, in
accordance with our protocol, the adversary is permitted to ask
\textit{just one} DDH query, i.e., whether $\elgPrivKey
\elgEphemeralPrivKey{} \testRingEquiv{\groupOrder}
\elgCiphertextLog{}$ for \elgPrivKey the second value in the array
(representing \elgPubKey) and \elgEphemeralPrivKey{} and
\elgCiphertextLog{} at specified indices in the array representing
$\groupGenerator^{\elgEphemeralPrivKey{}}$ and
$\groupGenerator^{\elgCiphertextLog{}}$, respectively.  This
corresponds to providing \adversary{2} with the answer to
$\resultCiphertext{}
\testIn \ciphertextSpace{\pubKey}(\groupIdentity)$, where $\pubKey =
\langle \groupGenerator, \elgPubKey\rangle$ and $\resultCiphertext{} =
\langle \groupGenerator^{\elgEphemeralPrivKey{}},
\groupGenerator^{\elgCiphertextLog{}}\rangle$.

\begin{proof}[Proof of \propref{prop:genericGroup}]
  Through the computation operations available to \adversary{}, every
  value stored in the oracle is of the form
  \begin{align}
  \elgPrivKeyCoeff \elgPrivKey + 
  \sum_{\bloomFilterBitIdx\in\residues{\bloomFilterSize}} \elgEphemeralPrivKeyCoeff{\bloomFilterBitIdx} \elgEphemeralPrivKey{\bloomFilterBitIdx} +
  \sum_{\bloomFilterBitIdx\in\residues{\bloomFilterSize}} \elgCiphertextLogCoeff{\bloomFilterBitIdx} \elgCiphertextLog{\bloomFilterBitIdx} +
  \residueConstant
  \label{eqn:genericGroupForm}
  \end{align}
  for constants \elgPrivKeyCoeff,
  $\{\elgEphemeralPrivKeyCoeff{\bloomFilterBitIdx}\}_{\bloomFilterBitIdx\in\residues{\bloomFilterSize}}$,
  $\{\elgCiphertextLogCoeff{\bloomFilterBitIdx}\}_{\bloomFilterBitIdx\in\residues{\bloomFilterSize}}$,
  $\residueConstant \in \ints{\groupOrder}$ known to
  \adversary{}.

  Each equality query tests whether $\groupGenerator^{\encExponent}
  \testEqual \groupGenerator^{\encExponentAlt}$ or, in other words,
  whether $\encExponent \testRingEquiv{\groupOrder} \encExponentAlt$,
  for values \encExponent and \encExponentAlt at the specified indices.
  These values are of the form in \eqnref{eqn:genericGroupForm}, i.e.,
  \begin{align*}
  \encExponent & \ringEquiv{\groupOrder} \elgPrivKeyCoeff \elgPrivKey +
  \sum_{\bloomFilterBitIdx\in\residues{\bloomFilterSize}} \elgEphemeralPrivKeyCoeff{\bloomFilterBitIdx} \elgEphemeralPrivKey{\bloomFilterBitIdx} +  
  \sum_{\bloomFilterBitIdx\in\residues{\bloomFilterSize}} \elgCiphertextLogCoeff{\bloomFilterBitIdx} \elgCiphertextLog{\bloomFilterBitIdx} +
  \residueConstant \\
  \encExponentAlt & \ringEquiv{\groupOrder} \elgPrivKeyCoeffAlt \elgPrivKey +
  \sum_{\bloomFilterBitIdx\in\residues{\bloomFilterSize}} \elgEphemeralPrivKeyCoeffAlt{\bloomFilterBitIdx} \elgEphemeralPrivKey{\bloomFilterBitIdx} +
  \sum_{\bloomFilterBitIdx\in\residues{\bloomFilterSize}} \elgCiphertextLogCoeffAlt{\bloomFilterBitIdx} \elgCiphertextLog{\bloomFilterBitIdx} +
  \residueConstantAlt
  \end{align*}
  and so the test $\encExponent \testRingEquiv{\groupOrder}
  \encExponentAlt$ is equivalent to
  \begin{align}
    & \left[(\elgPrivKeyCoeff - \elgPrivKeyCoeffAlt)
      + \sum_{\bloomFilterBitIdx\in\residues{\bloomFilterSize}\setminus\bloomFilterIndicesToSet{\requester}}(\elgCiphertextLogCoeff{\bloomFilterBitIdx} - \elgCiphertextLogCoeffAlt{\bloomFilterBitIdx}) \elgEphemeralPrivKey{\bloomFilterBitIdx}
\right] \elgPrivKey \nonumber\\
    & + \sum_{\bloomFilterBitIdx\in\residues{\bloomFilterSize}} (\elgEphemeralPrivKeyCoeff{\bloomFilterBitIdx} - \elgEphemeralPrivKeyCoeffAlt{\bloomFilterBitIdx}) \elgEphemeralPrivKey{\bloomFilterBitIdx}
      + \sum_{\bloomFilterBitIdx\in\bloomFilterIndicesToSet{\requester}} (\elgCiphertextLogCoeff{\bloomFilterBitIdx} - \elgCiphertextLogCoeffAlt{\bloomFilterBitIdx}) \elgCiphertextLog{\bloomFilterBitIdx}
    + (\residueConstant - \residueConstantAlt)
    \testRingEquiv{\groupOrder} 0
  \label{eqn:eqOracleTest}
\end{align}
  where \elgPrivKey,
  $\{\elgEphemeralPrivKey{\bloomFilterBitIdx}\}_{\bloomFilterBitIdx
    \in \residues{\bloomFilterSize}}$, and
  $\{\elgCiphertextLog{\bloomFilterBitIdx}\}_{\bloomFilterBitIdx \in
    \bloomFilterIndicesToSet{\requester}}$ are chosen independently at
  random from \ints{\groupOrder}.  As such, ignoring queries that
  return \boolTrue with probability $1$ (and so teach the adversary
  nothing), the probability that each oracle query returns \boolTrue
  is $1/\groupOrder$.  So, letting \someEqEvent denote the event that
  at least one equality query returns \boolTrue, if \adversary{} makes
  \equalityQueries equality queries, then
  \begin{align}
    \prob{\someEqEvent} & \le \frac{\equalityQueries}{\groupOrder}
    \label{eqn:prob-someEqEvent}
  \end{align}
  
  \adversary{2} is eventually provided the answer to whether
  $\resultCiphertext{}
  \testIn \ciphertextSpace{\pubKey}(\groupIdentity)$ for a
  \resultCiphertext{} of its choosing.  In this case
  \resultCiphertext{} is represented by a pair $\langle
  \elgEphemeralPrivKey{}, \elgCiphertextLog{}\rangle$ where
  \elgEphemeralPrivKey{} and \elgCiphertextLog{} are of the form in
  \eqnref{eqn:genericGroupForm}, i.e.,
  \begin{align*}
  \elgEphemeralPrivKey{} & \ringEquiv{\groupOrder} \elgPrivKeyCoeff \elgPrivKey +
  \sum_{\bloomFilterBitIdx\in\residues{\bloomFilterSize}} \elgEphemeralPrivKeyCoeff{\bloomFilterBitIdx} \elgEphemeralPrivKey{\bloomFilterBitIdx} +  
  \sum_{\bloomFilterBitIdx\in\residues{\bloomFilterSize}} \elgCiphertextLogCoeff{\bloomFilterBitIdx} \elgCiphertextLog{\bloomFilterBitIdx} +
  \residueConstant \\
  \elgCiphertextLog{} & \ringEquiv{\groupOrder} \elgPrivKeyCoeffAlt \elgPrivKey +
  \sum_{\bloomFilterBitIdx\in\residues{\bloomFilterSize}} \elgEphemeralPrivKeyCoeffAlt{\bloomFilterBitIdx} \elgEphemeralPrivKey{\bloomFilterBitIdx} +
  \sum_{\bloomFilterBitIdx\in\residues{\bloomFilterSize}} \elgCiphertextLogCoeffAlt{\bloomFilterBitIdx} \elgCiphertextLog{\bloomFilterBitIdx} +
  \residueConstantAlt
  \end{align*}
and the test $\resultCiphertext{}
\testIn \ciphertextSpace{\pubKey}(\groupIdentity)$ is equivalent to
$\elgPrivKey\elgEphemeralPrivKey{} \testRingEquiv{\groupOrder}
\elgCiphertextLog{}$ or, in other words,
\begin{align}
\elgPrivKeyCoeff \elgPrivKey^2 +
(\residueConstant - \elgPrivKeyCoeffAlt) \elgPrivKey
& + \sum_{\bloomFilterBitIdx\in\residues{\bloomFilterSize}} (\elgEphemeralPrivKeyCoeff{\bloomFilterBitIdx}\elgPrivKey - \elgEphemeralPrivKeyCoeffAlt{\bloomFilterBitIdx}) \elgEphemeralPrivKey{\bloomFilterBitIdx} \nonumber \\ 
& + \sum_{\bloomFilterBitIdx\in\residues{\bloomFilterSize}} (\elgCiphertextLogCoeff{\bloomFilterBitIdx}\elgPrivKey - \elgCiphertextLogCoeffAlt{\bloomFilterBitIdx}) \elgCiphertextLog{\bloomFilterBitIdx} - \residueConstantAlt
\testRingEquiv{\groupOrder} 0
  \label{eqn:ddhOracleTest}
\end{align}

Let $\bloomFilterIndicesToSet{} = \{\bloomFilterBitIdx \in
\residues{\bloomFilterSize} \mid
\elgCiphertextLogCoeff{\bloomFilterBitIdx}\elgPrivKey -
\elgCiphertextLogCoeffAlt{\bloomFilterBitIdx}
\not\ringEquiv{\groupOrder} 0\}$.  The number of possible sets
\bloomFilterIndicesToSet{\requester} such that
$\bloomFilterIndicesToSet{} \cap \bloomFilterIndicesToSet{\requester}
= \emptyset$ is ${{\bloomFilterSize -
    \setSize{\bloomFilterIndicesToSet{}}} \choose
  {\bloomFilterHashFns}}$, and so
\begin{align}
  \cprob{\big}{\bloomFilterIndicesToSet{} \cap \bloomFilterIndicesToSet{\requester} = \emptyset}{\neg\someEqEvent} = \frac{{{\bloomFilterSize - \setSize{\bloomFilterIndicesToSet{}}} \choose {\bloomFilterHashFns}}}{{\bloomFilterSize \choose \bloomFilterHashFns}}
\label{eqn:prob-noIntersection}
\end{align}
Since the answer to $\resultCiphertext{}
\testIn \ciphertextSpace{\pubKey}(\groupIdentity)$ is not computed
using \elgCiphertextLog{\bloomFilterBitIdx} for any
$\bloomFilterBitIdx \in \residues{\bloomFilterSize} \setminus
\bloomFilterIndicesToSet{}$, \adversary{2} must choose which of these
indices are in \bloomFilterIndicesToSet{\requester} blindly.  So, if
$\bloomFilterIndicesToSet{} \cap \bloomFilterIndicesToSet{\requester}
= \emptyset$, then
\begin{align}
\cprob{\big}{\bloomFilterIndicesToSet{\adversary{}} =
\bloomFilterIndicesToSet{\requester}}{\bloomFilterIndicesToSet{} \cap \bloomFilterIndicesToSet{\requester} = \emptyset \wedge \neg\someEqEvent} \le
\frac{1}{{{\bloomFilterSize - \setSize{\bloomFilterIndicesToSet{}}} \choose {\bloomFilterHashFns}}}
\label{eqn:prob-noIntersectionGuess}
\end{align}
On the other hand, now suppose $\bloomFilterIndicesToSet{} \cap
\bloomFilterIndicesToSet{\requester} \neq \emptyset$, and recall that
\elgCiphertextLog{\bloomFilterBitIdx} for each $\bloomFilterBitIdx \in
\bloomFilterIndicesToSet{} \cap \bloomFilterIndicesToSet{\requester}$
is distributed uniformly and independently in \ints{\groupOrder}.  In
the event $\neg\someEqEvent$, at least $\groupOrder -
\equalityQueries$ values remain equally possible from the adversary's
point of view for each \elgCiphertextLog{\bloomFilterBitIdx},
$\bloomFilterBitIdx \in \bloomFilterIndicesToSet{} \cap
\bloomFilterIndicesToSet{\requester}$, and so $\resultCiphertext{}
\in \ciphertextSpace{\pubKey}(\groupIdentity)$ with probability
\begin{align}
  \cprob{\big}{\resultCiphertext{} \in \ciphertextSpace{\pubKey}(\groupIdentity)}{\bloomFilterIndicesToSet{} \cap \bloomFilterIndicesToSet{\requester} \neq \emptyset \wedge \neg\someEqEvent} \le \frac{1}{\groupOrder - \equalityQueries}
    \label{eqn:prob-ddh}
\end{align}
Moreover, if $\resultCiphertext{}
\not\in \ciphertextSpace{\pubKey}(\groupIdentity)$, \adversary{2} can
succeed with choosing \bloomFilterIndicesToSet{\requester} with
probability
\begin{align}
  \cprob{\Big}
        {\bloomFilterIndicesToSet{\adversary{}} =
          \bloomFilterIndicesToSet{\requester}}
        {\begin{array}{@{\extracolsep{-0.5em}}l@{\extracolsep{0em}}}
            \resultCiphertext{} \not\in \ciphertextSpace{\pubKey}(\groupIdentity)~\wedge \\
            \bloomFilterIndicesToSet{} \cap \bloomFilterIndicesToSet{\requester} \neq \emptyset
            \wedge \neg\someEqEvent
            \end{array}
        }
        \le
\frac{1}{{{\bloomFilterSize} \choose {\bloomFilterHashFns}} -
      {{\bloomFilterSize - \setSize{\bloomFilterIndicesToSet{}}}
        \choose {\bloomFilterHashFns}}}
\label{eqn:prob-intersectionGuess}
\end{align}
So,
\begin{alignat*}{2}
& \mathrlap{\prob{\experiment{\responder{}}{\encScheme}(\adversary{}) = \boolTrue}} \\
  & = &&
  ~\prob{\bloomFilterIndicesToSet{\adversary{}} = \bloomFilterIndicesToSet{\requester}} \\
  & \le &&
  ~\cprob{\big}{\bloomFilterIndicesToSet{\adversary{}} = \bloomFilterIndicesToSet{\requester}}{\neg\someEqEvent} + \prob{\someEqEvent} \\
  & \le &&
  ~\cprob{\big}{\bloomFilterIndicesToSet{\adversary{}} =
  \bloomFilterIndicesToSet{\requester}}{\bloomFilterIndicesToSet{} \cap \bloomFilterIndicesToSet{\requester} = \emptyset \wedge \neg\someEqEvent}
  \cprob{\big}{\bloomFilterIndicesToSet{} \cap \bloomFilterIndicesToSet{\requester} = \emptyset}{\neg\someEqEvent} \\
  &     &&
  +~\cprob{\big}{\bloomFilterIndicesToSet{\adversary{}} =
  \bloomFilterIndicesToSet{\requester}}{\bloomFilterIndicesToSet{} \cap \bloomFilterIndicesToSet{\requester} \neq \emptyset \wedge \neg\someEqEvent}
  \cprob{\big}{\bloomFilterIndicesToSet{} \cap \bloomFilterIndicesToSet{\requester} \neq \emptyset}{\neg\someEqEvent} \\
  &     && +~\prob{\someEqEvent} \\
  & \le &&
  ~\cprob{\big}{\bloomFilterIndicesToSet{\adversary{}} =
  \bloomFilterIndicesToSet{\requester}}{\bloomFilterIndicesToSet{} \cap \bloomFilterIndicesToSet{\requester} = \emptyset \wedge \neg\someEqEvent}
  \cprob{\big}{\bloomFilterIndicesToSet{} \cap \bloomFilterIndicesToSet{\requester} = \emptyset}{\neg\someEqEvent} \\
  &     &&
  +~\left(\begin{array}{r}
    \left[\begin{array}{r}
        \cprob{\Big}
              {\bloomFilterIndicesToSet{\adversary{}} =
                \bloomFilterIndicesToSet{\requester}}
              {\begin{array}{@{\extracolsep{-0.5em}}l@{\extracolsep{0em}}}
                  \resultCiphertext{} \not\in \ciphertextSpace{\pubKey}(\groupIdentity)~\wedge \\
                  \bloomFilterIndicesToSet{} \cap \bloomFilterIndicesToSet{\requester} \neq \emptyset
                  \wedge \neg\someEqEvent
                \end{array}
              } \\
              +~\cprob{\big}{\resultCiphertext{} \in \ciphertextSpace{\pubKey}(\groupIdentity)}{\bloomFilterIndicesToSet{} \cap \bloomFilterIndicesToSet{\requester} \neq \emptyset \wedge \neg\someEqEvent}
      \end{array}\right] \\
    \cdot~\cprob{\big}{\bloomFilterIndicesToSet{} \cap \bloomFilterIndicesToSet{\requester} \neq \emptyset}{\neg\someEqEvent}
  \end{array}\right)\\
  &     && +~\prob{\someEqEvent} \\
  & \le &&
  ~\cprob{\big}{\bloomFilterIndicesToSet{\adversary{}} =
  \bloomFilterIndicesToSet{\requester}}{\bloomFilterIndicesToSet{} \cap \bloomFilterIndicesToSet{\requester} = \emptyset \wedge \neg\someEqEvent}
  \cprob{\big}{\bloomFilterIndicesToSet{} \cap \bloomFilterIndicesToSet{\requester} = \emptyset}{\neg\someEqEvent} \\
  &     &&
  +~\cprob{\Big}
  {\bloomFilterIndicesToSet{\adversary{}} =
    \bloomFilterIndicesToSet{\requester}}
  {\begin{array}{@{\extracolsep{-0.5em}}l@{\extracolsep{0em}}}
      \resultCiphertext{} \not\in \ciphertextSpace{\pubKey}(\groupIdentity)~\wedge \\
      \bloomFilterIndicesToSet{} \cap \bloomFilterIndicesToSet{\requester} \neq \emptyset
      \wedge \neg\someEqEvent
    \end{array}
  }
  \cprob{\big}{\bloomFilterIndicesToSet{} \cap \bloomFilterIndicesToSet{\requester} \neq \emptyset}{\neg\someEqEvent} \\
  &     &&
  +~\cprob{\big}{\resultCiphertext{} \in \ciphertextSpace{\pubKey}(\groupIdentity)}{\bloomFilterIndicesToSet{} \cap \bloomFilterIndicesToSet{\requester} \neq \emptyset \wedge \neg\someEqEvent}
  +~\prob{\someEqEvent}
\end{alignat*}
Filling in the values from \eqnref{eqn:prob-someEqEvent} and
\eqnsref{eqn:prob-noIntersection}{eqn:prob-intersectionGuess} gives
the result.
\end{proof}

\section{Resource Utilization Microbenchmarks}
\label{sec:microbenchmarks}

In this appendix we evaluate the resource utilization imposed by our
protocol in \figref{fig:protocol}.  \figref{fig:resources:computation}
shows the computational burden for computing the query
(\lineref{prot:msg:bloomFilter}) and response
(\lineref{prot:msg:result}) messages in the protocol of
\figref{fig:protocol}.  (In comparison, the computational cost on the
\requesterTerm to process the response is minimal and so is omitted
here.)  Recall from
\secref{sec:eval:implementation:precomputation} that our protocol
implementation leverages precomputation; precomputation costs are not
included in \figref{fig:resources:computation}.  Tor was not used in
these tests.  \figref{fig:resources:bandwidth} shows the size of the
query message (\msgref{prot:msg:bloomFilter}), which is the cost that
dominates the bandwidth use of the protocol, since the response
(\msgref{prot:msg:result}) is only a single ciphertext.

\begin{figure}[t]
  \centering
  \setlength\figureheight{2in}
  \begin{subfigure}[b]{\columnwidth}
    \input{figures/non-tor/sec_param_other.tex}
    \caption{Mean query and response computation times ($\relstddev \le 6\%$)}
    \label{fig:resources:computation}
    \vspace{1.5ex}
  \end{subfigure}
  \begin{subfigure}[b]{\columnwidth}
\begin{tikzpicture}

\definecolor{color0}{rgb}{0.129411764705882,0.380392156862745,0.549019607843137}

\pgfplotsset{every axis/.append style={
					xlabel={\nmbrSimilarPasswords},
					ylabel={Query message size (MB)},
					compat=1.3,
                    label style={font=\small},
                    tick label style={font=\small}  
                    }}

\begin{axis}[
xmin=-0.1, xmax=6,
ymin=0, ymax=10,
width=0.95\figurewidth,
height=\figureheight,
xtick={0.56,1.56,2.56,3.56,4.56,5.56},
xticklabels={$2^7$,$2^8$,$2^9$,$2^{10}$,$2^{11}$,$2^{12}$},
tick align=outside,
tick pos=left,
xmajorgrids,
x grid style={lightgray!92.026143790849673!black},
ymajorgrids,
y grid style={lightgray!92.026143790849673!black}
]
\addlegendimage{ybar,ybar legend,fill=color0,draw opacity=0};
\draw[fill=color0,draw opacity=0] (axis cs:0.07 + 0.1,0) rectangle (axis cs:0.23 + 0.1,0.15864804);
\draw[fill=color0,draw opacity=0] (axis cs:1.07 + 0.1,0) rectangle (axis cs:1.23 + 0.1,0.31683404);
\draw[fill=color0,draw opacity=0] (axis cs:2.07 + 0.1,0) rectangle (axis cs:2.23 + 0.1,0.6507889);
\draw[fill=color0,draw opacity=0] (axis cs:3.07 + 0.1,0) rectangle (axis cs:3.23 + 0.1,1.30112056);
\draw[fill=color0,draw opacity=0] (axis cs:4.07 + 0.1,0) rectangle (axis cs:4.23 + 0.1,2.58423172);
\draw[fill=color0,draw opacity=0] (axis cs:5.07 + 0.1,0) rectangle (axis cs:5.23 + 0.1,5.2558264);
\addlegendimage{ybar,ybar legend,fill=color0,draw opacity=0};
\draw[fill=color0,draw opacity=0] (axis cs:0.23 + 0.11,0) rectangle (axis cs:0.39 + 0.11,0.18694646);
\draw[fill=color0,draw opacity=0] (axis cs:1.23 + 0.11,0) rectangle (axis cs:1.39 + 0.11,0.3734321);
\draw[fill=color0,draw opacity=0] (axis cs:2.23 + 0.11,0) rectangle (axis cs:2.39 + 0.11,0.76711186);
\draw[fill=color0,draw opacity=0] (axis cs:3.23 + 0.11,0) rectangle (axis cs:3.39 + 0.11,1.5337539);
\draw[fill=color0,draw opacity=0] (axis cs:4.23 + 0.11,0) rectangle (axis cs:4.39 + 0.11,3.04631294);
\draw[fill=color0,draw opacity=0] (axis cs:5.23 + 0.11,0) rectangle (axis cs:5.39 + 0.11,6.19570576);
\addlegendimage{ybar,ybar legend,fill=color0,draw opacity=0};
\draw[fill=color0,draw opacity=0] (axis cs:0.39 + 0.12,0) rectangle (axis cs:0.55 + 0.12,0.21722644);
\draw[fill=color0,draw opacity=0] (axis cs:1.39 + 0.12,0) rectangle (axis cs:1.55 + 0.12,0.43387968);
\draw[fill=color0,draw opacity=0] (axis cs:2.39 + 0.12,0) rectangle (axis cs:2.55 + 0.12,0.89126564);
\draw[fill=color0,draw opacity=0] (axis cs:3.39 + 0.12,0) rectangle (axis cs:3.55 + 0.12,1.78195588);
\draw[fill=color0,draw opacity=0] (axis cs:4.39 + 0.12,0) rectangle (axis cs:4.55 + 0.12,3.53924822);
\draw[fill=color0,draw opacity=0] (axis cs:5.39 + 0.12,0) rectangle (axis cs:5.55 + 0.12,7.19827896);
\addlegendimage{ybar,ybar legend,fill=color0,draw opacity=0};
\draw[fill=color0,draw opacity=0] (axis cs:0.55 + 0.13,0) rectangle (axis cs:0.71 + 0.13,0.2466235);
\draw[fill=color0,draw opacity=0] (axis cs:1.55 + 0.13,0) rectangle (axis cs:1.71 + 0.13,0.49271286);
\draw[fill=color0,draw opacity=0] (axis cs:2.55 + 0.13,0) rectangle (axis cs:2.71 + 0.13,1.01217842);
\draw[fill=color0,draw opacity=0] (axis cs:3.55 + 0.13,0) rectangle (axis cs:3.71 + 0.13,2.02380438);
\draw[fill=color0,draw opacity=0] (axis cs:4.55 + 0.13,0) rectangle (axis cs:4.71 + 0.13,4.01972124);
\draw[fill=color0,draw opacity=0] (axis cs:5.55 + 0.13,0) rectangle (axis cs:5.71 + 0.13,8.17556104);

\node at (axis cs:0.21 + 0.172,0.30864804 + 0.722)[
  scale=0.5,
  anchor=south,
  text=black,
  rotate=89.9
]{\bfseries 160-bit};
\node at (axis cs:1.204 + 0.172,0.46683404 + 0.722)[
  scale=0.5,
  anchor=south,
  text=black,
  rotate=89.9
]{\bfseries 160-bit};
\node at (axis cs:2.198 + 0.172,0.8007889 + 0.722)[
  scale=0.5,
  anchor=south,
  text=black,
  rotate=89.9
]{\bfseries 160-bit};
\node at (axis cs:3.192 + 0.172,1.45112056 + 0.722)[
  scale=0.5,
  anchor=south,
  text=black,
  rotate=89.9
]{\bfseries 160-bit};
\node at (axis cs:4.186 + 0.172,2.73423172 + 0.722)[
  scale=0.5,
  anchor=south,
  text=black,
  rotate=89.9
]{\bfseries 160-bit};
\node at (axis cs:5.18 + 0.172,5.4058264 + 0.722)[
  scale=0.5,
  anchor=south,
  text=black,
  rotate=89.9
]{\bfseries 160-bit};
\node at (axis cs:0.35 + 0.172,0.33694646 + 0.722)[
  scale=0.5,
  anchor=south,
  text=black,
  rotate=89.9
]{\bfseries 192-bit};
\node at (axis cs:1.344 + 0.197,0.5234321 + 0.722)[
  scale=0.5,
  anchor=south,
  text=black,
  rotate=89.9
]{\bfseries 192-bit};
\node at (axis cs:2.338 + 0.197,0.91711186 + 0.722)[
  scale=0.5,
  anchor=south,
  text=black,
  rotate=89.9
]{\bfseries 192-bit};
\node at (axis cs:3.332 + 0.197,1.6837539 + 0.722)[
  scale=0.5,
  anchor=south,
  text=black,
  rotate=89.9
]{\bfseries 192-bit};
\node at (axis cs:4.326 + 0.197,3.19631294 + 0.722)[
  scale=0.5,
  anchor=south,
  text=black,
  rotate=89.9
]{\bfseries 192-bit};
\node at (axis cs:5.32 + 0.197,6.34570576 + 0.722)[
  scale=0.5,
  anchor=south,
  text=black,
  rotate=89.9
]{\bfseries 192-bit};
\node at (axis cs:0.49 + 0.197,0.36722644 + 0.722)[
  scale=0.5,
  anchor=south,
  text=black,
  rotate=89.9
]{\bfseries 224-bit};
\node at (axis cs:1.484 + 0.247,0.58387968 + 0.722)[
  scale=0.5,
  anchor=south,
  text=black,
  rotate=89.9
]{\bfseries 224-bit};
\node at (axis cs:2.478 + 0.247,1.04126564 + 0.722)[
  scale=0.5,
  anchor=south,
  text=black,
  rotate=89.9
]{\bfseries 224-bit};
\node at (axis cs:3.472 + 0.247,1.93195588 + 0.722)[
  scale=0.5,
  anchor=south,
  text=black,
  rotate=89.9
]{\bfseries 224-bit};
\node at (axis cs:4.466 + 0.247,3.68924822 + 0.722)[
  scale=0.5,
  anchor=south,
  text=black,
  rotate=89.9
]{\bfseries 224-bit};
\node at (axis cs:5.46 + 0.247,7.34827896 + 0.722)[
  scale=0.5,
  anchor=south,
  text=black,
  rotate=89.9
]{\bfseries 224-bit};
\node at (axis cs:0.63 + 0.272,0.3966235 + 0.722)[
  scale=0.5,
  anchor=south,
  text=black,
  rotate=89.9
]{\bfseries 256-bit};
\node at (axis cs:1.624 + 0.272,0.64271286 + 0.722)[
  scale=0.5,
  anchor=south,
  text=black,
  rotate=89.9
]{\bfseries 256-bit};
\node at (axis cs:2.618 + 0.272,1.16217842 + 0.722)[
  scale=0.5,
  anchor=south,
  text=black,
  rotate=89.9
]{\bfseries 256-bit};
\node at (axis cs:3.612 + 0.272,2.17380438 + 0.722)[
  scale=0.5,
  anchor=south,
  text=black,
  rotate=89.9
]{\bfseries 256-bit};
\node at (axis cs:4.606 + 0.272,4.16972124 + 0.722)[
  scale=0.5,
  anchor=south,
  text=black,
  rotate=89.9
]{\bfseries 256-bit};
\node at (axis cs:5.6 + 0.272,8.32556104 + 0.722)[
  scale=0.5,
  anchor=south,
  text=black,
  rotate=89.9
]{\bfseries 256-bit};
\end{axis}

\end{tikzpicture}
    \caption{Query message size ($\relstddev < 0.02\%$)}
    \label{fig:resources:bandwidth}
  \end{subfigure}
  \caption{Resource usage for various elliptic curves and
    numbers \nmbrSimilarPasswords of similar passwords}
  \label{fig:resources}
\end{figure}

We caution the reader in interpreting these figures that the resource
costs for large values of \nmbrSimilarPasswords are included for
completeness and to inform the optimization in
\secref{sec:eval:optimization}.  For reasons we discuss in
\secref{sec:eval:optimization}, such large values of \nmbrSimilarPasswords
will generally not be necessary in our protocol.

One peculiarity evident in \figref{fig:resources:computation} is that
the \responderTerm's computational cost is better when using the
256-bit elliptic curve than using the 224-bit one. This anomaly is
caused by the point compression technique (see
\secref{sec:eval:implementation:crypto}): to recover the points'
$y$ coordinates from received EC-ElGamal ciphertexts
$\{\bloomFilterBitCtext{\bloomFilterBitIdx}\}_{\bloomFilterBitIdx\in\residues{\bloomFilterSize}}$,
the \responderTerm needs to calculate square roots of $y^2 = x^3 + ax
+ b$ over the field \ints{\ecPrime} for prime \ecPrime.  If $\ecPrime
\ringEquiv{4} 3$, then $(y^2)^\frac{\ecPrime+1}{4}$ immediately gives
the solution. However, if $\ecPrime \ringEquiv{4} 1$, then one needs
to use other less efficient algorithms to find the solution and,
unfortunately, secp224r1 (NIST P-224) happens to be this case. Query
generation involves no point decompression and so is not subject to
this peculiarity.
\fi

\end{document}